\documentclass[10pt,twocolumn]{article}
\usepackage[utf8]{inputenc}

\usepackage{microtype}
\usepackage{times}        
\usepackage{geometry}
\usepackage{graphicx}
\usepackage{subcaption}
\usepackage{booktabs}
\usepackage{multirow}
\usepackage{adjustbox}
\usepackage{float}
\usepackage[table]{xcolor}
\usepackage{colortbl}

\usepackage{amsmath,amssymb,mathtools,amsthm}

\usepackage[round,comma]{natbib}
\usepackage[hidelinks]{hyperref}

\usepackage{soul}
\usepackage[textsize=tiny]{todonotes}
\usepackage[capitalize,noabbrev]{cleveref}
\DeclareUnicodeCharacter{03BC}{\ensuremath{\mu}}

\theoremstyle{plain}
\newtheorem{theorem}{Theorem}[section]
\newtheorem{proposition}[theorem]{Proposition}

\newtheorem{corollary}[theorem]{Corollary}
\theoremstyle{definition}

\theoremstyle{remark}
\newtheorem{remark}[theorem]{Remark}

\title{Coupled Integral PINN for Discontinuity}
\author{
  Yeping Wang \\
  \and
  Shihao Yang \\
}

\date{} 

\begin{document}

\maketitle
\begin{abstract}
Physics-Informed Neural Networks (PINNs) solve forward PDEs by minimizing residual losses from the governing equations with initial and boundary conditions, but they often struggle with discontinuities such as shocks. In contrast, finite volume methods (FVM) handle discontinuities by enforcing integral conservation, which admits weak solutions. Motivated by this, we propose a \emph{Coupled Integral PINN (CI-PINN)} that augments a standard PINN with an auxiliary network for integral potentials and coupled integral constraints. This improves robustness near shocks while avoiding meshing and the numerical flux integration/reconstruction used in classical schemes. We validate CI-PINN on forward benchmarks including Burgers, Buckley--Leverett, the Euler system, and the Shallow-Water equations.
\end{abstract}
\vspace{0.5cm}

\section{Introduction}
\label{submission}

Physics-Informed Neural Networks (PINNs)~\citep{raissi2019, raissi2017} have become a central building block in scientific machine learning (SciML), providing a flexible way to fuse neural function approximation with the structure of governing differential equations~\citep{karniadakis2021physics, Lu_2021}. PINN embeds the residual of a partial differential equation (PDE) into the training objective via automatic differentiation. While PINNs have proven effective in modeling diverse physical systems, they exhibit well-documented limitations when the underlying solution exhibits sharp fronts or discontinuities~\citep{krishnapriyan2021characterizingpossiblefailuremodes, fuks2020limitations}. (Despite PINN's success in many application areas, applying it to problems involving sudden changes or discontinuities presents significant challenges~\citep{karniadakis2021physics}.) In this work we try to theoretically investigate why PINN fails on PDE systems with discontinuities. To better understand the challenge these PDEs impose on PINN, we leave out the data loss in PINN and therefore focus on the forward problem without interior observational data. Our goal is to investigate why PINN cannot approximate PDE solutions directly from the governing equations and prescribed initial/boundary conditions for systems with discontinuities, and propose an approach that addresses the issue.

We argue that the root cause is a fundamental mismatch between the \emph{strong-form residual objective} and the \emph{hyperbolic nature of the target solution}, using a viewpoint from an optimization and function-approximation perspective. PINN training inherently fails to converge to the true solution in the presence of discontinuities. We formalize this failure mode through two key theoretical insights that PINN strong-form is not identifiable, and that PINN physics loss diverges near shocks.

Classical numerical analysis does not have this issue. This is because these regimes are handled by enforcing PDE structure through integral flux-balance principles (e.g., finite volume methods), which remain meaningful at shocks. This motivates us to propose integral-form PINN constraints that extend PINNs to shock-dominated problems, aiding strong-form PINN objectives so that PINN can behave in the discontinuous regime.


\subsection{Our Contribution}
In this work, we theoretically investigate why vanilla PINNs fail in discontinuity regime. Given the lessons, we propose to enforce the conservation law in its integral form via auxiliary networks, we naturally handle discontinuities without derivative singularities. We name our architecture \textbf{Coupled Integral PINN (CI-PINN)}, which explicitly decouples the state variable from its integral representation. Our specific contributions are:\\
(1) \textbf{Optimization analysis:} We provide a theoretical analysis of why strong-form PINNs fail at shocks, proving the existence of an optimization barrier that penalizes convergence to the entropy solution.\\
(2) \textbf{Algorithm Design:} We introduce a dual-network architecture that enforces the integral conservation law directly, ensuring that the loss function remains bounded and smooth even in the presence of sharp shocks, meanwhile because CI-PINN avoids any explicit local or global numerical quadrature, it can be seamlessly integrated with existing PINN enhancements—such as domain decomposition, adaptive weighting, or curriculum sampling—without additional algorithmic overhead. \\
(3) \textbf{Empirical Performance:} We demonstrate that CI-PINN significantly outperforms standard PINN variants on benchmark hyperbolic systems (Burgers, Buckley-Leverett, Euler), effectively capturing sharp shock interfaces and accurately recovering the ground truth entropy solution where traditional PINNs fail.


The proposed CI‑PINN offers three principal advantages:

\textbf{(1) Better accuracy.} On benchmark hyperbolic conservation laws, CI‑PINN captures both shock and rarefaction waves accurately, and achieve best accuracy on most of evaluation benchmark.\\
\textbf{(2) Automatic conservation.} By enforcing a pointwise integral form of the PDE during training, the network simultaneously approximates the integral, local, and global conservation statements once convergence is reached.\\
\textbf{(3) Algorithmic flexibility.} CI-PINN avoids any explicit local or global numerical quadrature. The loss are evaluated on mesh-less point, it can be seamlessly integrated with existing PINN enhancements without additional algorithmic overhead.

The remainder of this paper is organized as follows. Section~\ref{sec:analysis} introduces the mechanism of the shock waves, then delivers an argument why vanilla PINNs almost fail at every discontinuity and how CI-PINN overcome this.
Section~\ref{sec:metho} present the proposed dual network architecture and how the consistency between the neural networks and the true physical solution are enforced. Section~\ref{sec:result} presents experimental results on classical benchmark problems, highlighting the advantages of Coupled Integral PINNs over other methods. Finally, Section~\ref{sec:conclusion} concludes and outlines future work.

\subsection{Related Work}

The difficulty of applying strong-form PINNs to discontinuous solutions has been widely recognized. 
For inviscid Burgers’ equation, standard PINNs often converge to overly smooth surrogates rather than the correct shock profile, and similar failures occur in more complex hyperbolic settings with moving shocks and compound interactions~\citep{diab2021pinnssolutionhyperbolicbuckleyleverett}. 
These limitations affect \emph{both} forward and inverse problems: while additional data may partially constrain inverse solutions, it does not resolve the fundamental mismatch between pointwise residual minimization and the admissible weak/entropy solution concept~\citep{krishnapriyan2021characterizingpossiblefailuremodes,fuks2020limitations}.

A large body of work attempts to improve training stability without changing the strong-form objective. 
Domain decomposition methods (e.g., XPINNs and related multi-network or overlapping-subdomain strategies) localize approximation and improve scalability~\citep{jagtap2020extended,jagtap2020conservative}. 
Adaptive or clustered residual sampling increases collocation density in hard regions (often near shocks)~\citep{MAO2020112789}, and gradient-based weighting further rebalances training signals around discontinuities~\citep{liu2024discontinuity}. 
Curriculum/causal schedules and architectural modifications address temporal stiffness and gradient pathologies~\citep{wang2022respecting,wang2020understandingmitigatinggradientpathologies,wang2022randomweightfactorizationimproves}. 
Although these approaches can reduce variance and improve robustness, they typically do \emph{not} remove the core issue that the strong-form residual becomes a poor learning signal as interfaces sharpen, leading to smeared shocks or distorted plateau levels.

Another direction replaces pointwise residuals with weak/variational formulations that are better aligned with discontinuous solution concepts. 
Examples include Ritz--Galerkin and related variational solvers~\citep{yu2018deep}, mortar/domain-interface formulations~\citep{jagtap2020extended}, and Petrov--Galerkin or hp-variational PINNs~\citep{kharazmi2021hp,KHARAZMI2021113547}. 
Weak-PINNs further target weak residuals and error control~\citep{deryck2022wpinns,chaumet2023improving}. 
These methods soften derivative singularities by shifting differentiation onto smoother test functions, but their effectiveness depends on test-space design and numerical quadrature accuracy; moreover, for nonlinear conservation laws, weak satisfaction alone does not guarantee uniqueness without an admissibility mechanism (e.g., entropy selection), and multidimensional implementations can be computationally heavy.

In the specific context of conservation laws, many methods enforce integral flux balance by discretizing space--time and applying numerical quadrature (e.g., trapezoidal rules or finite-volume-style control volumes) during training~\citep{mei2024unified,jeon2021fvm,PATEL2022110754,li2023finite,li2024predicting}. 
These approaches substantially improve shock consistency, and cvPINN~\citep{PATEL2022110754} has become a representative \emph{standard} baseline for conservation enforcement via local control volumes and numerical quadrature. 
However, such formulations can require careful discretization, quadrature, and sometimes reconstruction choices, and they primarily enforce local balance rather than directly learning a global integral representation.

Our CI-PINN targets the \emph{forward} solution of nonlinear conservation laws in a mesh-free manner by fitting an auxiliary potential network and enforcing coupled integral constraints without explicit space--time discretization, quadrature, or flux reconstruction. 
This also distinguishes our setting from supervised operator learning (e.g., DeepONet~\citep{DBLP:journals/corr/abs-1910-03193} and neural operators~\citep{li2020fourier,gupta2021multiwavelet}) and hybrid operator learners (e.g., PINO~\citep{li2024physics,goswami2023physics}) and their variants, since we focus on unsupervised physics-based forward solving without interior observational data. 
In experiments, we compare against general optimization enhancements (Adaptive Learning Rate~\citep{wang2023experts}, Causal Training~\citep{wang2022respecting}, Modified MLP~\citep{wang2020understandingmitigatinggradientpathologies}, Random Weight Factorization~\citep{wang2022randomweightfactorizationimproves}) and conservation-specific formulations including cvPINN~\citep{PATEL2022110754}, IPINN~\citep{rajvanshi2024integral}, and a small-scale comparison with Bayesian ProbConserv~\citep{HANSEN2024133952} in ~\ref{sec:appendH}.

\section{Failure of PINNs for Discontinuous PDE Solutions in $L^2(\Omega)$
}
\label{sec:analysis}
We analyze a fundamental failure mode of strong-form PINNs when the target solution is discontinuous. In this section we will use $\mathbf{q}$ instead of the typical $\mathbf{u}$ to represent conservative quantities to split out from the primitive element in section~\ref{sec:metho}
\subsection{Preliminaries: Why shocks form}

Shocks arise most naturally in \emph{nonlinear hyperbolic conservation laws}, i.e., PDEs that admit the conservative form $\mathbf{q}_t + \nabla\!\cdot\mathbf{F}(\mathbf{q})=0$ (or $q_t + f(q)_x = 0$) and exhibit wave-like propagation.
For example, in the scalar conservation law $q_t + f(q)_x = 0$, smooth solutions propagate along characteristic curves with speed $f'(q)$.
When $f'(q)$ varies with the state, different parts of the wave travel at different speeds, leading to compression: faster characteristics overtake slower ones, the profile steepens, and the gradient $q_x$ blows up in finite time.
Physically, this corresponds to wave steepening in compressible media, where high-pressure (or high-density) regions propagate faster and ``pile up'' into a thin dissipative layer, producing an irreversible jump.
Beyond this point, classical (strong) solutions cease to exist and the physically relevant solution must be interpreted in the weak sense(integral). 
Conservation laws appear broadly in science and engineering, including fluid dynamics, transport, and porous-media flow.
Their integral formulation is obtained by integrating the differential form over a control volume $\Omega\subset\mathbb{R}^d$ and applying the divergence theorem:
\begin{equation}
\frac{\mathrm{d}}{\mathrm{d}t}\int_{\Omega} q(\mathbf{x},t)\,\mathrm{d}\Omega
\;=\;- \int_{\partial\Omega}\mathbf{F}(q)\cdot\mathbf{n}\,\mathrm{d}\Gamma\,,
\label{eq:cons-int}
\end{equation}
where $\partial\Omega$ is the boundary of $\Omega$ and $\mathbf{n}$ denotes the outward unit normal.
A detailed derivation is provided in Appendix~\ref{sec:appendB}.

\subsection{PDE and strong-form PINN objective}

Let $\mathbf{q}:\Omega\times(0,T)\to\mathbb{R}^m$ be the unknown field on $\Omega\subset\mathbb{R}^d$.
We consider nonlinear hyperbolic PDEs that admit discontinuous physical solutions, written in a standard flux-form operator where $\mathbf{F}$ is the flux operator:
\begin{equation}
\label{eq:hyperbolic_flux_form}
\partial_t \mathbf{q} + \nabla\cdot \mathbf{F}(\mathbf{q}) = \mathbf{0},
\qquad
\mathbf{F}(\mathbf{q})\in\mathbb{R}^{m\times d}.
\end{equation}
Strong-form PINNs typically minimize the squared residual
\begin{equation}
\begin{split}
\mathcal{L}_{\mathrm{strong}}(\mathbf{q}_\theta)
&= \left\|\partial_t \mathbf{q}_\theta + \nabla\cdot \mathbf{F}(\mathbf{q}_\theta)\right\|_{L^2(\Omega_T)}^2, \\
&\quad \Omega_T := \Omega \times (0,T),
\end{split}
\label{eq:strong_form_loss}
\end{equation}
where $\mathbf{q}_\theta$ is a smooth neural approximation.

\subsection{Core pathologies of strong-form training}

\textbf{Non-Uniqueness of Strong-Form Minimizers:} Since the true shock contains a derivative singularity, the strong-form loss is unbounded at the physical solution and have very large loss on the approximating solution. Consequently, neural networks are biased toward spurious, smooth approximations that yield deceptively low residuals but incorrect physics (see Proposition \ref{prop:spurious_minimizers}). From optimization perspective, this means physical solution is not global minimum. 

\textbf{The Optimization Barrier:} We reveal that the standard PINN loss is ``inverse-consistent'' near a shock—as a candidate solution approaches the true sharp interface, the residual loss strictly \textit{diverges}. This implies that the true solution is not a local minimum in the standard loss landscape; rather, the gradient actively repels the optimizer away from the shock (see Proposition \ref{prop:blowup_shock}).

\begin{proposition}[Spurious smooth low-loss solutions dominate the strong-form objective]
\label{prop:spurious_minimizers}
Let $\mathbf{q}^*$ be a physically relevant discontinuous weak solution of \eqref{eq:hyperbolic_flux_form}
containing at least one jump discontinuity.
Consider minimizing the strong-form residual objective over a smooth hypothesis class:
\[
\mathcal{L}_{\mathrm{strong}}(\tilde{\mathbf{q}})
:=
\left\|\partial_t \tilde{\mathbf{q}} + \nabla\!\cdot \mathbf{F}(\tilde{\mathbf{q}})\right\|_{L^2(\Omega_T)}^2,
\qquad \tilde{\mathbf{q}}\in C^1(\Omega_T).
\]
Then there exist smooth functions $\tilde{\mathbf{q}}$ with \emph{arbitrarily small} strong-form loss
even though $\tilde{\mathbf{q}}$ does not reproduce the discontinuous structure of $\mathbf{q}^*$
(e.g., it yields an overly smeared shock/front).
Consequently, minimizing $\mathcal{L}_{\mathrm{strong}}$ is non-selective within smooth classes and can favor diffused,
nonphysical smooth states.
\end{proposition}

\begin{remark}
Proposition~\ref{prop:spurious_minimizers} is an \emph{existence / non-selectivity} statement:
a small (even vanishing) interior strong-form residual does not certify that the learned solution matches the
physically relevant discontinuous weak solution.
Intuitively, the residual contains spatial derivatives; if a shock/front is spread over a wider layer, the gradients shrink,
and an $L^2$ objective can be made small even though the interface is severely smeared.

\end{remark}

\begin{proposition}[Strong-form blow-up as a discontinuity sharpens]
\label{prop:blowup_shock}
Let $\mathbf{q}^*$ be a discontinuous physical solution containing a jump of amplitude $\Delta \mathbf{q}\neq \mathbf{0}$ across an interface.
Let $\{\mathbf{q}^\varepsilon\}_{\varepsilon>0}$ be a family of smooth approximations that resolves the jump over a transition layer of thickness $\varepsilon$.
Assume $\mathbf{F}$ is continuously differentiable and its Jacobians remain bounded on the relevant state range.
Then the strong-form residual scales as
\begin{equation}
\label{eq:strong_blowup_rate}
\mathcal{L}_{\mathrm{strong}}(\mathbf{q}^\varepsilon)
=
\left\|\partial_t \mathbf{q}^\varepsilon + \nabla\cdot \mathbf{F}(\mathbf{q}^\varepsilon)\right\|_{L^2(\Omega_T)}^2
\;\gtrsim\;
\frac{C}{\varepsilon},
\ \varepsilon\to 0,
\end{equation}
for some $C>0$ depending on $\Delta\mathbf{q}$ and $\mathbf{F}$.
Consequently, sharpening an approximation toward the correct discontinuity (\emph{smaller} $\varepsilon$) \emph{increases} the strong-form penalty.
\end{proposition}

\begin{proof}
When a fixed jump of magnitude $\|\Delta \mathbf{q}\|$ between two levels must occur within a transition layer of thickness $\varepsilon$, the gradient (slope) of this transition necessarily scales as $\|\Delta \mathbf{q}\|/\varepsilon$. That is, within the $\mathcal{O}(\varepsilon)$-thick transition region, gradients scale as
$
\|\nabla \mathbf{q}^\varepsilon\|
\sim
\mathcal{O}\!\left(\frac{\|\Delta \mathbf{q}\|}{\varepsilon}\right).
$
By the chain rule,
\[
\nabla\cdot \mathbf{F}(\mathbf{q}^\varepsilon)
=
\sum_{j=1}^d \partial_{x_j}\mathbf{F}_j(\mathbf{q}^\varepsilon)
=
\sum_{j=1}^d \mathbf{J}_{\mathbf{F}_j}(\mathbf{q}^\varepsilon)\,\partial_{x_j}\mathbf{q}^\varepsilon,
\]
so bounded Jacobians imply
$\|\nabla\cdot \mathbf{F}(\mathbf{q}^\varepsilon)\|\sim \mathcal{O}(1/\varepsilon)$
inside the layer. Hence,
$
\|\partial_t \mathbf{q}^\varepsilon+\nabla\cdot \mathbf{F}(\mathbf{q}^\varepsilon)\|
\sim \mathcal{O}(1/\varepsilon)
$
over a region of spatial measure $\mathcal{O}(\varepsilon)$ and time measure $\mathcal{O}(1)$, yielding
$
\mathcal{L}_{\mathrm{strong}}(\mathbf{q}^\varepsilon)
\;\gtrsim\;
(\varepsilon)\cdot \left(\frac{1}{\varepsilon}\right)^2
=
\frac{1}{\varepsilon}.
$
\end{proof}

This completes the proof of  proposition~\ref{prop:blowup_shock} and ~\ref{prop:spurious_minimizers}, we provide a detailed proof sketch in appendix~\ref{appendix:proof}

\begin{corollary}[The Optimization barrier]
\label{cor:optimization_paradox}
Combining Propositions \ref{prop:spurious_minimizers} and \ref{prop:blowup_shock} establishes a fundamental pathology in the strong-form optimization landscape. Specifically, $\mathcal{L}_{\mathrm{strong}}$ is \emph{inverse-consistent} with discontinuity sharpness:
\textbf{Divergence at the Solution ($\varepsilon \to 0$):} As the approximation $\mathbf{q}^\varepsilon$ converges to the physical sharp shock $\mathbf{q}^*$, the loss diverges: $\mathcal{L}_{\mathrm{strong}} \to \infty$.
\textbf{Convergence to Artifacts ($\varepsilon \to \infty$):} As the approximation deviates from physics via excessive smoothing, the loss converges to a global minimum: $\mathcal{L}_{\mathrm{strong}} \to 0$.

This creates an optimization barrier: minimizing the strong-form residual inherently penalizes physical accuracy and drives the optimizer toward non-physical, smeared solutions.
\end{corollary} 

\begin{remark}
\label{remark:ichold}
(Adding an initial-condition penalty does not remove the strong-form pathology).
Adding an initial-condition (IC) loss anchors the state at $t=0$, but it does not alter the interior scaling mechanism in
Proposition~\ref{prop:blowup_shock}: for any smooth approximation that must realize a fixed jump by a transition layer of
thickness $\varepsilon$, the strong-form contribution still satisfies
$
\mathcal{L}_{\mathrm{strong}}(\mathbf{q}^\varepsilon)\ \gtrsim\ \frac{1}{\varepsilon}\qquad (\varepsilon\to 0).
$
Hence IC supervision does not remove the blow-up barrier associated with sharpening discontinuities, and the optimizer may still
prefer overly diffused profiles at later times.
\end{remark}
\subsection{Why the Coupled Integral PINN resolves the failure mode}

\begin{remark}[How CI-PINN avoids strong-form shock blow-up]
The key difficulty for classical PINNs on hyperbolic problems is that a sharp jump in the state $\mathbf{q}$ forces derivatives to scale like $1/\varepsilon$ inside an $\mathcal{O}(\varepsilon)$ shock layer, which can make strong-form residual penalties grow as the discontinuity sharpens (Proposition~\ref{prop:blowup_shock}).
CI-PINN avoids this failure mode by enforcing the PDE through \emph{coupled integral (potential) variables} rather than directly squaring shock-amplified derivatives of the primitive/state fields.

Concretely, we learn a primitive network $\mathbf{u}_\theta$ and a potential network $\mathbf{S}_\phi=\{\mathbf{S}_{\phi,j}\}_{j=1}^d$, define $\mathbf{q}_\theta=\mathbf{q}(\mathbf{u}_\theta)$, and impose the coupled constraints
\[
\mathbf{q}_\theta \approx \nabla\cdot \mathbf{S}_\phi,
\qquad
\partial_t \mathbf{S}_{\phi,j} + \mathbf{F}_j(\mathbf{q}_\theta) \approx \mathbf{0}, \;\; j=1,\dots,d,
\]
which correspond exactly to the \emph{coupled-loss} and \emph{physical-loss} terms in our objective.
Because $\mathbf{S}_\phi$ acts as an antiderivative of $\mathbf{q}_\theta$ (through $\nabla\cdot \mathbf{S}_\phi \approx \mathbf{q}_\theta$), a jump in $\mathbf{q}$ typically manifests as a much milder non-smoothness in $\mathbf{S}$ (often a continuous function with a kink), so the training signal does not require differentiating the discontinuous state itself.
In turn, the dominant contributions in the physical-loss arise from $\partial_t \mathbf{S}_{\phi,j}$ and the flux evaluation $\mathbf{F}_j(\mathbf{q}_\theta)$, rather than from squaring $1/\varepsilon$-scale spatial derivatives of $\mathbf{q}$.
This removes the main driver behind strong-form blow-up and makes shock sharpening compatible with optimization. Finally, other loss terms involved are enforced to be stable.
\end{remark}

\noindent\textbf{On the remaining kink in $\mathbf{S}$.}
Even if $\mathbf{S}$ is only continuous with a corner at the shock, this non-smoothness is mild: $\mathbf{S}$ remains Lipschitz and is smooth almost everywhere, with the kink confined to a codimension-one interface (measure zero in $\Omega_T$). 
As a result, $L^2$ losses involving only first derivatives of $\mathbf{S}$ stay finite and well-behaved, unlike strong-form losses on $\mathbf{q}$ whose gradients can scale like $\mathcal{O}(1/\varepsilon)$ when a jump is represented by a vanishing-thickness layer.


\section{Methodology}
\label{sec:metho}

\subsection{The CI-PINN method}

\begin{figure}[t]
    \centering
    \includegraphics[width=1\linewidth]{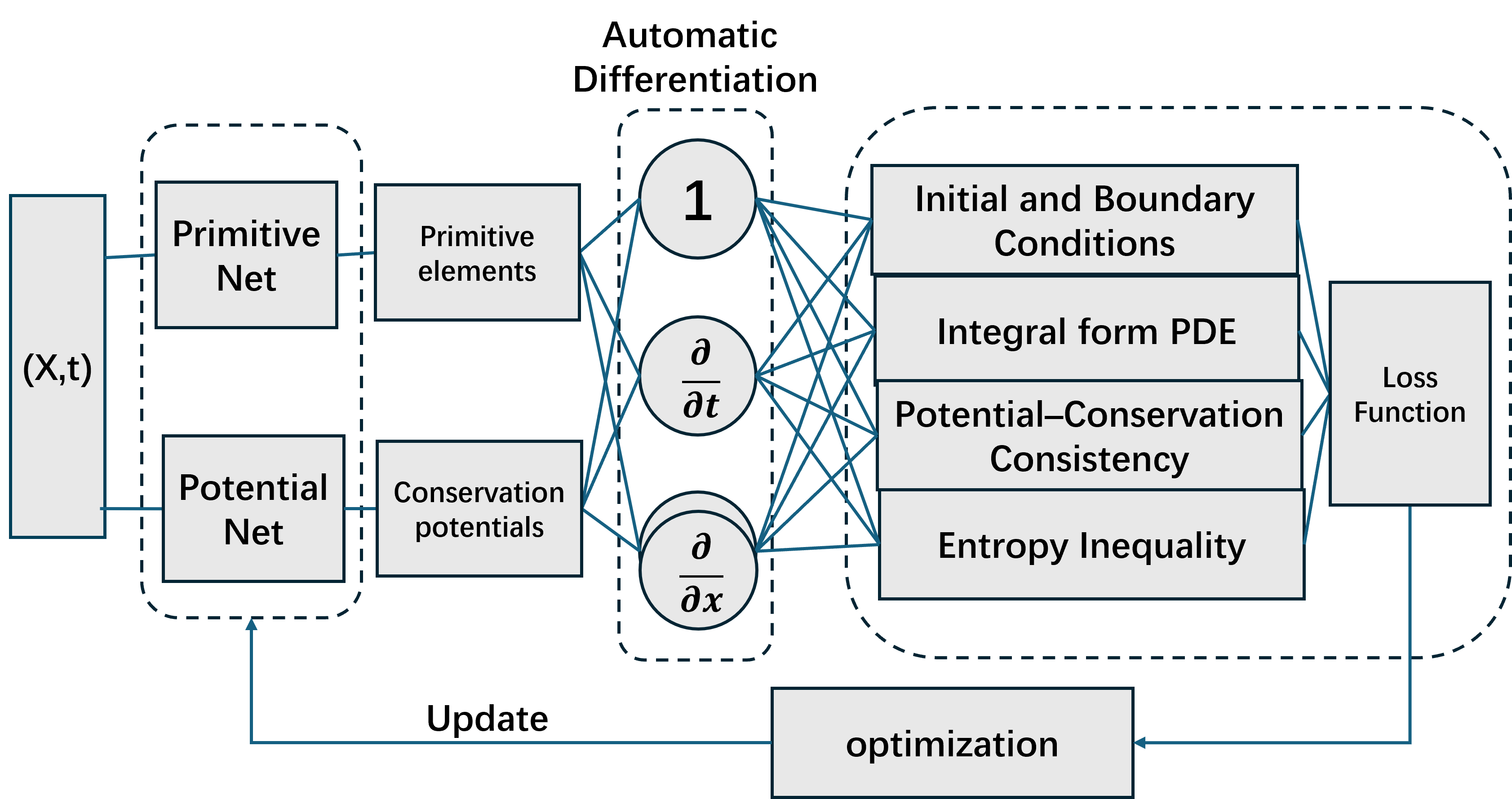}
    \caption{CI-PINN architecture to solve the forward problem of nonlinear PDEs.}
    \label{fig:enter1}
\end{figure}

Inspired by the integral form and the corresponding FVM in numerical methods, the Coupled Integral PINN (CI-PINN) employs two fully connected neural networks (see Figure~\ref{fig:enter1}). The first network, \(\tilde{\mathbf{u}}(x,t)\), approximates the \emph{primitive} state variables of the solution (e.g., density, velocity, pressure in compressible flow). Since the governing PDE is typically written in conservative form, we define the corresponding \emph{conserved} variables through a deterministic mapping
\[
\mathbf{q}(x,t) = \mathbf{q}(\mathbf{u}(x,t)),
\qquad 
\tilde{\mathbf{q}}(x,t) := \mathbf{q}(\tilde{\mathbf{u}}(x,t)).
\]
The second network learns a potential (integral) representation \(\tilde{\mathbf{S}}(x,t)=\{\tilde{\mathbf{S}}_j(x,t)\}_{j=1}^d\) associated with the conserved quantities. In the general $d$-dimensional setting, CI-PINN enforces the coupled relations
\[
\tilde{\mathbf{q}}(x,t)\approx \nabla\cdot \tilde{\mathbf{S}}(x,t),
\]
\[
\partial_t \tilde{\mathbf{S}}_j(x,t) + \mathbf{F}_j(\tilde{\mathbf{q}}(x,t)) \approx \mathbf{0}, \;\; j=1,\dots,d,
\]
which correspond to an integral-form relaxation of the conservation law
\(\partial_t \mathbf{q} + \nabla\cdot \mathbf{F}(\mathbf{q}) = \mathbf{0}\).
The required derivatives (e.g., \(\nabla\cdot \tilde{\mathbf{S}}\) and \(\partial_t\tilde{\mathbf{S}}\)) are computed through automatic differentiation. Because differentiation is applied to the learned potential rather than directly to discontinuous primitive states, the proposed method remains compatible with shock solutions while retaining a simple neural architecture.

As a special case in 1D scalar conservation laws, the primitive and conserved variables coincide ($\mathbf{q}=u$). The potential reduces to a scalar function $S(x,t)$ defined as the spatial antiderivative of the solution, implying $u(x,t) = \partial_x S(x,t)$. For instance, in the inviscid Burgers' equation (where flux $f(u) = u^2/2$), the \textbf{Primitive Net} maps spacetime coordinates $(x,t)$ to the state variable $u$, while the \textbf{Potential Net} maps $(x,t)$ to the potential $S$ (the integral of the conserved quantity). These networks are trained jointly to enforce consistency ($\partial_x S \approx u$) and the integral conservation law ($\partial_t S + u^2/2 \approx 0$).

Training CI-PINN amounts to enforcing four conditions that mirror the four constraint blocks in Figure~\ref{fig:enter1}. 
First, the primitive network output \(\tilde{\mathbf{u}}(x,t)\) must satisfy the prescribed initial and boundary data. 
Second, the learned potential \(\tilde{\mathbf{S}}(x,t)\) is required to be consistent with the conserved variables through \(\nabla\!\cdot\tilde{\mathbf{S}}(x,t)\approx \tilde{\mathbf{q}}(x,t)\), where \(\tilde{\mathbf{q}}=\mathbf{q}(\tilde{\mathbf{u}})\). 
Third, conservation is enforced in integral form by requiring \(\partial_t\tilde{\mathbf{S}}_j(x,t)+\mathbf{F}_j(\tilde{\mathbf{q}}(x,t))\approx \mathbf{0}\) for \(j=1,\dots,d\). 
Finally, to select the physically admissible weak solution in the presence of shocks, we impose the entropy inequality. When jointly enforced, these four constraints are sufficient to recover the unique entropy solution almost everywhere; removing the consistency constraint leads to architectural degeneracy while removing any of the other three leads to non-uniqueness. 

Figure~\ref{fig:enter1} illustrates the architecture of CI-PINN, where normalized spatial and temporal coordinates serve as inputs to the two MLPs. The outputs, \(\tilde{\mathbf{u}}(x,t)\) and \(\tilde{\mathbf{S}}(x,t)\), are optimized to satisfy 4 conditions mentioned above. Under the PINN setting, these four conditions in Figure~\ref{fig:enter1} are realized by optimizing the summation of four losses.

The total loss \(L\) is given by:
\begin{equation}
\begin{split}
L
&= \lambda_{\mathrm{ibc}}\underbrace{\frac{1}{N_b}\sum_{j=1}^{N_b}
\big\|\mathbf{u}_{\mathrm{IBC}}(x_j,t_j)-\tilde{\mathbf{u}}(x_j,t_j)\big\|_2^2}_{\text{initial-boundary loss}}
\\
&\quad + \lambda_{\mathrm{phy}}\underbrace{\frac{1}{N_f\,d}\sum_{i=1}^{N_f}\sum_{k=1}^{d}
\big\|\partial_t \tilde{\mathbf{S}}_{k}(x_i,t_i)+\mathbf{F}_{k}\!\big(\tilde{\mathbf{q}}(x_i,t_i)\big)\big\|_2^2}_{\text{integral-form (physical) loss}}
\\
&\quad + \lambda_{\mathrm{cpl}}\underbrace{\frac{1}{N_f}\sum_{i=1}^{N_f}
\big\|\nabla\!\cdot\!\tilde{\mathbf{S}}(x_i,t_i)-\tilde{\mathbf{q}}(x_i,t_i)\big\|_2^2}_{\text{coupling loss}}
\\
&\quad + \lambda_{\mathrm{ent}}\underbrace{\frac{1}{N_f}\sum_{i=1}^{N_f}
\Big[\max\!\big(0,\ \partial_t \eta(\tilde{\mathbf{q}})+\nabla\!\cdot\!\boldsymbol{\phi}(\tilde{\mathbf{q}})\big)\Big]^2}_{\text{entropy admissibility loss}}
\\
&\quad + \lambda_{\mathrm{str}}\underbrace{\frac{1}{N_f}\sum_{i=1}^{N_f}
\big(1-w_i\big)\,\big\|\partial_t \tilde{\mathbf{q}}(x_i,t_i)+\nabla\!\cdot\mathbf{F}\!\big(\tilde{\mathbf{q}}(x_i,t_i)\big)\big\|_2^2}_{\text{adaptive strong-form loss}}.
\end{split}
\end{equation}

\textbf{Initial-Boundary Loss:} 
This component ensures that the solution \(\tilde{\mathbf{u}}(x,t)\) satisfies the initial and boundary conditions of the PDE,it minimizing the difference between \(\tilde{\mathbf{u}}(x,t)\) and the true values at boundary and initial conditions; This loss term often referred to as the "data loss" in PINNs, aligns the predicted solution with the provided boundary and initial condition.
    
\textbf{Physical Loss:} 
This component enforces the integral-form PDE constraint through the potential variable by penalizing the mismatch in \(\partial_t\tilde{\mathbf{S}}_j(x,t) + \mathbf{F}_j(\tilde{\mathbf{q}}(x,t))\). This formulation enforces conservation in integral form while avoiding direct differentiation of discontinuous primitive states. This necessitates the addition of the entropy loss to ensure the uniqueness and physical realism of the solution.

\textbf{Coupled Loss:}
The coupled loss enforces the correct relationship between the conserved-state representation \(\tilde{\mathbf{q}}(x,t)=\mathbf{q}(\tilde{\mathbf{u}}(x,t))\) and the potential \(\tilde{\mathbf{S}}(x,t)\) by constraining \(\nabla\cdot\tilde{\mathbf{S}}(x,t)\) to match \(\tilde{\mathbf{q}}(x,t)\). It is not necessary to fix the absolute values of \(\tilde{\mathbf{S}}(x,t)\); instead, the focus is on constraining its divergence, ensuring consistency between the conserved quantities and their potential representation.

\textbf{Entropy Admissibility Loss:}
While the integral form of a hyperbolic conservation law enforces Rankine--Hugoniot jump conditions, it does not guarantee uniqueness and may admit non-physical weak solutions~\cite{dafermos2005hyperbolic}.
To select the physically relevant solution, one imposes an entropy condition, e.g., via vanishing viscosity limits~\cite{LeVeque1992,bianchini2005vanishing}.
In CI-PINN, we enforce entropy admissibility using a convex entropy function
$\eta:\mathbb{R}^N\to\mathbb{R}$ and an associated entropy flux
$\boldsymbol{\phi}:\mathbb{R}^N\to\mathbb{R}^d$, defined in the conserved variables $\mathbf{q}$.
For each spatial direction $i\in\{1,\dots,d\}$ with flux $\mathbf{F}^{(i)}(\mathbf{q})$,
the entropy pair satisfies the compatibility relation
$
\nabla_{\mathbf{q}}\boldsymbol{\phi}^{(i)}(\mathbf{q})
=
\nabla_{\mathbf{q}}\eta(\mathbf{q})^\top \, \nabla_{\mathbf{q}}\mathbf{F}^{(i)}(\mathbf{q}).
$
The entropy solution satisfies the entropy inequality
$
\partial_t \eta(\mathbf{q}) + \nabla\cdot \boldsymbol{\phi}(\mathbf{q}) \le 0.
$

\textbf{Adaptive Strong Loss:}
While the strong-form residual is ill-posed at discontinuities (Section~\ref{sec:analysis}), it remains a highly informative and computationally efficient training signal in regions where the solution is smooth and classical derivatives exist. We therefore couple the differential and integral objectives through an adaptive shock mask \(w_i\in[0,1]\) and apply the strong-form penalty only where the local flow is estimated to be smooth.

Concretely, we construct \(w_i\) from a compression indicator based on the \emph{negative spatial divergence} of the predicted velocity field:
$
w_i
=\sigma\!\Big(k\,\mathrm{ReLU}\big(-\nabla_{\mathbf{x}}\!\cdot\tilde{\mathbf{v}}(\mathbf{x}_i,t_i)\big)\Big),
$$
\sigma(z)=\frac{1}{1+e^{-z}}.
$
The term \(\mathrm{ReLU}(-\nabla_{\mathbf{x}}\!\cdot\tilde{\mathbf{v}})\) activates only in compressive regions, which are well known to precede shock formation in hyperbolic dynamics. The sharpness parameter \(k>0\) controls the transition of the mask;

\section{Experiments}
\label{sec:result}

\textbf{Benchmarks \& Metrics.}
We evaluate CI-PINN on canonical hyperbolic PDE benchmarks spanning both scalar laws and systems.
Our 1D benchmarks include the inviscid Burgers' equation, the Buckley--Leverett (BL) equation, and the Euler shock-tube problems (Sod and Lax).
We further consider two 2D benchmarks with complex multi-wave interactions and geometric structure: a 2D Euler Riemann problem and the 2D shallow-water equations (SWE).

To assess performance near discontinuities in \emph{1D}, we partition the space--time domain $\Omega\times[0,T]$ into a \textit{shock neighborhood} ($\Omega_s$), a \textit{rarefaction region} ($\Omega_r$), and the \textit{full domain} ($\Omega$) (see Appendix~\ref{sec:appendG} for definitions).

\subsection{Main result on 1D Benchmarks}

The quantitative results for all 1D benchmarks (Tables~\ref{tab:burger}--\ref{tab:lax}) suggest that in the majority of tested cases, CI-PINN achieves the lowest errors in shock and global regions while maintaining high accuracy in smooth rarefaction zones.

\begin{figure}[t]
  \centering

    
    
  \includegraphics[width=\linewidth]{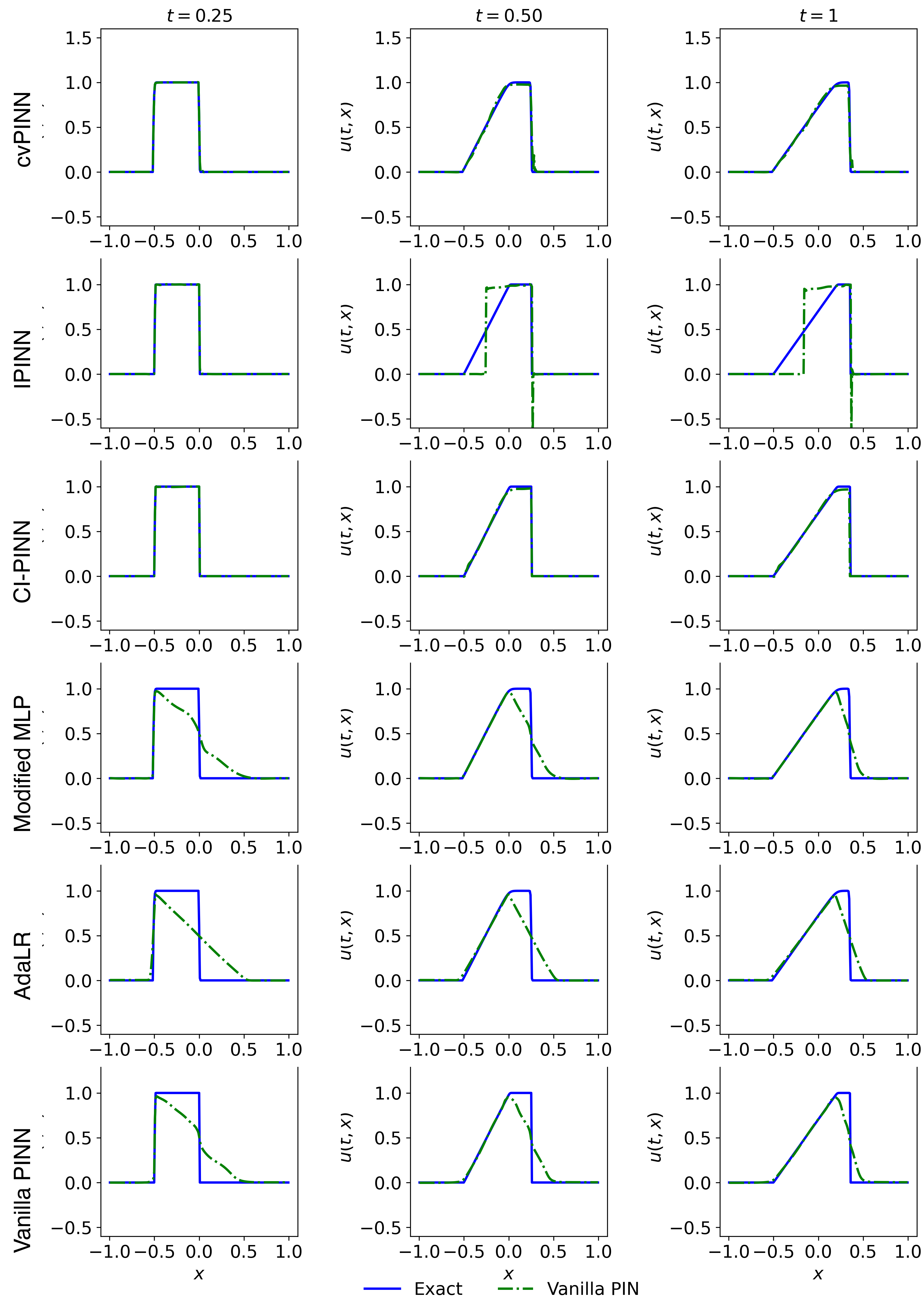}

  \caption{We utilize a square-wave initial condition ($u(x,0)=1$ for $x\in(-0.5,0)$ as in the left-most panels), which naturally induces both a shock (where characteristics pile up, ie., sudden changes in right-most panels) and a rarefaction wave (where characteristics fan apart, ie., linear slope in right-most panels.) 
}
  \label{fig:burgers_split}
\end{figure}

\paragraph{Shock and Rarefaction Analysis.}
Figure~\ref{fig:burgers_split} highlights a known failure mode of strong-form training: the Vanilla PINN smooths the discontinuity and fails to represent the shock sharply.
In contrast, integral-form training (CI-PINN, cvPINN, IPINN) is substantially more shock-consistent and yields sharper interfaces, consistent with enforcing conservation in a way that aligns with Rankine--Hugoniot-type jump behavior.
However, integral formulations can be less stable in smooth rarefaction regions, where non-physical oscillations or biased profiles may appear if admissibility is not sufficiently controlled.
Between the integral baselines, CI-PINN provides the best balance across regimes: it matches the shock location while reducing rarefaction-region artifacts, leading to the lowest global $L_1$ error in Table~\ref{tab:burger}.
Similar trends hold for the BL equation and the 1D Euler shock tubes (Tables~\ref{tab:bl}--\ref{tab:lax}).

\begin{table}[t]
\centering
\caption{Error metrics for Burgers equation across models (values in $10^{-3}$).}
\label{tab:burger}

\setlength{\tabcolsep}{2.2pt}   
\renewcommand{\arraystretch}{1.05}

\scriptsize
\begin{adjustbox}{max width=\columnwidth}
\begin{tabular}{llcccccccc}
\toprule
\textbf{Region} & \textbf{Metric}
& \cellcolor{green!10}\textbf{CI-PINN}
& \textbf{cvPINN} & \textbf{IPINN}
& \textbf{Mod.\ MLP} & \textbf{Causal} & \textbf{RWF} & \textbf{AdaLR} & \textbf{Vanilla} \\
\midrule

\multirow{3}{*}{Global}
& \ $L_2$
& \cellcolor{green!10}\textbf{47.04}
& \underline{96.15} & 153.23 & 120.62 & 282.44 & 131.00 & 146.20 & 123.87 \\
& \ $L_1$
& \cellcolor{green!10}\textbf{7.55}
& \underline{18.18} & 65.39 & 54.59 & 109.26 & 58.74 & 71.94 & 56.58 \\
& $L_\infty$
& \cellcolor{green!10}1012.71
& 997.44 & 1594.20 & \underline{750.43} & 1268.73 & 875.97 & \textbf{497.03} & 915.87 \\
\midrule

\multirow{3}{*}{Shock}
& \ $L_2$
& \cellcolor{green!10}\textbf{103.67}
& 211.89 & \underline{166.24} & 258.13 & 561.66 & 280.65 & 307.54 & 266.31 \\
& \ $L_1$
& \cellcolor{green!10}\textbf{21.26}
& 62.34 & \underline{39.80} & 217.81 & 325.96 & 232.72 & 265.51 & 222.87 \\
& $L_\infty$
& \cellcolor{green!10}1012.71
& 997.44 & 1594.20 & \underline{750.43} & 1262.88 & 875.97 & \textbf{497.03} & 915.87 \\
\midrule

\multirow{3}{*}{Raref.}
& \ $L_2$
& \cellcolor{green!10}25.47
& 60.45 & 225.02 & \textbf{7.40} & 112.72 & 37.09 & 22.34 & \underline{18.92} \\
& \ $L_1$
& \cellcolor{green!10}10.95
& 19.54 & 159.59 & \textbf{3.62} & 99.60 & \underline{5.91} & 14.89 & 7.45 \\
& $L_\infty$
& \cellcolor{green!10}1012.71
& 997.44 & 1000.35 & 390.70 & \textbf{215.62} & 875.97 & \underline{374.24} & 915.87 \\
\bottomrule
\end{tabular}
\end{adjustbox}
\end{table}

\begin{table}[t]
\centering
\caption{Error metrics for BL equation across models (values in $10^{-3}$).}
\label{tab:bl}

\includegraphics[width=\linewidth]{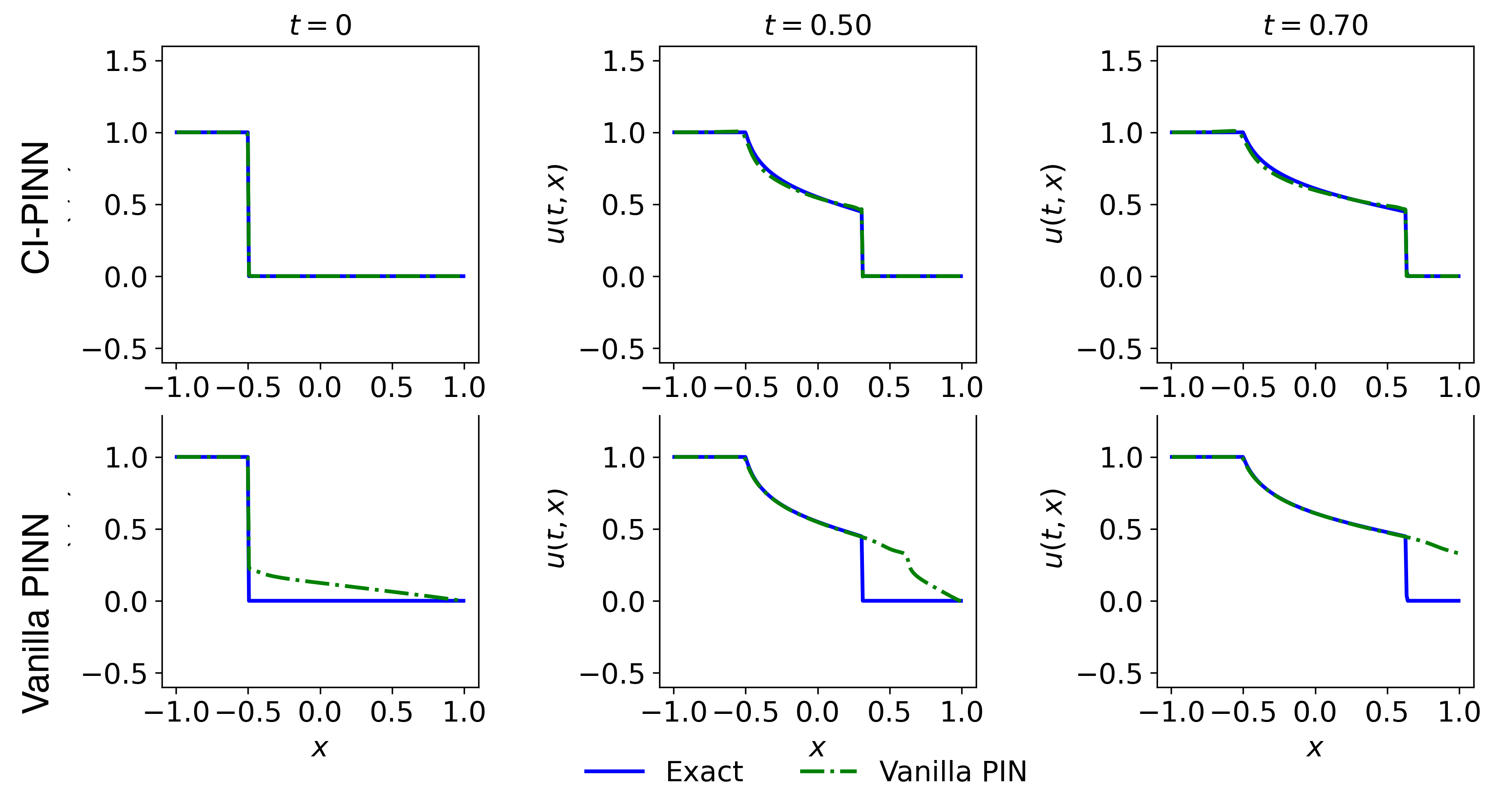}\\
\setlength{\tabcolsep}{2.2pt}
\renewcommand{\arraystretch}{1.05}

\scriptsize
\begin{adjustbox}{max width=\columnwidth}
\begin{tabular}{llcccccccc}
\toprule
\textbf{Region} & \textbf{Metric}
& \cellcolor{green!10}\textbf{CI-PINN}
& \textbf{cvPINN} & \textbf{IPINN}
& \textbf{Mod.\ MLP} & \textbf{Causal} & \textbf{RWF} & \textbf{AdaLR} & \textbf{Vanilla} \\
\midrule

\multirow{3}{*}{Global}
& \ $L_2$
& \cellcolor{green!10}\textbf{17.15}
& 83.02 & 239.59 & 149.87 & \underline{69.51} & 138.67 & 118.75 & 147.83 \\
& \ $L_1$
& \cellcolor{green!10}\textbf{6.76}
& \underline{26.76} & 133.14 & 80.28 & 35.96 & 64.16 & 60.65 & 78.17 \\
& $L_\infty$
& \cellcolor{green!10}\underline{444.08}
& 794.39 & 1479.38 & 445.39 & 697.07 & 447.82 & \textbf{440.21} & 451.45 \\
\midrule

\multirow{3}{*}{Shock}
& \ $L_2$
& \cellcolor{green!10}\textbf{28.64}
& 181.74 & 337.91 & 263.95 & \underline{131.11} & 260.39 & 235.03 & 261.74 \\
& \ $L_1$
& \cellcolor{green!10}\textbf{8.52}
& \underline{91.77} & 227.89 & 182.55 & 94.32 & 180.45 & 163.71 & 180.62 \\
& $L_\infty$
& \cellcolor{green!10}\underline{444.08}
& 794.39 & 1479.38 & 445.39 & 697.07 & 447.82 & \textbf{440.21} & 451.45 \\
\midrule

\multirow{3}{*}{Raref.}
& \ $L_2$
& \cellcolor{green!10}25.07
& 42.75 & 420.00 & \textbf{4.32} & 70.88 & \underline{7.13} & 16.95 & \underline{5.68} \\
& \ $L_1$
& \cellcolor{green!10}22.52
& 21.56 & 396.93 & \textbf{2.24} & 61.84 & \underline{3.85} & 10.96 & 4.41 \\
& $L_\infty$
& \cellcolor{green!10}\textbf{57.16}
& 794.39 & 1479.38 & 71.89 & 304.96 & 171.17 & 128.16 & \underline{62.54} \\
\bottomrule
\end{tabular}
\end{adjustbox}
\end{table}
\paragraph{Buckley--Leverett.}
In the Buckley--Leverett problem (Table~\ref{tab:bl}), the non-convex flux $f(u)$ produces a compound wave in which a rarefaction and a shock appear in close proximity.
This regime is challenging for standard PINNs and can be sensitive to flux non-convexity.
CI-PINN resolves the sharp interface with the lowest shock-region $L_1$ error, without requiring auxiliary flux convexification, indicating robustness to more complex flux geometries.

\begin{table}[ht]
\centering
\caption{Averaged error metrics (across $u, p, \rho$) for Sod Shock Tube (values in $10^{-3}$). Best is bold; second best is underlined. For full diagram please refer to appendix~\ref{fulltable}}
\label{tab:sod}
\setlength{\tabcolsep}{2.5pt} 
\renewcommand{\arraystretch}{1.1}
\scriptsize
\begin{adjustbox}{max width=\columnwidth}
\begin{tabular}{llcccccccc}
\toprule
\textbf{Region} & \textbf{Metric} 
& \cellcolor{green!10}\textbf{CI-PINN} 
& \textbf{cvPINN} & \textbf{IPINN} & \textbf{Mod.MLP} & \textbf{Causal} & \textbf{RWF} & \textbf{AdaLR} & \textbf{Vanilla} \\
\midrule
\multirow{3}{*}{Global}
& $L_1$     & \cellcolor{green!10}\textbf{6.77} & \underline{11.84} & 360.28 & 143.78 & 203.26 & 140.32 & 180.60 & 209.07 \\
& $L_2$     & \cellcolor{green!10}\textbf{25.94} & \underline{32.04} & 558.38 & 201.35 & 369.30 & 200.31 & 255.16 & 392.55 \\
& $L_\infty$  & \cellcolor{green!10}\underline{946.43} & 1131.47 & 4378.85 & \textbf{927.25} & 1210.98 & 1094.87 & 1005.75 & 952.66 \\
\midrule
\multirow{3}{*}{Shock}
& $L_1$     & \cellcolor{green!10}\textbf{15.36} & \underline{30.79} & 534.18 & 277.17 & 387.46 & 281.46 & 267.74 & 327.86 \\
& $L_2$     & \cellcolor{green!10}\textbf{49.84} & \underline{61.07} & 698.27 & 324.45 & 523.87 & 330.20 & 321.00 & 465.65 \\
& $L_\infty$  & \cellcolor{green!10}\underline{946.43} & 1131.47 & 3994.40 & \textbf{927.25} & 1199.11 & 1094.87 & 1005.75 & 952.61 \\
\midrule
\multirow{3}{*}{Raref.}
& $L_1$     & \cellcolor{green!10}\textbf{42.93} & \underline{46.35} & 660.08 & 171.76 & 217.06 & 184.58 & 297.09 & 480.75 \\
& $L_2$     & \cellcolor{green!10}\textbf{76.78} & \underline{80.73} & 714.11 & 213.95 & 264.37 & 211.56 & 358.06 & 560.09 \\
& $L_\infty$  & \cellcolor{green!10}\textbf{810.77} & \underline{865.67} & 1616.80 & 927.25 & 1039.96 & 1094.87 & 1005.75 & 952.61 \\
\bottomrule
\end{tabular}
\end{adjustbox}
\end{table}

\begin{figure}[h]

  \centering  \includegraphics[width=1.02\linewidth]{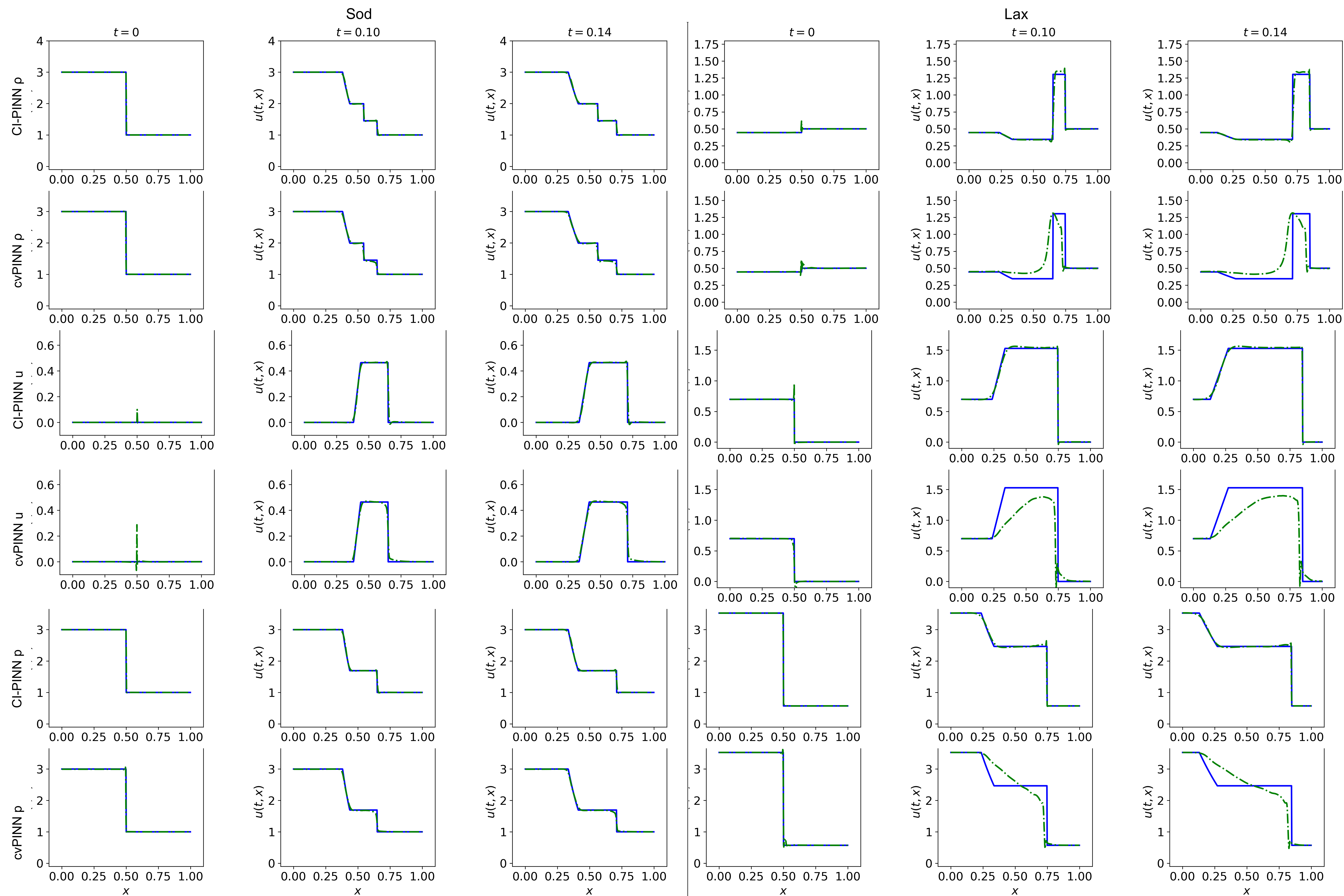}
  \caption{In these shock-tube profiles, the \emph{contact discontinuity} appears as a sharp change in density while the velocity and pressure remain essentially unchanged across the interface. The \emph{plateau states} are the flat segments between the smooth expansion region and the sharp jumps, and their constant levels are fixed by conservation across the discontinuities (Rankine--Hugoniot) together with the expansion relations.}
  \label{fig:1deuler}
\end{figure}
\paragraph{Euler System (Sod \& Lax Shock Tubes).}
For the 1D Euler system, CI-PINN and cvPINN consistently outperform the other baselines in Tables~\ref{tab:sod} and~\ref{tab:lax}.
Figure~\ref{fig:1deuler} visualizes representative profiles and highlights the key physical structures: a smooth expansion fan, a contact discontinuity (sharp density change with nearly unchanged $(u,p)$), a shock (jump in $(\rho,u,p)$), and the plateau states between them. In the Sod case, both methods recover the qualitative pattern, but cvPINN exhibits mild smearing around the contact and shock, leading to visibly perturbed plateau levels; The Lax case is substantially more challenging due to the stronger shock and steeper gradients: CI-PINN maintains sharp interfaces and accurate plateau states, indicating that the integral-form residual (enforcing Rankine--Hugoniot consistency) and the coupled consistency are effectively controlled even in the shock-dominated regime. In contrast, cvPINN often attains a reasonable shock location yet degrades the remaining structures—most notably the contact and plateau states—consistent with a smooth-surrogate solution that can reduce pointwise residuals but incurs a large accumulated mismatch in integral balance and admissibility near discontinuities. Overall, the results support the conclusion that integral-form enforcement is critical for correctly capturing shock/contact physics.

\begin{table}[t]
\label{tab:lax}
\centering
\caption{Averaged error metrics (across $u, p, \rho$) for the Lax shock tube (values in $10^{-3}$). Best is bold; second best is underlined. For full diagram please refer to appendix~\ref{fulltable}}
\label{tab:lax}

\setlength{\tabcolsep}{2.5pt}
\renewcommand{\arraystretch}{1.1}
\scriptsize

\begin{adjustbox}{max width=\columnwidth}
\begin{tabular}{llcccccccc}
\toprule
\textbf{Region} & \textbf{Metric}
& \cellcolor{green!10}\textbf{CI-PINN}
& \textbf{cvPINN} & \textbf{IPINN}
& \textbf{Mod.\ MLP} & \textbf{Causal} & \textbf{RWF}
& \textbf{AdaLR} & \textbf{Vanilla} \\
\midrule

\multirow{3}{*}{Global}
& \ $L_2$
& \cellcolor{green!10}\textbf{55.89}  
& \underline{283.63} 
& 789.22 
& 389.87 
& 404.06 
& 421.49 
& 383.89 
& 934.98 \\ 

& \ $L_1$
& \cellcolor{green!10}\textbf{14.83}  
& \underline{142.00} 
& 600.61 
& 245.62 
& 237.60 
& 249.23 
& 286.91 
& 828.75 \\ 

& $L_\infty$
& \cellcolor{green!10}\underline{1279.37} 
& 1630.57 
& 2091.30 
& 1443.45 
& \textbf{1301.50} 
& 1519.52 
& 1388.93 
& 1738.78 \\ 

\midrule

\multirow{3}{*}{Shock}
& \ $L_2$
& \cellcolor{green!10}\textbf{111.55} 
& \underline{492.82} 
& 1061.01 
& 818.64 
& 652.92 
& 640.17 
& 907.60 
& 940.05 \\ 

& \ $L_1$
& \cellcolor{green!10}\textbf{42.18}  
& \underline{282.83} 
& 646.66 
& 497.16 
& 451.16 
& 525.91 
& 451.15 
& 938.51 \\ 

& $L_\infty$
& \cellcolor{green!10}\underline{1279.37} 
& 1630.57 
& 2091.30 
& 1443.45 
& \textbf{1301.50} 
& 1519.52 
& 1388.93 
& 1736.84 \\ 

\midrule

\multirow{3}{*}{Raref.}
& \ $L_2$
& \cellcolor{green!10}\textbf{168.17} 
& \underline{338.42} 
& 532.36 
& 569.93 
& 362.03 
& 555.72 
& 480.68 
& 779.72 \\ 

& \ $L_1$
& \cellcolor{green!10}\textbf{55.74}  
& \underline{268.49} 
& 309.37 
& 488.70 
& 266.65 
& 555.12 
& 443.96 
& 852.34 \\ 

& $L_\infty$
& \cellcolor{green!10}\underline{1379.37} 
& \textbf{1001.55} 
& 2091.30 
& 1504.43 
& 1301.50 
& 1519.53 
& 1388.93 
& 1738.77 \\ 

\bottomrule
\end{tabular}
\end{adjustbox}
\end{table}


\subsection{2D Complex Flow Benchmarks}

\begin{table}[ht]
\label{tab:2d}
\centering
\caption{Global space-time error metrics (values in $10^{-3}$). Top: 2D Euler Riemann problem. Bottom: 2D SWE. Best is \textbf{bold}; another is \underline{underlined}.}
\label{tab:global_metrics_euler_swe_matrix}
\vspace{-1.2mm}

\setlength{\tabcolsep}{3.2pt}
\renewcommand{\arraystretch}{1.08}
\scriptsize

\begin{adjustbox}{max width=\linewidth}
\begin{tabular}{@{}l|cc|cc|cc@{}}
\toprule
\textbf{Var} 
& \multicolumn{2}{c|}{\textbf{$L_2$}}
& \multicolumn{2}{c|}{\textbf{$L_1$}}
& \multicolumn{2}{c}{\textbf{$L_\infty$}} \\
\cmidrule(lr){2-3}\cmidrule(lr){4-5}\cmidrule(lr){6-7}
& \cellcolor{green!10}\textbf{CI-PINN} & \textbf{cvPINN}
& \cellcolor{green!10}\textbf{CI-PINN} & \textbf{cvPINN}
& \cellcolor{green!10}\textbf{CI-PINN} & \textbf{cvPINN} \\
\midrule

$\rho$
& \cellcolor{green!10}\textbf{82.46}    & \underline{99.55}
& \cellcolor{green!10}\textbf{25.76}   & \underline{42.47}
& \cellcolor{green!10}\textbf{1964.00} & \underline{2019.33} \\

$u$
& \cellcolor{green!10}\underline{54.59} & \textbf{53.29}
& \cellcolor{green!10}\textbf{14.13}   & \underline{19.41}
& \cellcolor{green!10}\textbf{997.63}  & \underline{1235.11} \\

$v$
& \cellcolor{green!10}\underline{47.64} & \textbf{47.33}
& \cellcolor{green!10}\textbf{13.75}   & \underline{20.85}
& \cellcolor{green!10}\underline{874.52} & \textbf{602.83} \\

$p$
& \cellcolor{green!10}\textbf{16.12}    & \underline{29.50}
& \cellcolor{green!10}\textbf{9.54}    & \underline{17.66}
& \cellcolor{green!10}\textbf{238.25}  & \underline{496.03} \\

\bottomrule
\end{tabular}
\end{adjustbox}

\vspace{0.5mm}
{\footnotesize (a) Euler 2D Riemann}

\vspace{4mm} 

\begin{adjustbox}{max width=\linewidth}
\begin{tabular}{@{}l|cc|cc|cc@{}}
\toprule
\textbf{Var} 
& \multicolumn{2}{c|}{\textbf{$L_2$}}
& \multicolumn{2}{c|}{\textbf{$L_1$}}
& \multicolumn{2}{c}{\textbf{$L_\infty$}} \\
\cmidrule(lr){2-3}\cmidrule(lr){4-5}\cmidrule(lr){6-7}
& \cellcolor{green!10}\textbf{CI-PINN} & \textbf{cvPINN}
& \cellcolor{green!10}\textbf{CI-PINN} & \textbf{cvPINN}
& \cellcolor{green!10}\textbf{CI-PINN} & \textbf{cvPINN} \\
\midrule

$h$
& \cellcolor{green!10}\textbf{26.10}   & \underline{225.81}
& \cellcolor{green!10}\textbf{9.181}   & \underline{127.875}
& \cellcolor{green!10}\textbf{981.002} & \underline{1063.928} \\

$u$
& \cellcolor{green!10}\textbf{16.37}   & \underline{154.01}
& \cellcolor{green!10}\textbf{5.365}   & \underline{79.312}
& \cellcolor{green!10}\textbf{223.043} & \underline{566.177} \\

$v$
& \cellcolor{green!10}\textbf{16.25}   & \underline{152.97}
& \cellcolor{green!10}\textbf{5.522}   & \underline{77.358}
& \cellcolor{green!10}\textbf{185.416} & \underline{566.113} \\

\bottomrule
\end{tabular}
\end{adjustbox}

\vspace{0.5mm}
{\footnotesize (b) SWE 2D}

\vspace{-2mm}
\end{table}

Compared to 1D, these 2D cases incur substantially higher training cost and do not admit a clean region-wise shock/rarefaction partition; we therefore report global space--time metrics and focus on comparisons against cvPINN as a strong baseline.

\begin{figure}[ht]
    \centering
    \begin{subfigure}[b]{0.48\textwidth}
        \centering
        \includegraphics[width=\textwidth]{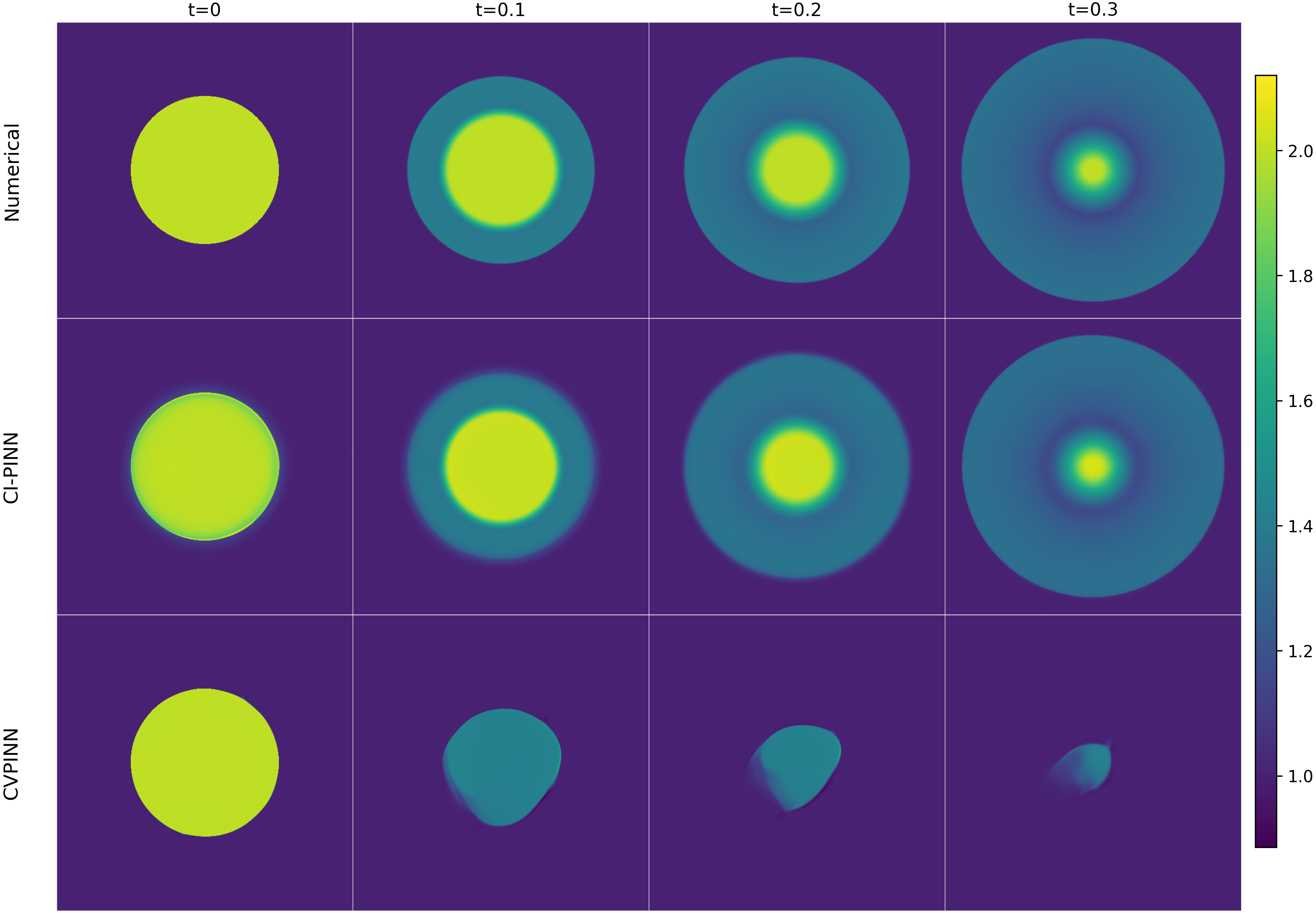}
        \caption{Height $h$ in SWE}
    \end{subfigure}
    \hfill
    \begin{subfigure}[b]{0.48\textwidth}
        \centering
        \includegraphics[width=\textwidth]{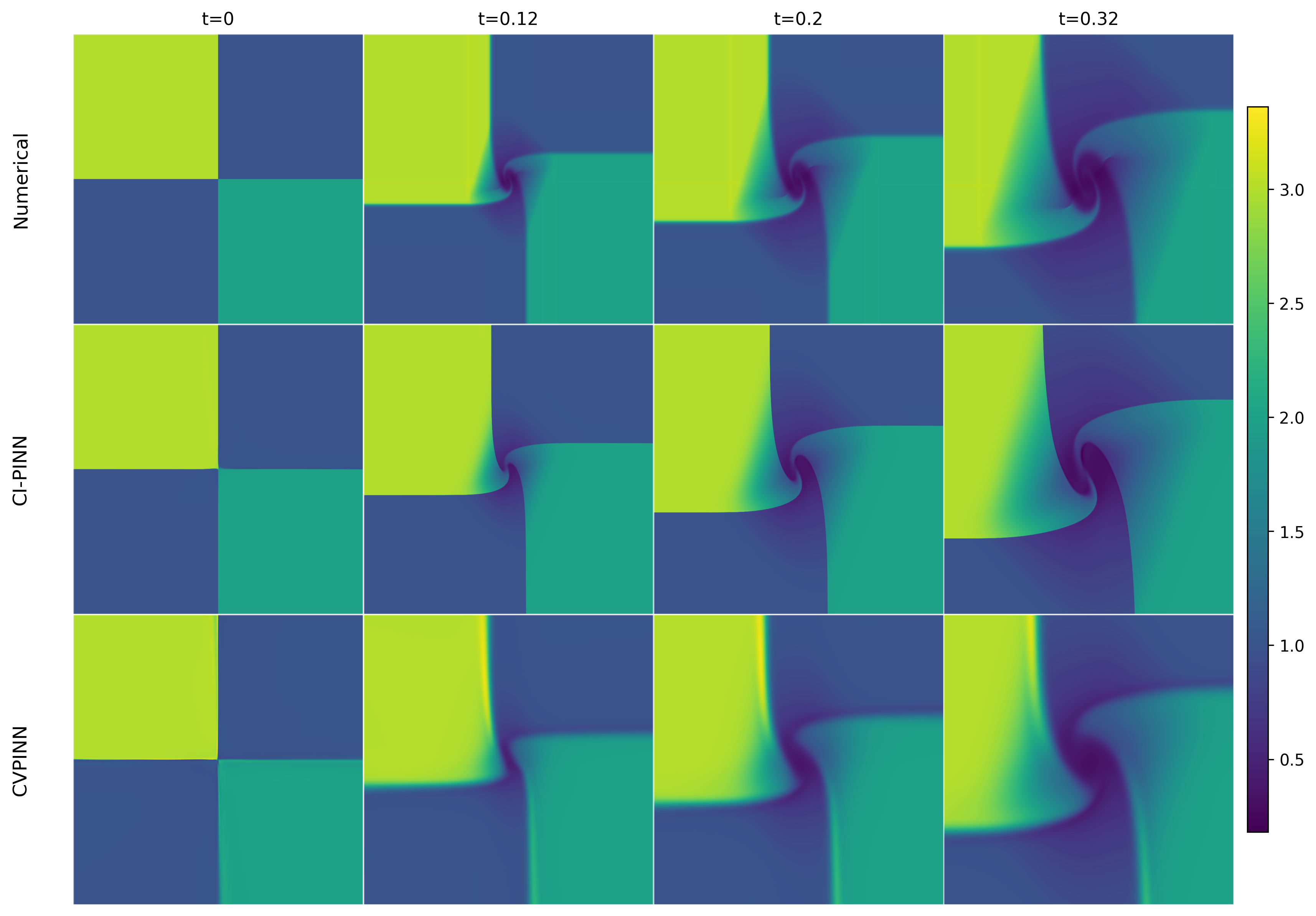}
        \caption{Density $\rho$ in Euler system}
    \end{subfigure}

    \caption{Selective component on 2D system, full result please refer to appendix~\ref{Appendix:2dexp}}
    \label{fig:2dselective}
\end{figure}

As shown in Figure~\ref{fig:2dselective}, CI-PINN preserves the expected 2D dynamics. 
For SWE, the initial height hump develops into a coherent, radially propagating gravity-wave pattern. 
In contrast, cvPINN is visibly more diffusive: the wave amplitude is strongly damped and the ring structure collapses over time. 
This failure is already apparent by $t\approx 0.2$, where the wavefront is largely smeared out, producing severe physical inaccuracies. 
CI-PINN, on the other hand, maintains circular symmetry and sustained outward propagation throughout the window $t\in[0,0.3]$.

These qualitative differences are reflected quantitatively in Table~\ref{tab:global_metrics_euler_swe_matrix}(b). 
CI-PINN reduces the $L_2$ error by $88.4\%$ for $h$ (225.81 $\to$ 26.10) and by $89.4\%$ for $(u,v)$ (154.01 $\to$ 16.37; 152.97 $\to$ 16.25).%

For the 2D Euler Riemann problem, CI-PINN more faithfully preserves sharp interfaces and the associated plateau states in the interacting-flow pattern (Figure~\ref{fig:2dselective}), yielding clear gains in the key thermodynamic variables (Table~\ref{tab:global_metrics_euler_swe_matrix}(a)). 
In particular, $\rho$ errors improve in $L_1$ by $39.3\%$ (42.47 $\to$ 25.76), with a $31.4\%$ reduction in squared $L_2$ error (i.e., MSE) implied by 99.55 $\to$ 82.46. 
For pressure, the improvement is stronger: $L_2$ decreases by $45.3\%$ (29.50 $\to$ 16.12), $L_1$ by $46.0\%$ (17.66 $\to$ 9.54), and $L_\infty$ by $52.0\%$ (496.03 $\to$ 238.25). 
Velocity errors are mixed in terms of $L_2$ but consistently better in $L_1$ (MAE reduced by $27.2\%$ for $u$ and $34.1\%$ for $v$), while $v$-$L_\infty$ can be dominated by a localized feature, leading to a larger maximum despite improved global fidelity.


\section{Conclusion and Future Work}
\label{sec:conclusion}
In this work, we identified that strong-form PINNs fail at discontinuities because residuals diverge, creating an optimization barrier. We proposed the Coupled Integral PINN (CI-PINN) to circumvent this via dual-network integral conservation and entropy constraints. Evaluation on 1D and 2D benchmarks confirms that CI-PINN significantly outperforms baselines, faithfully capturing shocks and contact discontinuities without mesh-based reconstruction. These results establish that integral enforcement is essential for physical fidelity in discontinuous flows. Future work includes: \emph{(i) Constrained optimization,} replacing penalties with explicit feasibility constraints (e.g., via augmented Lagrangian) to reduce hyperparameter sensitivity; \emph{(ii) Inverse problem,} extending the framework to real world inverse problem. 

\section*{Impact Statement}
This work advances physics-informed learning methods for hyperbolic PDEs by improving the reliability of neural PDE solvers in the presence of discontinuities (e.g., shocks). More accurate and stable surrogate solvers can reduce the cost of simulation and enable faster analysis in applications such as fluid dynamics, geophysics, and engineering design, which may have downstream benefits for safety and efficiency. 

Potential risks mainly stem from misuse or over-trust: learned solvers can be applied outside their validity range (e.g., different regimes, boundary conditions, or resolutions) and may produce plausible-looking but incorrect solutions, which could misinform high-stakes decisions if used without verification. We mitigate this by emphasizing that the method is a numerical surrogate—not a guarantee of physical fidelity—and by encouraging standard validation against established baselines, reporting uncertainty/error metrics, and careful documentation of training domains and constraints. We do not anticipate direct negative societal impacts beyond the general concerns associated with applying machine-learning models to scientific and engineering workflows.

\bibliography{example_paper}


\setcounter{section}{0}

\renewcommand{\thesection}{A\arabic{section}}
\setcounter{figure}{0}
\renewcommand{\thefigure}{A\arabic{figure}}
\setcounter{table}{0}
\renewcommand{\thetable}{A\arabic{table}}
\newpage

\setcounter{page}{1} 

\newpage

\appendix
\onecolumn
\bigskip
\begin{center}
{\huge\bf Appendix}
\end{center}

\section{Proof sketch for propositions}
\label{appendix:proof}
\textbf{Proposition}(Strong-form blow-up as a discontinuity sharpens)
Let $\mathbf{q}^*$ be a discontinuous physical solution containing a jump of amplitude
$\Delta \mathbf{q}\neq \mathbf{0}$ across a smooth interface.
Let $\{\mathbf{q}^\varepsilon\}_{\varepsilon>0}$ be a family of smooth approximations that resolves the jump
over a transition layer of thickness $\varepsilon$.
Assume $\mathbf{F}\in C^1$ and its Jacobians are bounded on the relevant state range.
Assume additionally that the jump produces a nontrivial normal flux change, i.e.
for a unit normal $n$ to the interface,
\[
\Delta \mathbf{F}_n := \mathbf{F}(\mathbf{q}_R)\cdot n - \mathbf{F}(\mathbf{q}_L)\cdot n \neq \mathbf{0}.
\]
Then the strong-form residual satisfies
\begin{equation}
\label{eq:strong_blowup_rate}
\mathcal{L}_{\mathrm{strong}}(\mathbf{q}^\varepsilon)
=
\left\|\partial_t \mathbf{q}^\varepsilon + \nabla\cdot \mathbf{F}(\mathbf{q}^\varepsilon)\right\|_{L^2(\Omega_T)}^2
\;\gtrsim\;
\frac{C}{\varepsilon},
\qquad \varepsilon\to 0,
\end{equation}
for some constant $C>0$ independent of $\varepsilon$.

\begin{proof}[Proof (normal-line mean-value argument)]
Fix a time $t\in(0,T)$.
Let $\Gamma(t)\subset\Omega$ denote the shock interface (a $C^1$ hypersurface), and pick a point $x_0\in\Gamma(t)$.
Let $n=n(x_0)$ be a unit normal at $x_0$.
Assume the discontinuous (limit) solution has one-sided traces
$
\mathbf q_L := \lim_{\substack{x\to x_0\\ (x-x_0)\cdot n<0}} \mathbf q^*(x,t),
$
$
\mathbf q_R := \lim_{\substack{x\to x_0\\ (x-x_0)\cdot n>0}} \mathbf q^*(x,t),
$
$
\Delta \mathbf q := \mathbf q_R-\mathbf q_L\neq 0.
$
Let $\mathbf q^\varepsilon\in C^1(\Omega_T)$ be a smooth approximation that resolves this jump over a transition layer
of thickness $\sim \varepsilon$ around $\Gamma(t)$.
Concretely, along the normal line $x(s):=x_0+s\,n$ with $|s|\lesssim \varepsilon$,
the state transitions from a value near $\mathbf q_L$ to a value near $\mathbf q_R$.

\paragraph{Step 1: a $1/\varepsilon$ normal derivative appears (MVT).}
Define the 1D profile along the normal line
\[
\phi(s):=\mathbf q^\varepsilon(x(s),t)\in\mathbb R^m.
\]
By the transition-layer assumption, there exist $s_-<s_+$ with $|s_\pm|\lesssim \varepsilon$ such that
$
\phi(s_-)\approx \mathbf q_L, \qquad \phi(s_+)\approx \mathbf q_R,
\qquad\text{hence}\qquad
\|\phi(s_+)-\phi(s_-)\|\ \sim\ \|\Delta\mathbf q\|.
$
Since $\phi$ is $C^1$, the (1D) mean value theorem applied componentwise gives a point $\xi\in(s_-,s_+)$ with
\[
\|\phi'(\xi)\|\ \gtrsim\ \frac{\|\phi(s_+)-\phi(s_-)\|}{|s_+-s_-|}
\ \sim\ \frac{\|\Delta\mathbf q\|}{\varepsilon}
\ \sim\ \frac{1}{\varepsilon}.
\]
But $\phi'(s)=\nabla \mathbf q^\varepsilon(x(s),t)\,n$, i.e. the directional derivative along the normal.
Thus at some point in the layer,
\[
\|\partial_n \mathbf q^\varepsilon\|\ :=\ \|\nabla \mathbf q^\varepsilon\,n\|\ \sim\ \frac{1}{\varepsilon}.
\]

\paragraph{Step 2: why this forces $\nabla\!\cdot \mathbf F(\mathbf q^\varepsilon)\sim 1/\varepsilon$.}
Write the flux as $\mathbf F(\mathbf q)=(\mathbf F_1(\mathbf q),\dots,\mathbf F_d(\mathbf q))$ with
$\mathbf F_j(\mathbf q)\in\mathbb R^m$.
Introduce the \emph{normal flux vector}
$
\mathbf F_n(\mathbf q)\ :=\ \sum_{j=1}^d n_j\,\mathbf F_j(\mathbf q)\ \in\mathbb R^m,
$
and its Jacobian $\mathbf J_{F_n}(\mathbf q)=\sum_{j=1}^d n_j\,\mathbf J_{\mathbf F_j}(\mathbf q)$.

A standard local-coordinate decomposition (normal $n$ + tangential directions $\tau_1,\dots,\tau_{d-1}$)
gives, near $x_0$,
\begin{equation}
\label{eq:div_decomp}
\nabla\!\cdot \mathbf F(\mathbf q^\varepsilon)
\;=\;
\partial_n \mathbf F_n(\mathbf q^\varepsilon)
\;+\;
\sum_{k=1}^{d-1}\partial_{\tau_k}\mathbf F_{\tau_k}(\mathbf q^\varepsilon)
\;+\;(\text{curvature terms}),
\end{equation}
where $\mathbf F_{\tau_k}(\mathbf q):=\sum_{j=1}^d (\tau_k)_j\,\mathbf F_j(\mathbf q)$.
The key point is: \emph{only the $\partial_n$ term differentiates across the thin layer}.
By assumption, $\mathbf q^\varepsilon$ varies sharply only in the normal direction (thickness $\sim\varepsilon$)
and remains $O(1)$-smooth tangentially; hence the tangential/curvature terms are $O(1)$, while
$
\partial_n \mathbf F_n(\mathbf q^\varepsilon)
=
\mathbf J_{F_n}(\mathbf q^\varepsilon)\,\partial_n \mathbf q^\varepsilon
\qquad\text{(chain rule in the normal direction)}.
$
Since the Jacobians are bounded and the jump in normal flux is $O(1)$ whenever $\Delta\mathbf q\neq 0$,
one generically has $\|\mathbf J_{F_n}(\mathbf q^\varepsilon)\|=O(1)$ in the layer and thus
\[
\|\partial_n \mathbf F_n(\mathbf q^\varepsilon)\|
\ \sim\
\|\partial_n \mathbf q^\varepsilon\|
\ \sim\
\frac{1}{\varepsilon}.
\]
Plugging into \eqref{eq:div_decomp} yields
\[
\|\nabla\!\cdot \mathbf F(\mathbf q^\varepsilon)\|\ \sim\ \frac{1}{\varepsilon}
\quad\text{on (part of) the layer.}
\]
\paragraph{Step 3: integrate over a set of space--time measure $O(\varepsilon)$.}
The transition layer is an $\varepsilon$-thick band around a codimension-$1$ interface, so its space--time measure scales like $O(\varepsilon)$.
On this band, the residual $\mathbf r^\varepsilon:=\partial_t\mathbf q^\varepsilon+\nabla\!\cdot \mathbf F(\mathbf q^\varepsilon)$ satisfies
\[
\|\mathbf r^\varepsilon\|\ \ge\ \|\nabla\!\cdot \mathbf F(\mathbf q^\varepsilon)\|-\|\partial_t\mathbf q^\varepsilon\|.
\]
If $\|\partial_t\mathbf q^\varepsilon\|=O(1)$ in the layer, it is dominated by the $O(1/\varepsilon)$ divergence term as $\varepsilon\to 0$;
if $\|\partial_t\mathbf q^\varepsilon\|$ also scales like $1/\varepsilon$, then $\|\mathbf r^\varepsilon\|\sim 1/\varepsilon$ in magnitude
(up to non-generic cancellations).
Therefore $\|\mathbf r^\varepsilon\|\sim 1/\varepsilon$ on a set of measure $O(\varepsilon)$, and hence
\[
\mathcal L_{\mathrm{strong}}(\mathbf q^\varepsilon)
=
\int_{\Omega_T}\|\mathbf r^\varepsilon\|^2\,dx\,dt
\ \gtrsim\
O(\varepsilon)\cdot \left(\frac{1}{\varepsilon}\right)^2
\ =\
\frac{C}{\varepsilon},
\]
for some $C>0$ independent of $\varepsilon$ which conclude the proof for proposition 2.3 and 2.1.

\end{proof}

\section{Benchmarks}
\label{sec:Classical}
We compare CI-PINN against several advances methods in physics-informed neural networks. \textbf{Adaptive learning rate (Adaptive lr)}~\citep{wang2023experts} dynamically adjusts optimizer step sizes based on gradient magnitudes, ensuring balanced contributions from each loss component and improving convergence stability. \textbf{Causal training (Causal)}~\citep{wang2022respecting} introduces a temporally structured curriculum that gradually incorporates the residual loss to align with the causal nature of time-dependent PDEs, thereby enhancing long-horizon stability. \textbf{Modified MLP}~\citep{wang2020understandingmitigatinggradientpathologies} increases model expressiveness by processing inputs through parallel subnetworks and aggregating outputs via gated mechanisms, significantly boosting accuracy with minimal computational overhead. \textbf{Random Weight Factorization (RWF)}~\citep{wang2022randomweightfactorizationimproves} reparameterizes neural network weights as products of random matrices, mitigating spectral bias and smoothing the optimization landscape in PINN training.

Among conservation-based formulations, \textbf{cvPINN}~\citep{PATEL2022110754} integrates conservation laws over local control volumes using numerical quadrature, inspired by finite-volume methods, to enforce local conservation properties. In contrast, \textbf{IPINN}~\citep{rajvanshi2024integral} directly approximates the spatial integral of the solution using neural networks, enforcing global conservation without explicit discretization. These integral-form methods provide more robust handling of shocks and discontinuities than standard differential-form PINNs. Additionally, we include a small-scale comparison with the \textbf{ProbConserv} Bayesian framework~\citep{HANSEN2024133952}, which imposes global conservation via probabilistic posterior updates on the predicted solution. Although this method requires problem-specific boundary treatment and is less generalizable to complex settings. We report the formulations and results of \textbf{cvPINN} and \textbf{IPINN} in the main text, while the performance of all other mentioned methods is provided in Appendix~\ref{sec:appendH} as a point of reference.

\section{Rigorous derivation of the Integral Form of Scalar Conservation Law}
\label{sec:appendB}
In nonlinear conservation law, the shocks appear when the characteristic lines of different values intersect with each other \cite{LeVeque1992}. The properties of these shocks are governed by the weak form of the PDE (i.e. Rankine-Hugoniot jump condition). The original conservation law can be written as: 
\begin{equation}
u_t+f(u)_x=0
\end{equation}
where  $u$ is conserved quantity, $u_t$ denotes the partial derivative of $u$ with respect to time $t$, $f(u)$ is  is a flux function that describes how the quantity $u$ flows or moves through space, $f(u)_x$ is denotes the partial derivative of $f(u)$ with respect to space.

For the weak form of a conservation law, we multiply the equation by a test function \(\phi: \mathbb{R} \to \mathbb{R}\) and integrate over space and time: 
\begin{equation}
\int_{0}^{\infty}\int_{-\infty}^{+\infty} [\phi u_t+\phi f_x] \,dx\,dt =0
\end{equation}
Due to the conservation form of the PDE, this equation must hold for any test function \(\phi\). To derive the integral form, we select a specific test function \(\phi\) defined as: 
\begin{equation}
\phi(x,t) = \begin{cases} 
 1 & \text{for $(x,t)\in[x_1,x_2]\times[t_1,t_2]$ } \\  
 0 & \text{for $(x,t)\notin[x_1-\epsilon,x_2+\epsilon]\times[t_1-\epsilon,t_2+\epsilon]$} 
 \end{cases} 
 \end{equation}
 with $\phi$ smooth in the transition layer of width $\epsilon$. Letting $\epsilon\to 0$ yields the conservative integral balance \cite{LeVeque1992}
\begin{equation}
\begin{split}
\int_{x_1}^{x_2} u(x,t_2)\,dx - \int_{x_1}^{x_2} u(x,t_1)\,dx \\
+\; \int_{t_1}^{t_2} \Big(f(u(x_2,t)) - f(u(x_1,t))\Big)\,dt &= 0,
\end{split}
\label{eq:integral_balance}
\end{equation}
or, equivalently,
\begin{equation}
\frac{d}{dt}\int_{x_1}^{x_2} u(x,t)\,dx
=
f(u(x_1,t)) - f(u(x_2,t)).
\label{eq:integral_form_1d}
\end{equation}
which holds for all \(x_1\), \(x_2\), and \(t\). This integral form is the basis for Finite Volume Methods (FVM). For smooth solutions, differentiating the integral balance with respect to the control volume recovers the differential form. For discontinuous solutions, the integral/weak form remains meaningful while the strong form does not. This method can also be extended to vector space later
\paragraph{Rankine--Hugoniot jump condition (shock speed).}
Suppose $u$ has a single shock located at $x=s(t)$ that separates left and right states
$u_L$ and $u_R$. Consider a thin space-time control volume that straddles the shock curve.
Applying the conservation balance \eqref{eq:integral_balance} and letting the thickness shrink to zero yields the Rankine--Hugoniot condition:
\begin{equation}
\dot{s}(t)\,[u] = [f(u)],
\
[u] := u_R - u_L, 
\ 
[f(u)] := f(u_R) - f(u_L),
\end{equation}
so that the shock speed is
\begin{equation}
\dot{s}(t) = \frac{f(u_R)-f(u_L)}{u_R-u_L}.
\label{eq:RH_speed}
\end{equation}
This jump condition is the statement that the discontinuity moves in a way that preserves the integral conservation law.

To extend to vector form, given in
Equation~\eqref{eq:integral_form}:
\begin{equation}\label{eq:integral_form}
\int_\Omega u(t,x)\,\mathrm{d}\Omega
=\int_\Omega h(x)\,\mathrm{d}\Omega
-\int_0^t \!\!\int_{\Gamma} F(u)\!\cdot\!n\,\mathrm{d}\Gamma\,\mathrm{d}t,
\end{equation}
we first integrate the differential form of the conservation law, given in
Equation~\eqref{eq:differential_form}:
\begin{equation}\label{eq:differential_form}
\mathcal{F}u \;=\; u_t \;+\;\nabla\!\cdot\!F(u),
\end{equation}
over the spatial domain~$\Omega$.  From this, we obtain an expression for the
rate of change in time of the total conserved quantity:
\begin{subequations}\label{eq:rate_of_change}
\begin{align}
\frac{\mathrm{d}}{\mathrm{d}t}\int_\Omega u(t,x)\,\mathrm{d}\Omega
&= \int_\Omega u_t(t,x)\,\mathrm{d}\Omega, \label{eq:roc_a}\\
&= -\int_\Omega \nabla\!\cdot\!F(u)\,\mathrm{d}\Omega, \label{eq:roc_b}\\
&= -\int_{\Gamma} F(u)\!\cdot\!n\,\mathrm{d}\Gamma. \label{eq:roc_c}
\end{align}
\end{subequations}
Next, integrating Equation~\eqref{eq:roc_a} in time from $0$ to $t$ gives
\begin{equation}\label{eq:time_integral}
\int_{0}^{t}\!\!\int_\Omega u_{\tau}(\tau,x)\,\mathrm{d}\Omega\,\mathrm{d}\tau
= \int_\Omega u(t,x)\,\mathrm{d}\Omega
- \int_\Omega u(0,x)\,\mathrm{d}\Omega,
\end{equation}
where $u(0,x)=h(x)$ is the prescribed initial condition.  Equating the right‐hand side
of~\eqref{eq:time_integral} with the temporal integral of~\eqref{eq:roc_c}
recovers the integral form~\eqref{eq:integral_form}.

\section{Mechanism of Shocks and Rarefactions}
\label{sec:appendD}

We consider a nonlinear hyperbolic conservation law in \(\Omega\subset\mathbb{R}^d\),
\begin{equation}
\label{eq:cons_law_vec}
\partial_t \mathbf{q}(\mathbf{x},t)\;+\;\nabla\!\cdot \mathbf{F}\!\bigl(\mathbf{q}(\mathbf{x},t)\bigr)\;=\;\mathbf{0},
\end{equation}
where \(\mathbf{q}(\mathbf{x},t)\in\mathbb{R}^m\) denotes the conserved variables and
\(\mathbf{F}(\mathbf{q})=[\mathbf{F}_1(\mathbf{q}),\dots,\mathbf{F}_d(\mathbf{q})]\) is the flux with \(\mathbf{F}_j(\mathbf{q})\in\mathbb{R}^m\).
Hyperbolicity means that, for any unit direction \(\mathbf{n}\in\mathbb{S}^{d-1}\), the directional flux Jacobian
\begin{equation}
\label{eq:dir_jacobian}
\mathbf{A}(\mathbf{q};\mathbf{n})
\;:=\;
\sum_{j=1}^{d} n_j\,\nabla_{\mathbf{q}}\mathbf{F}_j(\mathbf{q})
\;\in\;\mathbb{R}^{m\times m}
\end{equation}
is real diagonalizable, admitting eigenpairs \(\{(\lambda_i(\mathbf{q};\mathbf{n}),\mathbf{r}_i(\mathbf{q};\mathbf{n}))\}_{i=1}^m\).

To describe the local wave pattern, consider the planar Riemann initial data in direction \(\mathbf{n}\),
\begin{equation}
\label{eq:riemann_planar}
\mathbf{q}(\mathbf{x},0)=
\begin{cases}
\mathbf{q}_L, & \mathbf{n}\cdot\mathbf{x}<0,\\[2pt]
\mathbf{q}_R, & \mathbf{n}\cdot\mathbf{x}>0.
\end{cases}
\end{equation}
For each characteristic family \(i\in\{1,\dots,m\}\), the solution consists of elementary waves determined by the ordering of the characteristic speeds across the jump.

\begin{itemize}
  \item \textbf{Shock waves (compressive).}
  If
  \(\lambda_i(\mathbf{q}_L;\mathbf{n})>\lambda_i(\mathbf{q}_R;\mathbf{n})\),
  then \(i\)-characteristics enter the discontinuity and the jump persists as a shock.
  The shock propagates with speed \(s\) along \(\mathbf{n}\) and satisfies the \emph{Rankine--Hugoniot} condition
  \begin{equation}
  \label{eq:RH_vec}
  s\,[\![\mathbf{q}]\!]
  \;=\;
  [\![\mathbf{F}(\mathbf{q})]\!]\mathbf{n}
  \;=\;
  \sum_{j=1}^{d} n_j\,[\![\mathbf{F}_j(\mathbf{q})]\!],
  \qquad
  [\![\mathbf{q}]\!]:=\mathbf{q}_R-\mathbf{q}_L,
  \end{equation}
  together with the \emph{Lax entropy condition}
  \begin{equation}
  \label{eq:lax_entropy}
  \lambda_i(\mathbf{q}_R;\mathbf{n}) \;<\; s \;<\; \lambda_i(\mathbf{q}_L;\mathbf{n}),
  \end{equation}
  which selects the physically admissible (compressive) shock.

  \item \textbf{Rarefaction waves (expansive).}
  If
  \(\lambda_i(\mathbf{q}_L;\mathbf{n})<\lambda_i(\mathbf{q}_R;\mathbf{n})\),
  then \(i\)-characteristics diverge and the discontinuity spreads into a continuous rarefaction fan.
  The rarefaction admits a self-similar form \(\mathbf{q}(\mathbf{x},t)=\boldsymbol{\Phi}(\xi)\) with similarity variable
  \(\xi=\frac{\mathbf{n}\cdot\mathbf{x}}{t}\).
  In smooth regions, \(\boldsymbol{\Phi}\) satisfies the ODE
  \begin{equation}
  \label{eq:rarefaction_ode}
  \frac{d\boldsymbol{\Phi}}{d\xi}
  \;\parallel\;
  \mathbf{r}_i\!\bigl(\boldsymbol{\Phi};\mathbf{n}\bigr),
  \qquad
  \xi=\lambda_i\!\bigl(\boldsymbol{\Phi};\mathbf{n}\bigr),
  \end{equation}
  with boundary states \(\boldsymbol{\Phi}(\lambda_i(\mathbf{q}_L;\mathbf{n}))=\mathbf{q}_L\) and
  \(\boldsymbol{\Phi}(\lambda_i(\mathbf{q}_R;\mathbf{n}))=\mathbf{q}_R\).
\end{itemize}

The above characterization explains how discontinuities and smooth fans emerge from the same conservation structure: shocks correspond to compressive wave interactions, while rarefactions correspond to expansive wave spreading.
A detailed treatment can be found in standard references on hyperbolic conservation laws~\cite{LeVeque1992shock,LeVeque1992rare}.

\newpage
\section{Full Table of Euler system}
\label{fulltable}



\begin{table}[H]
\centering
\caption{Error metrics for Sod and Lax shock tubes (values shown in $10^{-3}$ units). Each entry reports $(\text{Normalized }L_2,\ \text{Normalized }L_1,\ L_\infty)$. Best is \textbf{bold}; second best is \underline{underlined}.}
\label{tab:sod_lax_compact}
\scriptsize
\renewcommand{\arraystretch}{1.08}
\setlength{\tabcolsep}{3pt}

\begin{adjustbox}{max width=\textwidth}
\begin{tabular}{l l l l cccccccc}
\toprule
\textbf{Case} & \textbf{Comp.} & \textbf{Region} & \textbf{Metric}
& \cellcolor{green!10}\textbf{CI-PINN}
& \textbf{cvPINN} & \textbf{IPINN}
& \textbf{Modified\_MLP} & \textbf{Causal} & \textbf{RWF}
& \textbf{Adaptive lr} & \textbf{Vanilla PINN} \\
\midrule
\multirow{9}{*}{\textbf{Sod}}
& \multirow{3}{*}{\textbf{$u$}} & Global & \shortstack{Norm.\ $L_2$\\Norm.\ $L_1$\\$L_\infty$}
& \cellcolor{green!10}\shortstack{\textbf{20.68}\\\textbf{4.42}\\\underline{412.73}}
& \shortstack{\underline{20.94}\\\underline{6.86}\\443.43}
& \shortstack{243.46\\138.43\\689.79}
& \shortstack{206.73\\147.04\\695.54}
& \shortstack{226.93\\126.42\\638.42}
& \shortstack{209.91\\147.20\\790.95}
& \shortstack{149.47\\113.01\\\textbf{407.81}}
& \shortstack{243.91\\133.38\\464.31} \\
&  & Shock & \shortstack{Norm.\ $L_2$\\Norm.\ $L_1$\\$L_\infty$}
& \cellcolor{green!10}\shortstack{\underline{45.02}\\\textbf{12.99}\\\underline{412.73}}
& \shortstack{\textbf{44.92}\\\underline{22.15}\\443.43}
& \shortstack{316.16\\218.10\\689.79}
& \shortstack{361.16\\313.58\\695.54}
& \shortstack{218.41\\155.21\\602.79}
& \shortstack{376.38\\326.31\\790.95}
& \shortstack{212.58\\189.30\\\textbf{407.81}}
& \shortstack{314.88\\214.45\\464.31} \\
&  & Rarefaction & \shortstack{Norm.\ $L_2$\\Norm.\ $L_1$\\$L_\infty$}
& \cellcolor{green!10}\shortstack{\underline{57.97}\\\textbf{24.80}\\\underline{412.73}}
& \shortstack{\textbf{53.44}\\\underline{26.03}\\443.43}
& \shortstack{275.83\\232.65\\559.87}
& \shortstack{150.17\\110.49\\695.54}
& \shortstack{96.94\\76.96\\418.82}
& \shortstack{147.68\\160.69\\790.95}
& \shortstack{195.19\\163.70\\\textbf{407.81}}
& \shortstack{274.90\\232.01\\464.31} \\
\cmidrule(lr){2-12}
& \multirow{3}{*}{\textbf{$p$}} & Global & \shortstack{Norm.\ $L_2$\\Norm.\ $L_1$\\$L_\infty$}
& \cellcolor{green!10}\shortstack{\textbf{30.75}\\\textbf{8.24}\\1213.53}
& \shortstack{\underline{37.32}\\\underline{14.78}\\1517.11}
& \shortstack{435.16\\268.63\\6311.43}
& \shortstack{214.04\\152.58\\\textbf{995.76}}
& \shortstack{388.12\\223.48\\1297.68}
& \shortstack{205.84\\141.56\\\underline{1016.46}}
& \shortstack{387.37\\258.91\\1451.26}
& \shortstack{508.67\\270.93\\1322.80} \\
&  & Shock & \shortstack{Norm.\ $L_2$\\Norm.\ $L_1$\\$L_\infty$}
& \cellcolor{green!10}\shortstack{\textbf{61.12}\\\textbf{18.37}\\1213.53}
& \shortstack{\underline{74.45}\\\underline{38.80}\\1517.11}
& \shortstack{462.70\\334.03\\5158.10}
& \shortstack{343.41\\288.09\\\textbf{995.76}}
& \shortstack{540.51\\382.49\\1297.68}
& \shortstack{340.50\\280.34\\\underline{1016.46}}
& \shortstack{457.73\\360.26\\1451.26}
& \shortstack{502.94\\336.94\\1322.65} \\
&  & Rarefaction & \shortstack{Norm.\ $L_2$\\Norm.\ $L_1$\\$L_\infty$}
& \cellcolor{green!10}\shortstack{\textbf{92.06}\\\textbf{55.51}\\\underline{806.56}}
& \shortstack{\underline{94.89}\\\underline{58.92}\\\textbf{719.72}}
& \shortstack{358.70\\296.01\\956.57}
& \shortstack{282.70\\237.44\\995.76}
& \shortstack{297.40\\248.81\\1144.46}
& \shortstack{273.91\\229.05\\1016.46}
& \shortstack{585.86\\482.13\\1451.26}
& \shortstack{795.28\\685.69\\1322.63} \\
\cmidrule(lr){2-12}
& \multirow{3}{*}{$\boldsymbol{\rho}$} & Global & \shortstack{Norm.\ $L_2$\\Norm.\ $L_1$\\$L_\infty$}
& \cellcolor{green!10}\shortstack{\textbf{26.40}\\\textbf{7.64}\\1213.02}
& \shortstack{\underline{37.87}\\\underline{13.89}\\1433.86}
& \shortstack{996.53\\673.78\\6135.32}
& \shortstack{183.28\\131.72\\\underline{1090.46}}
& \shortstack{492.86\\259.88\\1696.85}
& \shortstack{185.19\\132.21\\1477.19}
& \shortstack{228.64\\169.88\\1158.17}
& \shortstack{425.07\\222.89\\\textbf{1070.88}} \\
&  & Shock & \shortstack{Norm.\ $L_2$\\Norm.\ $L_1$\\$L_\infty$}
& \cellcolor{green!10}\shortstack{\textbf{43.38}\\\textbf{14.71}\\1213.02}
& \shortstack{\underline{63.85}\\\underline{31.41}\\1433.86}
& \shortstack{1315.96\\1050.42\\6135.32}
& \shortstack{268.77\\229.84\\\underline{1090.46}}
& \shortstack{812.70\\624.69\\1696.85}
& \shortstack{273.72\\237.73\\1477.19}
& \shortstack{292.70\\253.67\\1158.17}
& \shortstack{579.14\\432.20\\\textbf{1070.88}} \\
&  & Rarefaction & \shortstack{Norm.\ $L_2$\\Norm.\ $L_1$\\$L_\infty$}
& \cellcolor{green!10}\shortstack{\textbf{80.32}\\\textbf{48.48}\\1213.02}
& \shortstack{\underline{93.87}\\\underline{54.09}\\1433.86}
& \shortstack{1507.81\\1451.58\\3333.95}
& \shortstack{208.97\\167.36\\\underline{1090.46}}
& \shortstack{398.77\\325.41\\1556.61}
& \shortstack{213.09\\163.99\\1477.19}
& \shortstack{293.12\\245.43\\1158.17}
& \shortstack{610.09\\524.54\\\textbf{1070.88}} \\
\midrule
\multirow{9}{*}{\textbf{Lax}}
& \multirow{3}{*}{\textbf{$u$}} & Global & \shortstack{Norm.\ $L_2$\\Norm.\ $L_1$\\$L_\infty$}
& \cellcolor{green!10}\shortstack{\textbf{44.13}\\\textbf{16.28}\\\underline{1157.08}}
& \shortstack{\underline{282.17}\\158.04\\1736.74}
& \shortstack{629.87\\496.63\\1385.39}
& \shortstack{410.98\\285.26\\1352.67}
& \shortstack{298.65\\\underline{149.48}\\1326.55}
& \shortstack{443.50\\282.50\\1523.93}
& \shortstack{453.70\\372.99\\\textbf{1084.90}}
& \shortstack{887.05\\747.72\\1318.13} \\
&  & Shock & \shortstack{Norm.\ $L_2$\\Norm.\ $L_1$\\$L_\infty$}
& \cellcolor{green!10}\shortstack{\textbf{83.66}\\\textbf{22.13}\\\underline{1157.08}}
& \shortstack{\underline{452.16}\\\underline{255.35}\\1736.74}
& \shortstack{776.69\\694.10\\1385.39}
& \shortstack{644.51\\553.33\\1352.67}
& \shortstack{568.98\\360.00\\1326.55}
& \shortstack{719.75\\580.34\\1523.93}
& \shortstack{610.56\\564.90\\\textbf{1084.90}}
& \shortstack{930.47\\752.99\\1318.13} \\
&  & Rarefaction & \shortstack{Norm.\ $L_2$\\Norm.\ $L_1$\\$L_\infty$}
& \cellcolor{green!10}\shortstack{\textbf{124.42}\\\textbf{57.73}\\1457.08}
& \shortstack{390.88\\319.74\\1736.74}
& \shortstack{552.93\\486.86\\1385.39}
& \shortstack{625.57\\562.44\\1535.63}
& \shortstack{\underline{349.88}\\\underline{264.66}\\1326.55}
& \shortstack{715.72\\633.13\\1523.94}
& \shortstack{559.44\\541.40\\\textbf{1084.90}}
& \shortstack{920.95\\879.62\\\underline{1318.11}} \\
\cmidrule(lr){2-12}
& \multirow{3}{*}{\textbf{$p$}} & Global & \shortstack{Norm.\ $L_2$\\Norm.\ $L_1$\\$L_\infty$}
& \cellcolor{green!10}\shortstack{\textbf{70.24}\\\textbf{20.09}\\1875.14}
& \shortstack{\underline{357.60}\\\underline{177.78}\\2184.09}
& \shortstack{689.18\\495.92\\\textbf{1703.83}}
& \shortstack{554.36\\349.94\\1908.40}
& \shortstack{642.40\\401.85\\\underline{1809.39}}
& \shortstack{605.80\\354.85\\1900.42}
& \shortstack{493.73\\392.65\\2186.46}
& \shortstack{1054.43\\929.45\\2211.95} \\
&  & Shock & \shortstack{Norm.\ $L_2$\\Norm.\ $L_1$\\$L_\infty$}
& \cellcolor{green!10}\shortstack{\textbf{146.60}\\\textbf{40.79}\\1875.14}
& \shortstack{643.03\\\underline{355.99}\\2184.09}
& \shortstack{1090.38\\837.22\\\textbf{1703.83}}
& \shortstack{878.63\\682.19\\1908.40}
& \shortstack{959.83\\651.30\\\underline{1809.39}}
& \shortstack{956.80\\718.03\\1900.42}
& \shortstack{\underline{636.79}\\522.61\\2186.46}
& \shortstack{1310.55\\988.26\\2206.13} \\
&  & Rarefaction & \shortstack{Norm.\ $L_2$\\Norm.\ $L_1$\\$L_\infty$}
& \cellcolor{green!10}\shortstack{\textbf{292.06}\\\textbf{78.35}\\1875.14}
& \shortstack{496.41\\405.21\\\textbf{297.01}}
& \shortstack{\underline{436.28}\\\underline{336.09}\\\underline{1703.83}}
& \shortstack{875.24\\793.96\\1908.40}
& \shortstack{522.34\\426.60\\1809.39}
& \shortstack{732.43\\908.75\\1900.42}
& \shortstack{737.43\\709.77\\2186.46}
& \shortstack{1145.92\\996.15\\2211.95} \\
\cmidrule(lr){2-12}
& \multirow{3}{*}{$\boldsymbol{\rho}$} & Global & \shortstack{Norm.\ $L_2$\\Norm.\ $L_1$\\$L_\infty$}
& \cellcolor{green!10}\shortstack{\textbf{53.29}\\\textbf{8.12}\\\underline{805.88}}
& \shortstack{211.11\\\underline{90.19}\\970.89}
& \shortstack{1048.62\\809.30\\3184.67}
& \shortstack{204.26\\101.65\\1069.27}
& \shortstack{271.12\\161.48\\\textbf{768.57}}
& \shortstack{215.17\\110.34\\1134.22}
& \shortstack{\underline{204.25}\\95.09\\895.44}
& \shortstack{863.43\\809.09\\1686.26} \\
&  & Shock & \shortstack{Norm.\ $L_2$\\Norm.\ $L_1$\\$L_\infty$}
& \cellcolor{green!10}\shortstack{\textbf{104.38}\\\textbf{63.61}\\\underline{805.88}}
& \shortstack{383.28\\\underline{237.16}\\970.89}
& \shortstack{1315.96\\408.67\\3184.67}
& \shortstack{932.79\\255.95\\1069.27}
& \shortstack{429.95\\342.17\\\textbf{768.57}}
& \shortstack{\underline{243.97}\\279.35\\1134.22}
& \shortstack{1475.46\\265.95\\895.44}
& \shortstack{579.13\\1074.29\\1686.26} \\
&  & Rarefaction & \shortstack{Norm.\ $L_2$\\Norm.\ $L_1$\\$L_\infty$}
& \cellcolor{green!10}\shortstack{\textbf{88.03}\\\textbf{31.13}\\\underline{805.88}}
& \shortstack{\underline{127.97}\\\underline{80.51}\\970.89}
& \shortstack{607.86\\105.17\\3184.67}
& \shortstack{208.97\\109.69\\1069.27}
& \shortstack{213.86\\108.68\\\textbf{768.57}}
& \shortstack{219.02\\123.48\\1134.22}
& \shortstack{145.16\\80.71\\895.44}
& \shortstack{272.28\\681.26\\1686.26} \\
\bottomrule
\end{tabular}
\end{adjustbox}
\end{table}

\newpage
\section{Clips plots}
\label{sec:appendF}


\subsection{Burgers Equation}
\label{sec:appendFbur}

\begin{figure}[!h]
    \centering
    \small 
    
    \includegraphics[width=0.48\linewidth]{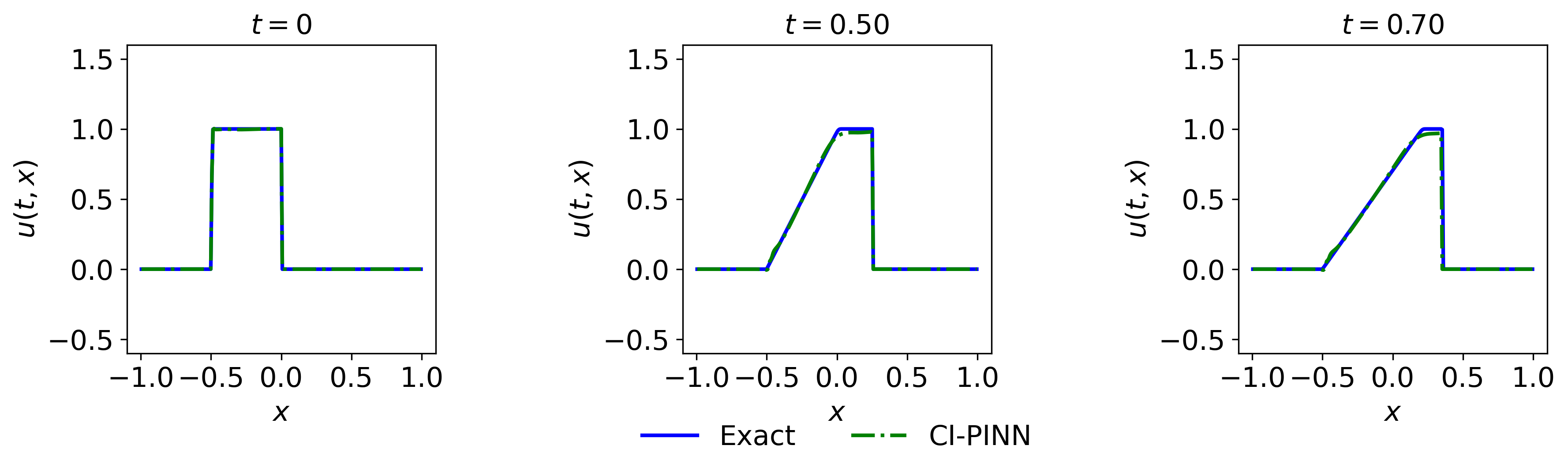} \hfill
    \includegraphics[width=0.48\linewidth]{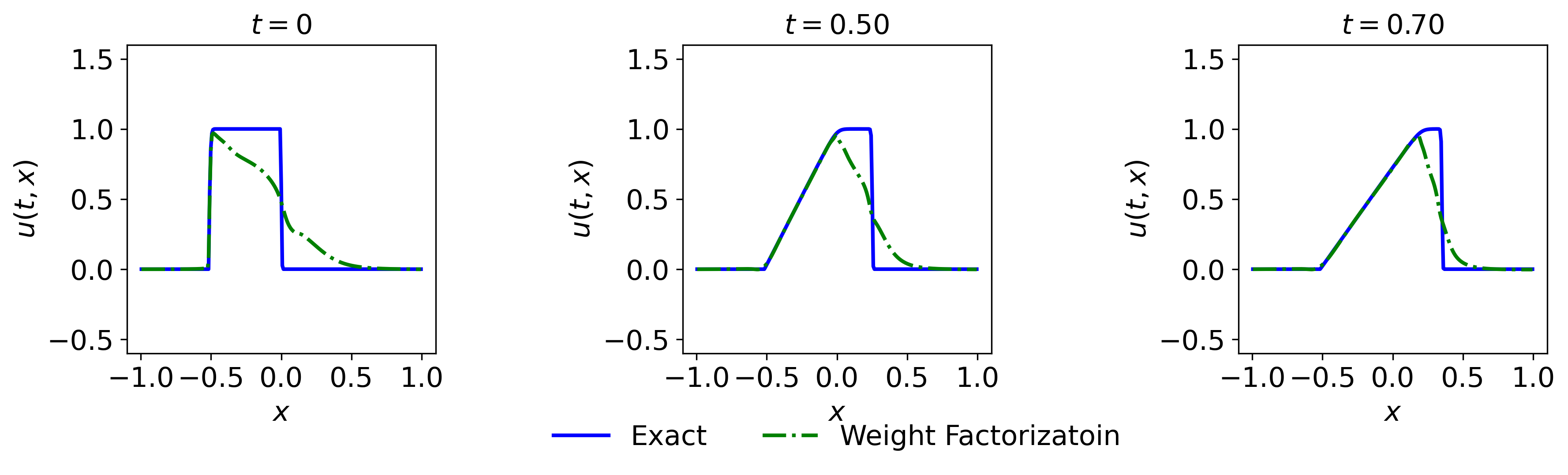}
    \vspace{1mm}

    \includegraphics[width=0.48\linewidth]{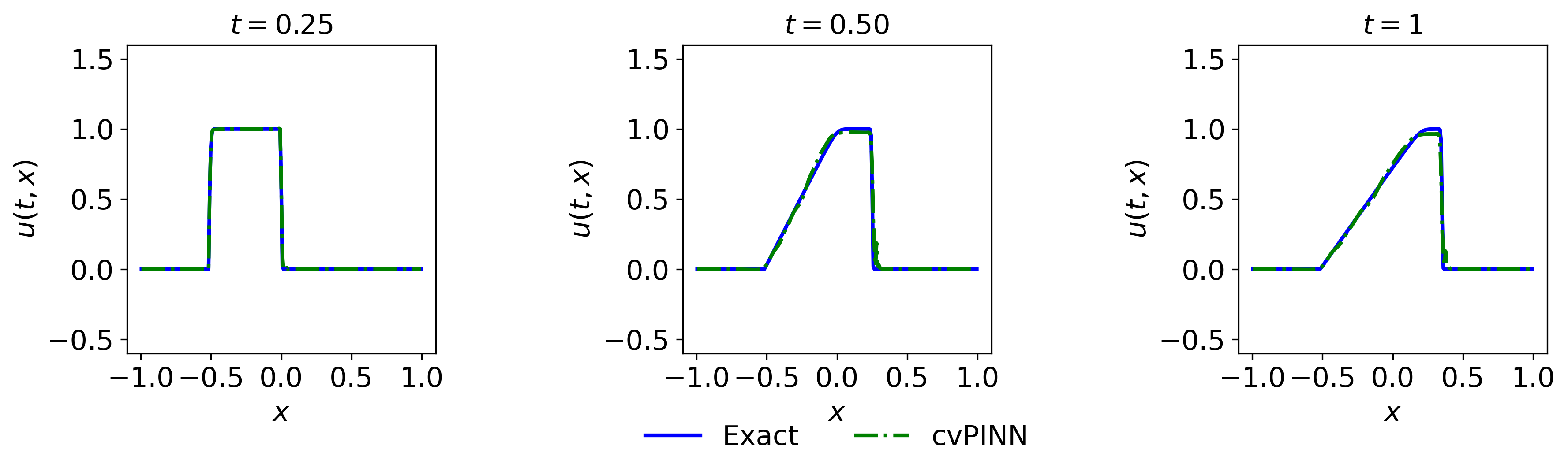} \hfill
    \includegraphics[width=0.48\linewidth]{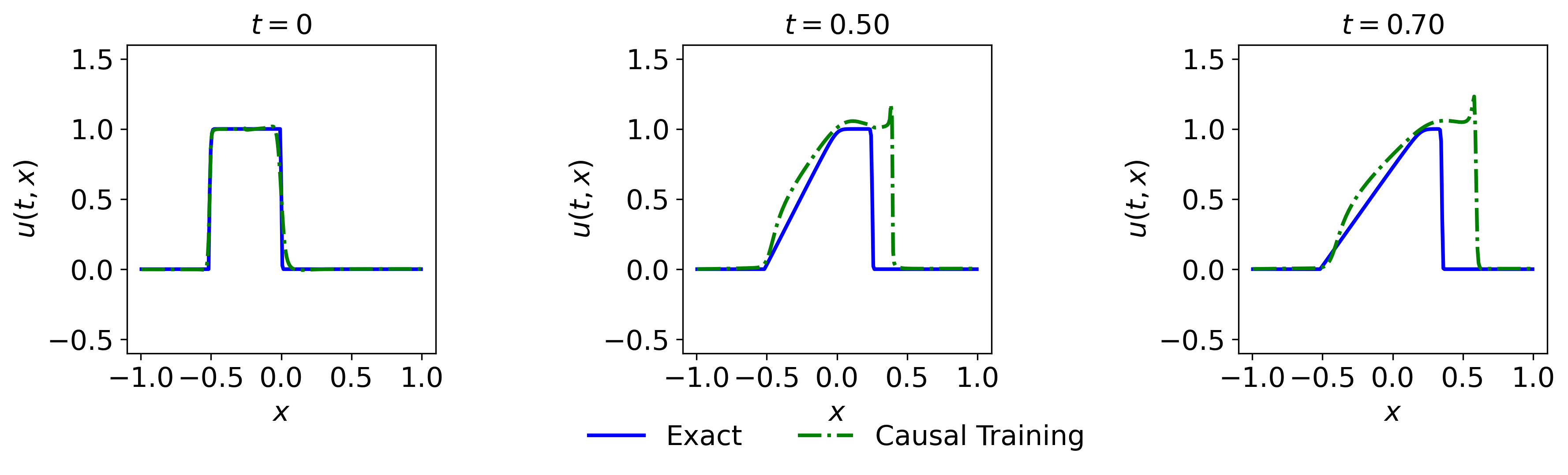}
    \vspace{1mm}

    \includegraphics[width=0.48\linewidth]{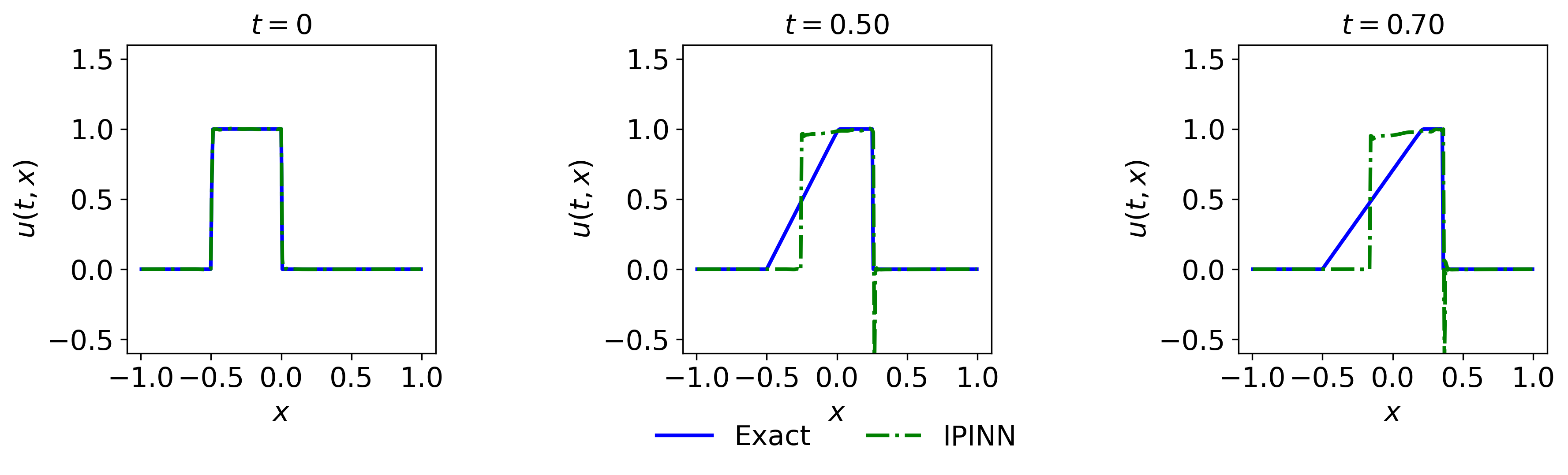} \hfill
    \includegraphics[width=0.48\linewidth]{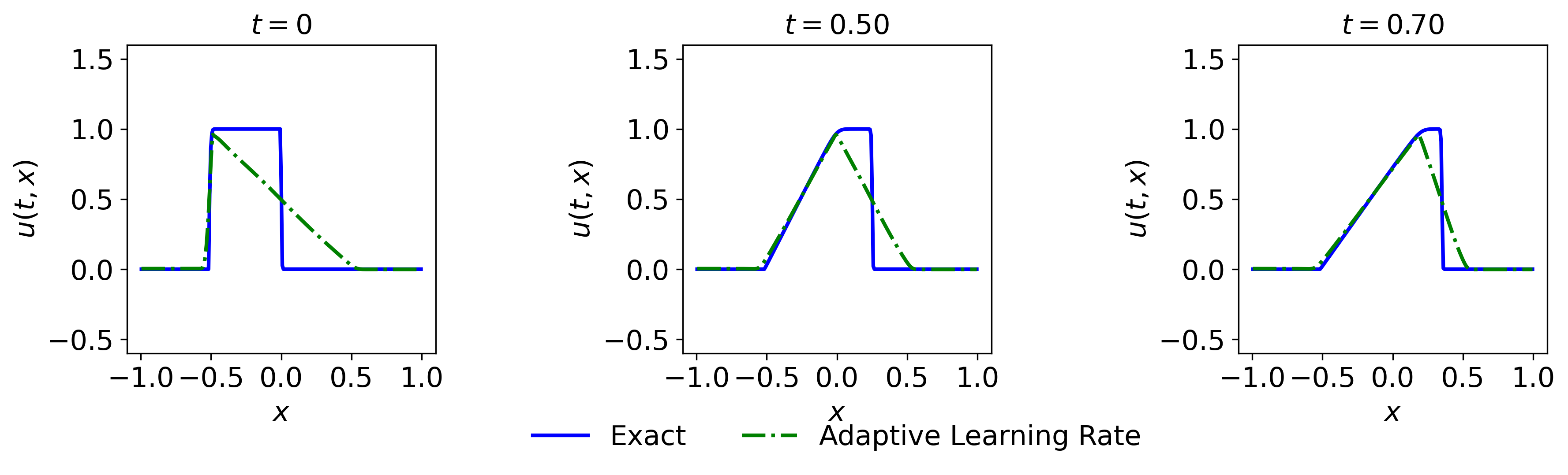}
    \vspace{1mm}

    \includegraphics[width=0.48\linewidth]{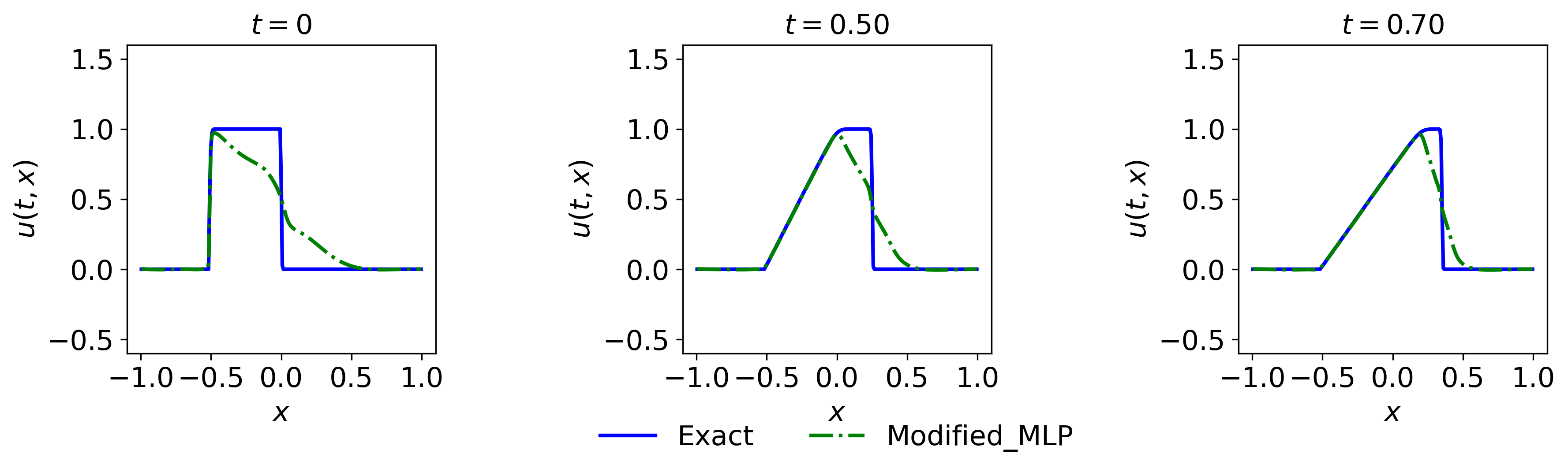} \hfill
    \includegraphics[width=0.48\linewidth]{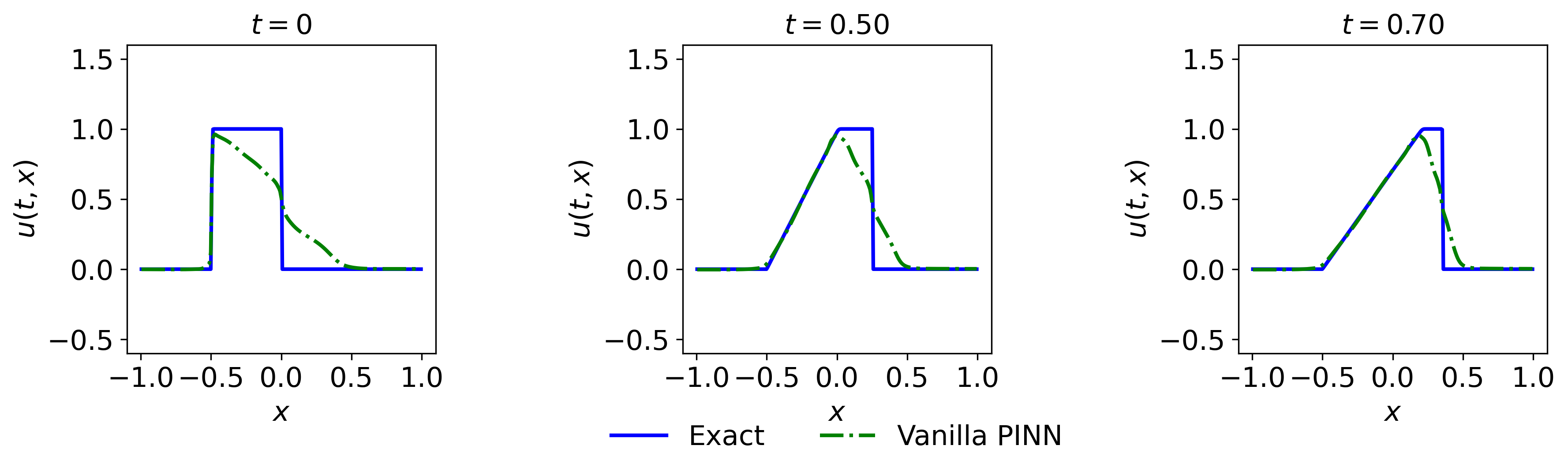}

    \vspace{-2mm}
    \caption{Comparison of predictions for Burger problem.}
    \label{fig:burgerclip}
\end{figure}

\vspace{-4mm} 

\subsection{BL Equation}
\label{sec:appendFbl}

\begin{figure}[H]
    \centering
    \small 
    \includegraphics[width=0.48\linewidth]{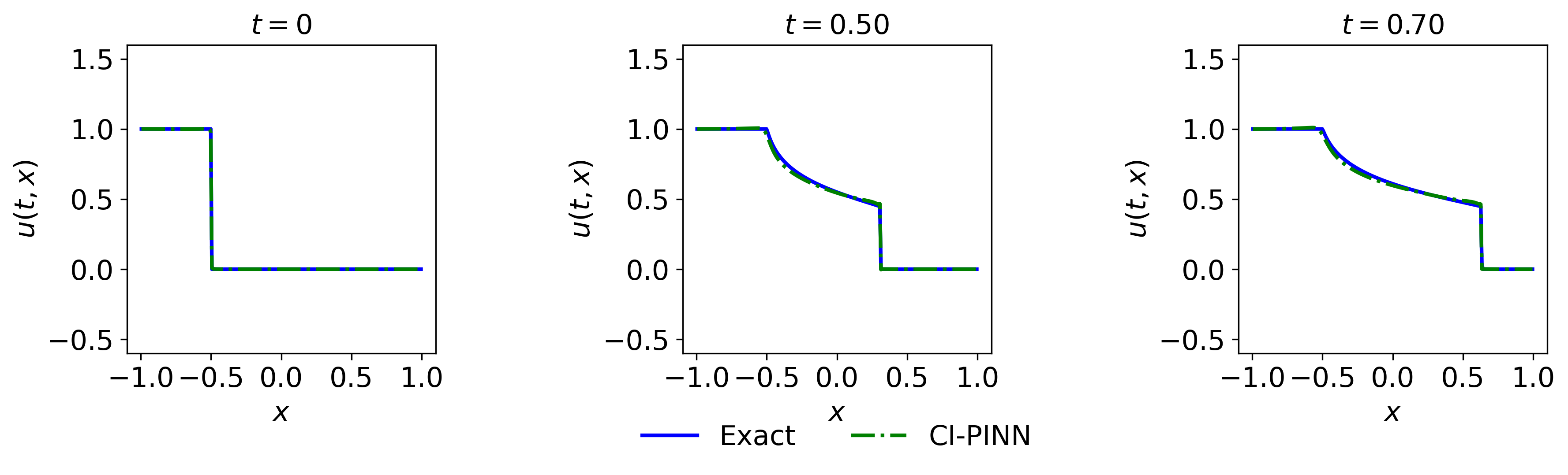} \hfill
    \includegraphics[width=0.48\linewidth]{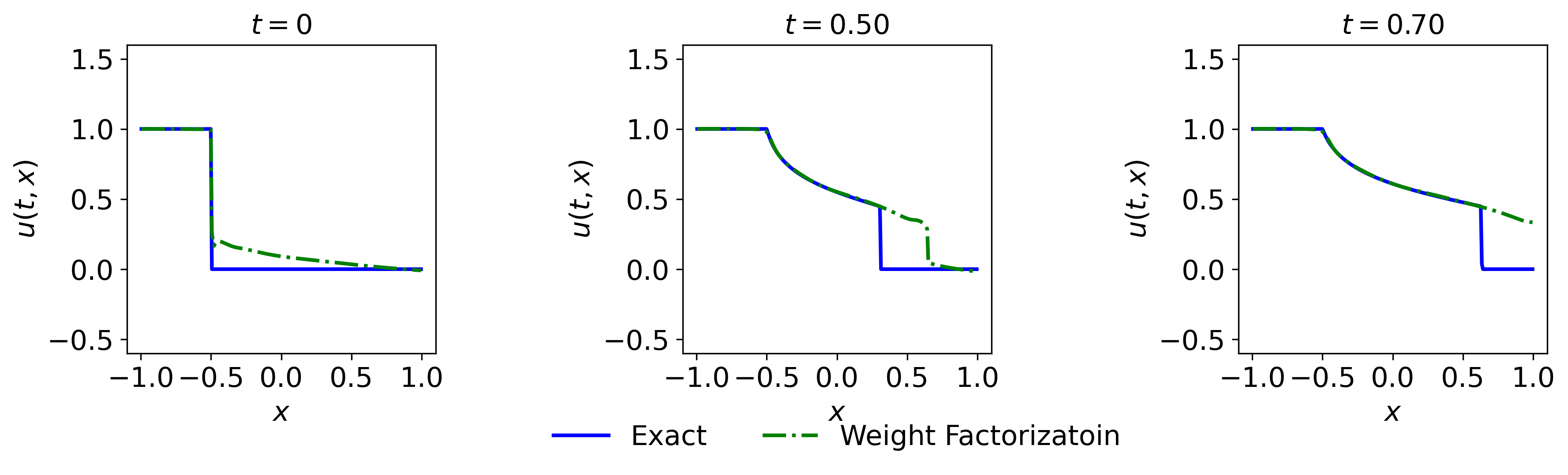}
    \vspace{1mm}
    \includegraphics[width=0.48\linewidth]{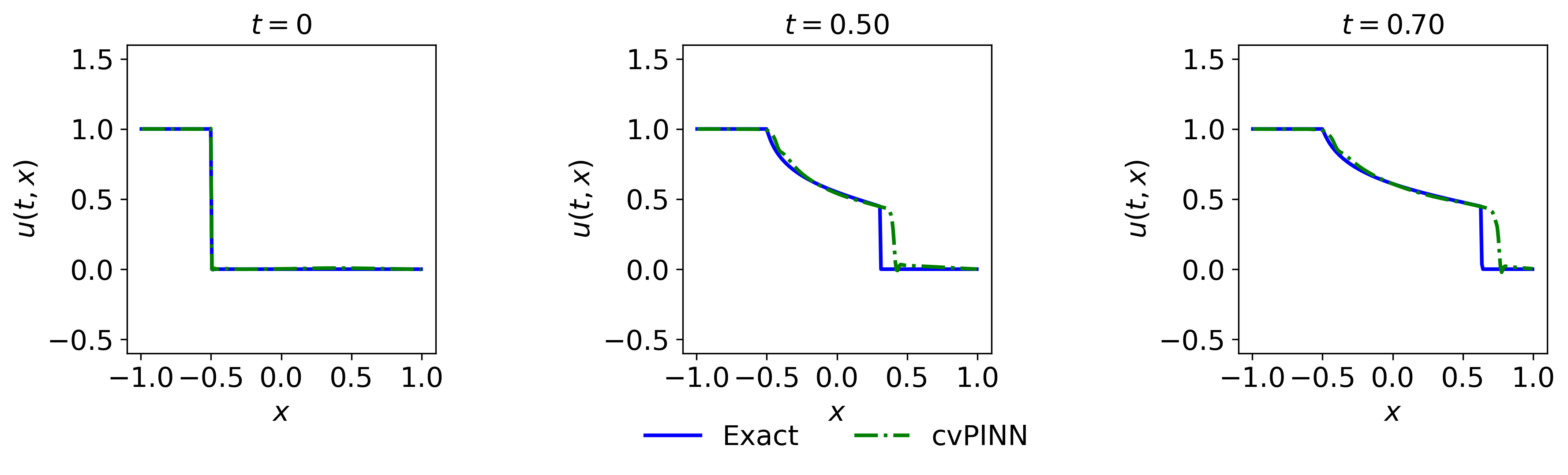} \hfill
    \includegraphics[width=0.48\linewidth]{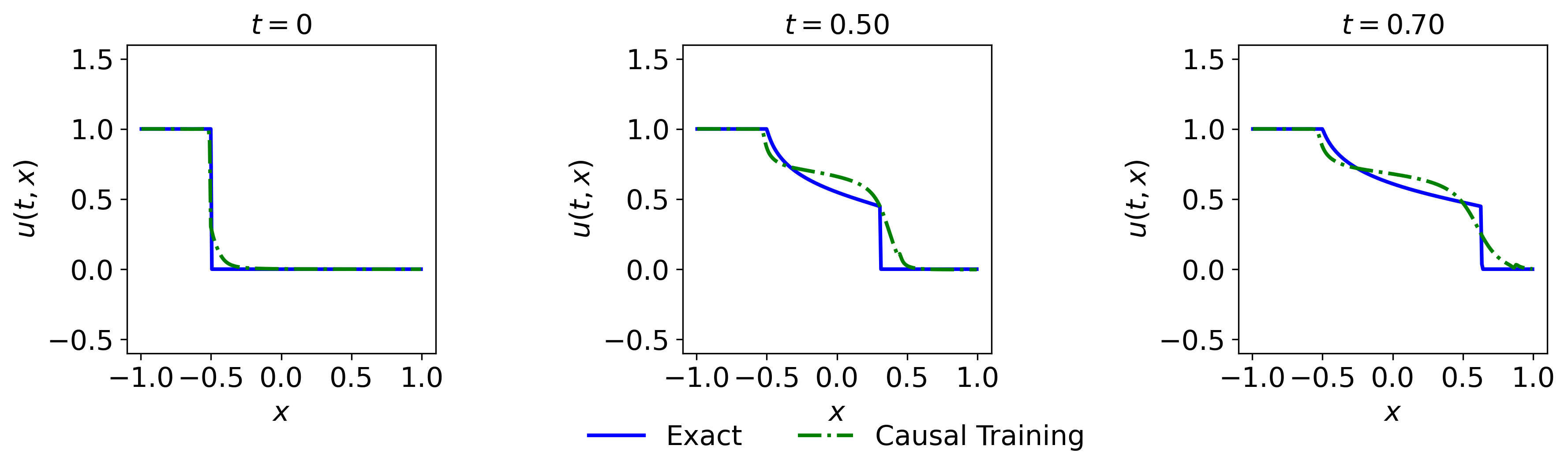}
    \vspace{1mm}
    \includegraphics[width=0.48\linewidth]{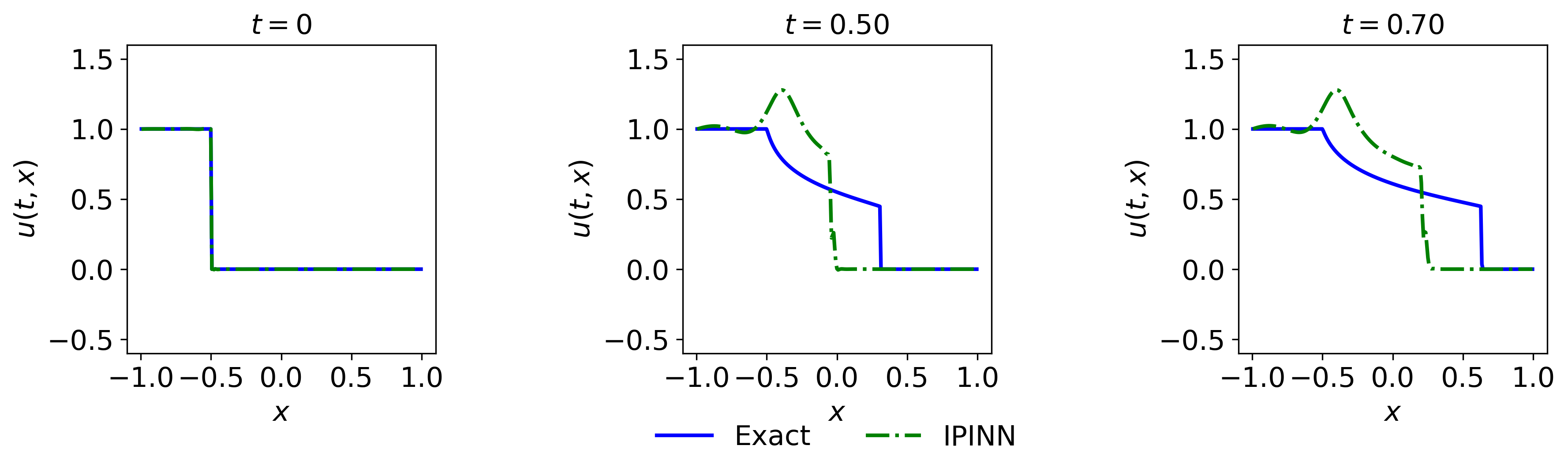} \hfill
    \includegraphics[width=0.48\linewidth]{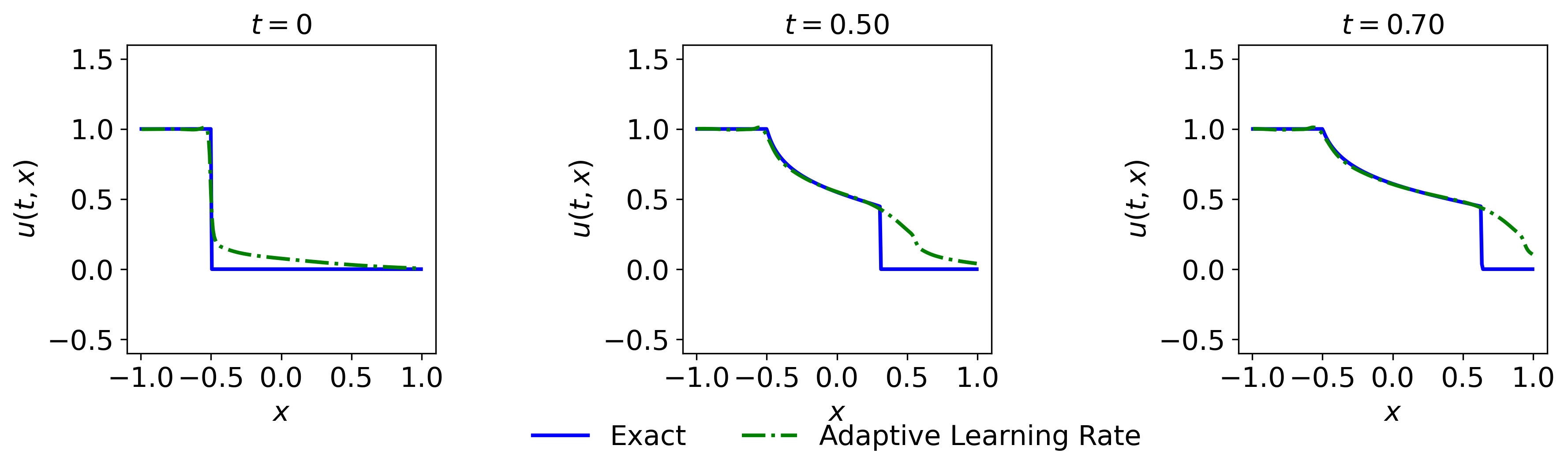}
    \vspace{1mm}
    \includegraphics[width=0.48\linewidth]{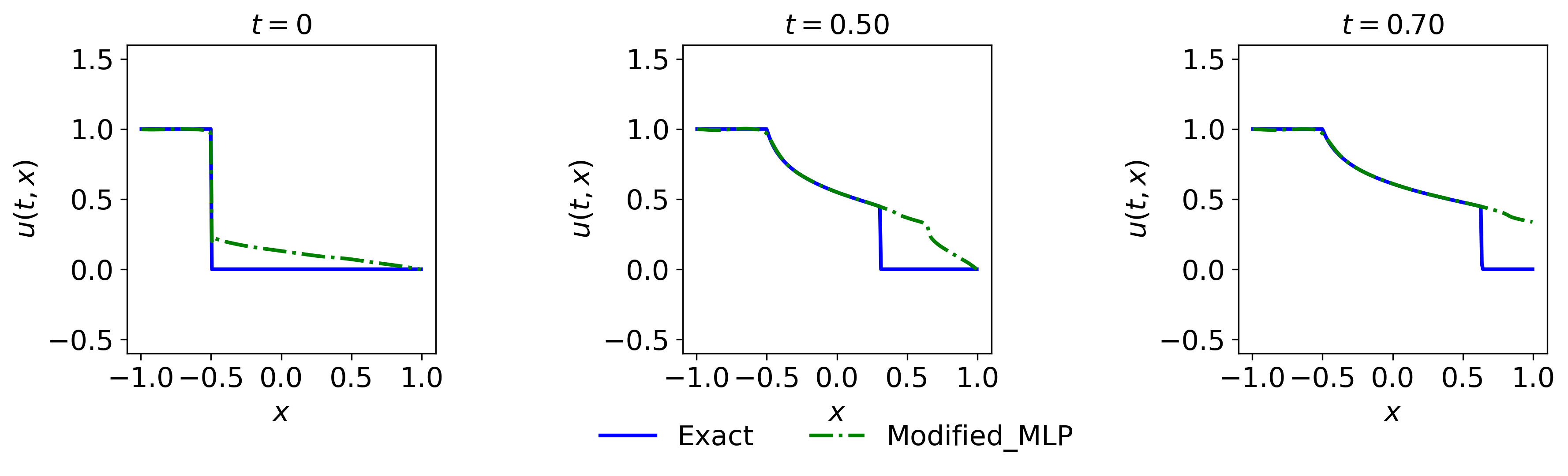} \hfill
    \includegraphics[width=0.48\linewidth]{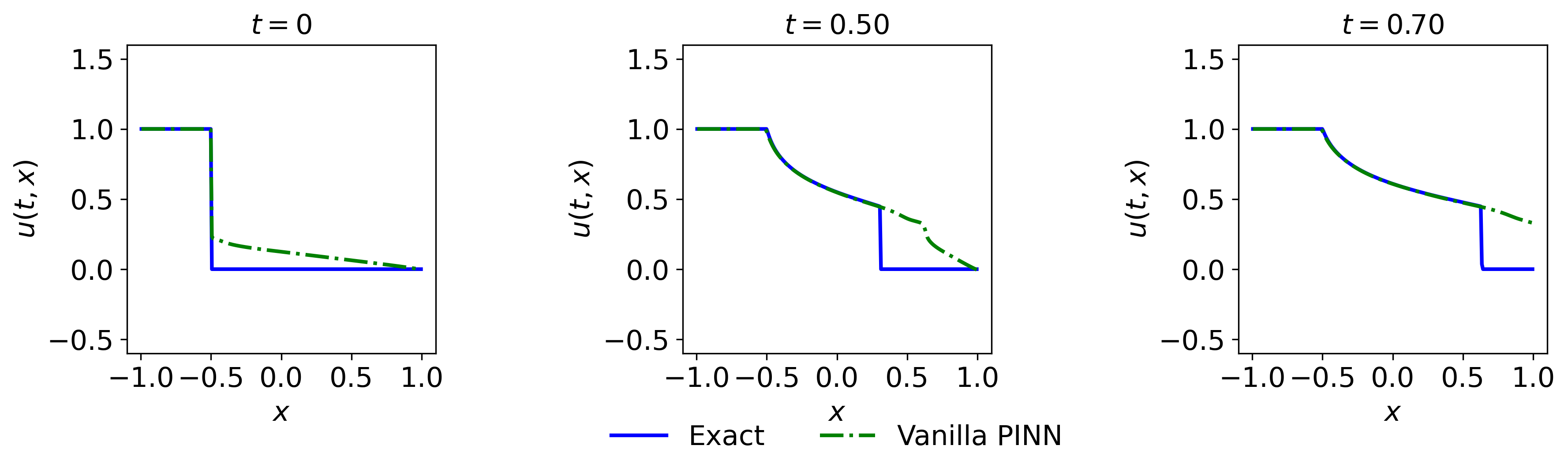}
    \vspace{-2mm}
    \caption{Comparison of predictions for BL problem.}
    \label{fig:blclip}
\end{figure}
\clearpage 
\subsection{Sod Problem}
\label{sec:appendFsod}
\begin{figure}[!h]
    \centering
    \small
    \includegraphics[width=0.48\linewidth]{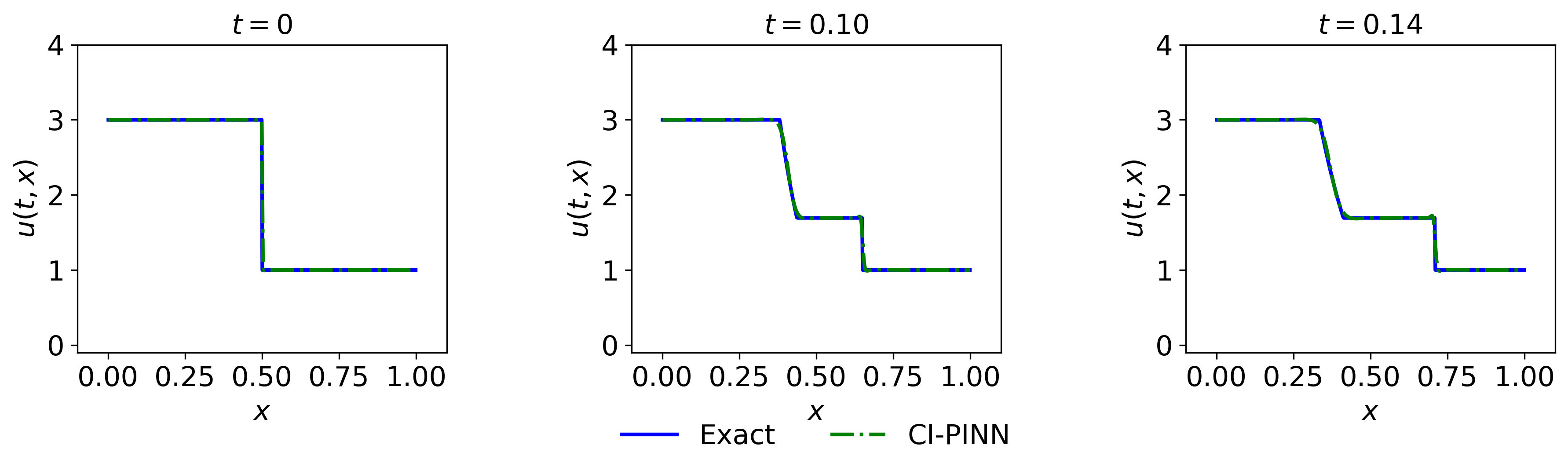} \hfill
    \includegraphics[width=0.48\linewidth]{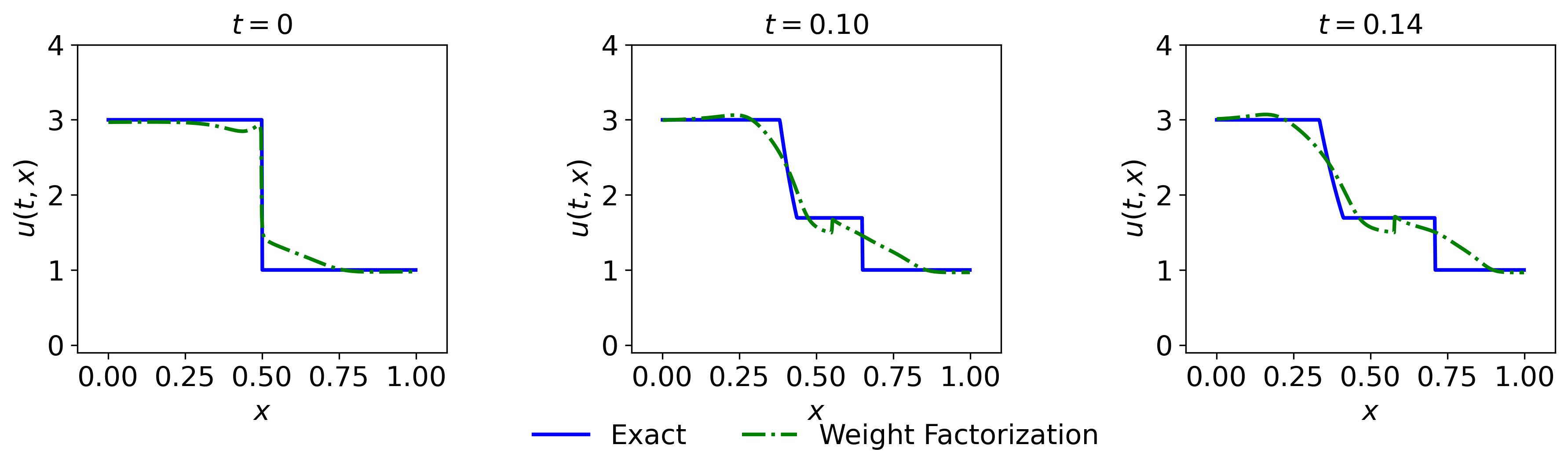}
    \vspace{1mm}
    \includegraphics[width=0.48\linewidth]{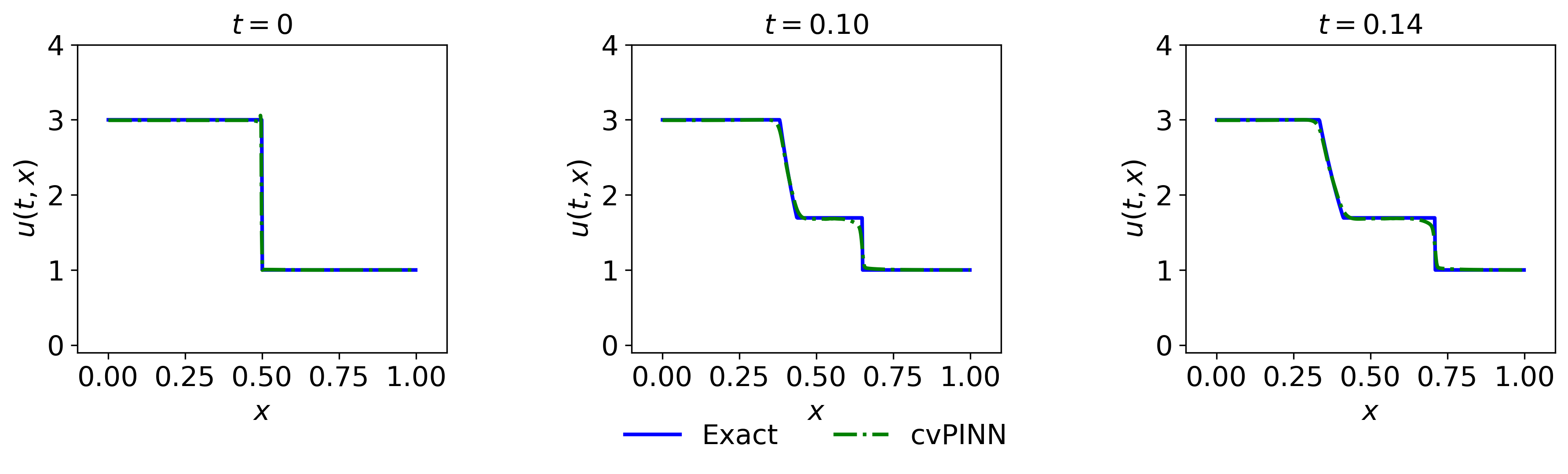} \hfill
    \includegraphics[width=0.48\linewidth]{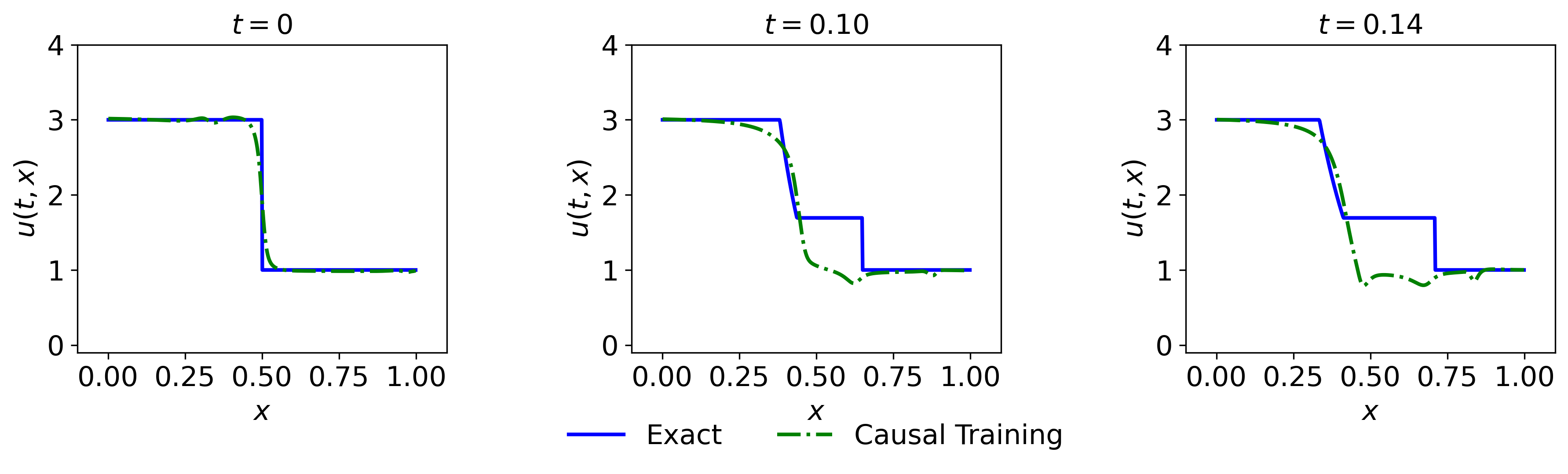}
    \vspace{1mm}
    \includegraphics[width=0.48\linewidth]{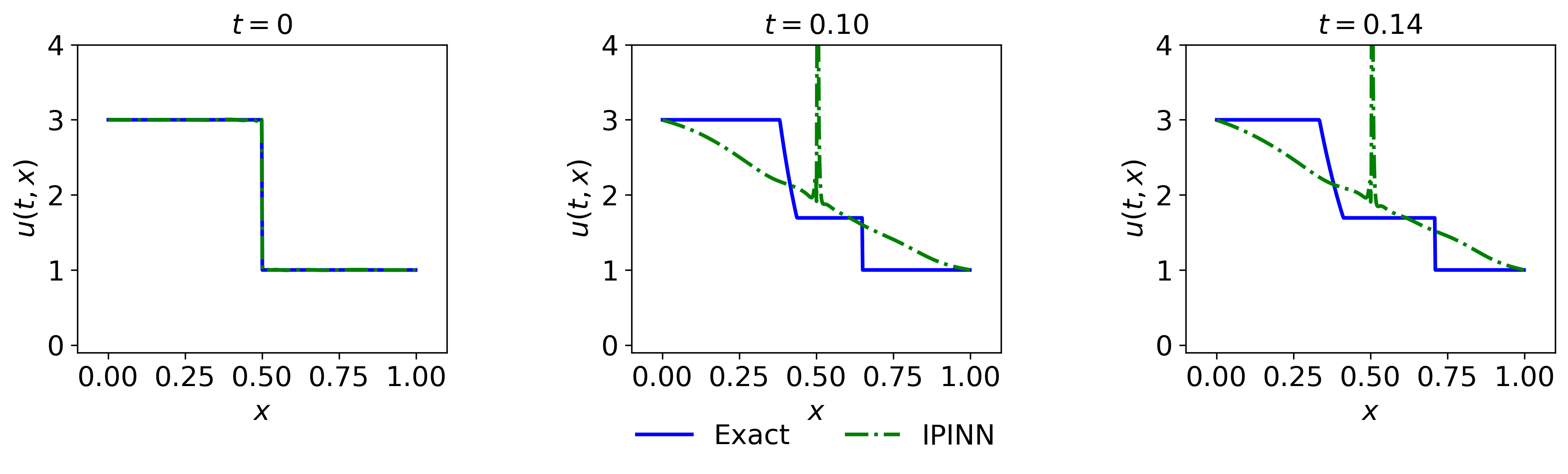} \hfill
    \includegraphics[width=0.48\linewidth]{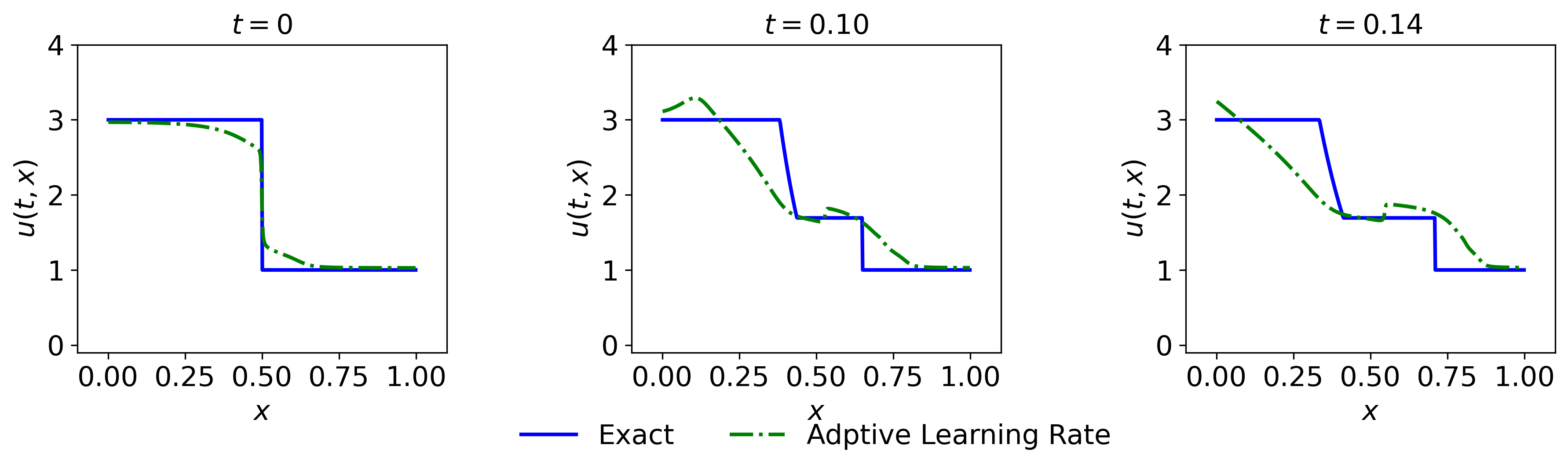}
    \vspace{1mm}
    \includegraphics[width=0.48\linewidth]{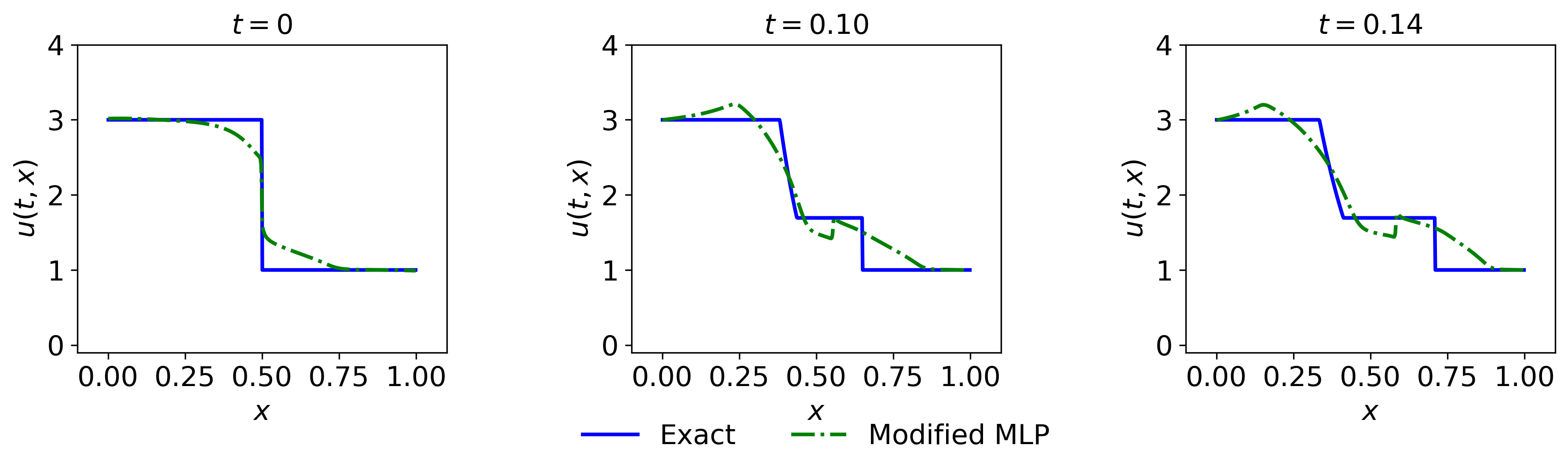} \hfill
    \includegraphics[width=0.48\linewidth]{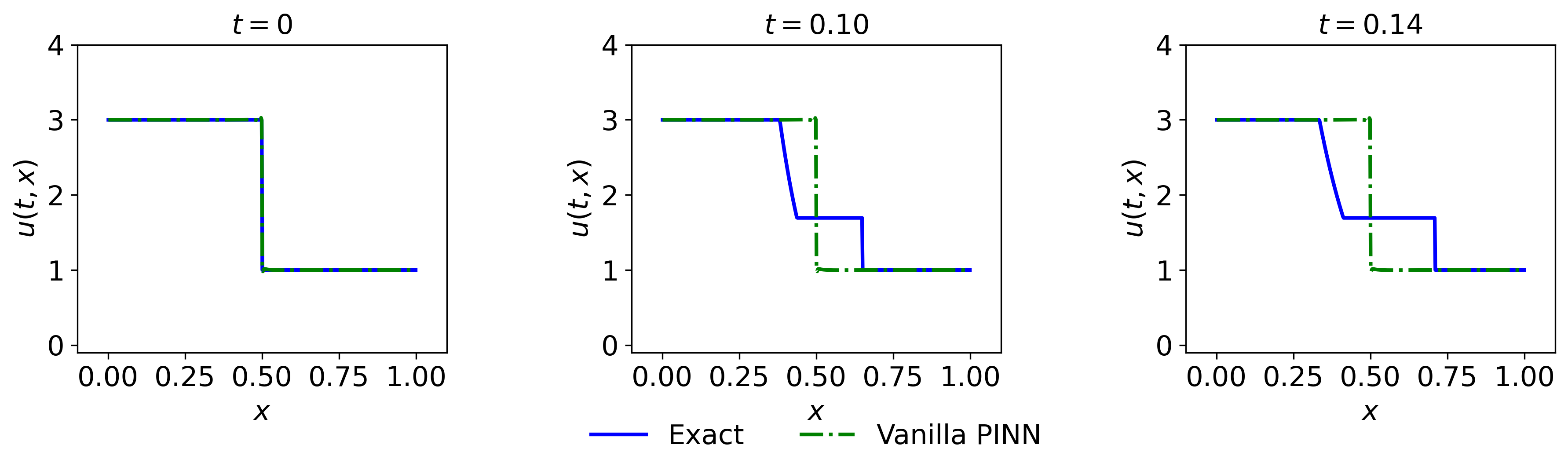}
    \vspace{-2mm}
    \caption{Sod problem: $p$ component.}
    \label{fig:sod_p}
\end{figure}
\vspace{-4mm}

\begin{figure}[!h]
    \centering
    \small
    \includegraphics[width=0.48\linewidth]{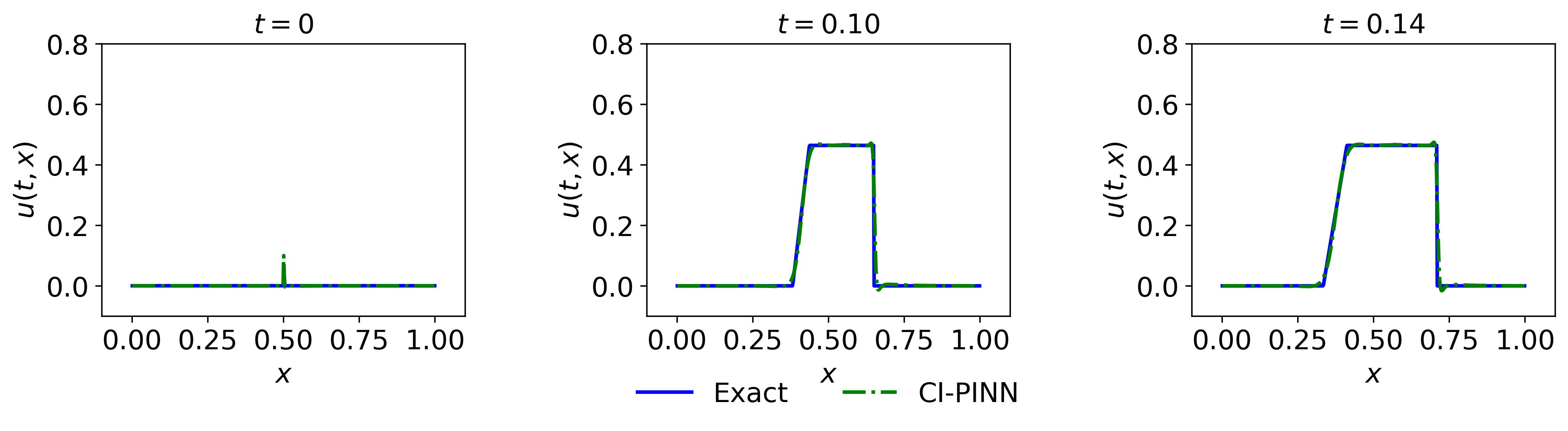} \hfill
    \includegraphics[width=0.48\linewidth]{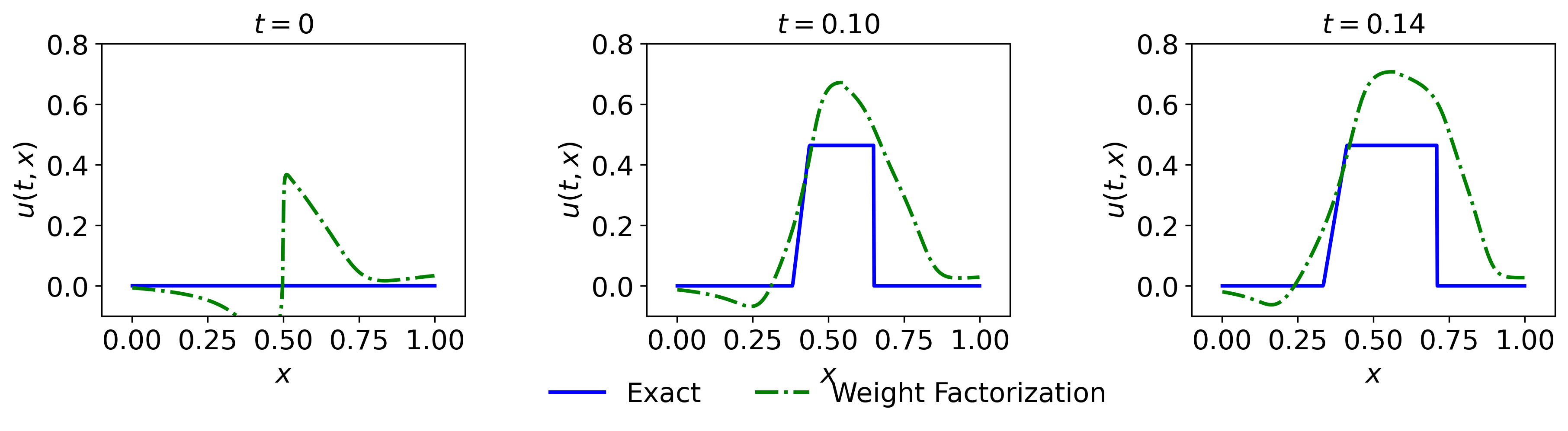}
    \vspace{1mm}
    \includegraphics[width=0.48\linewidth]{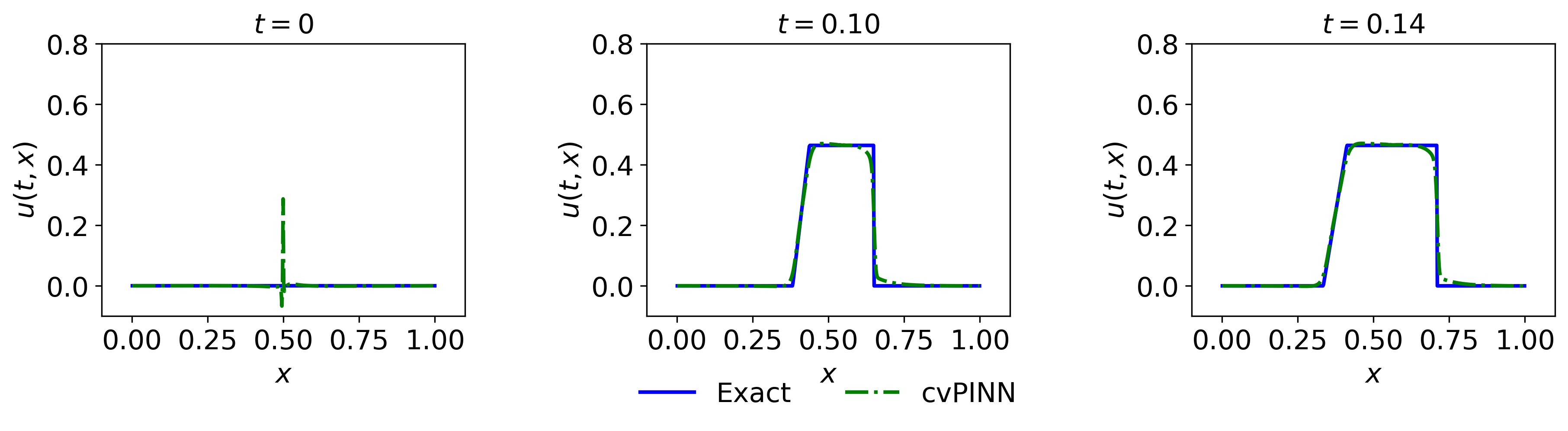} \hfill
    \includegraphics[width=0.48\linewidth]{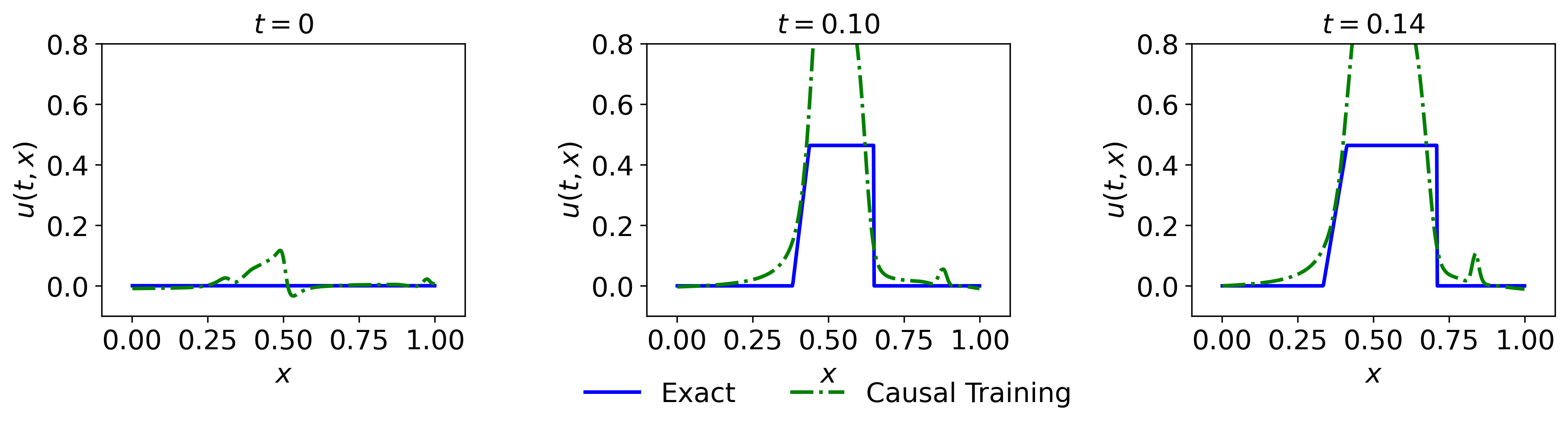}
    \vspace{1mm}
    \includegraphics[width=0.48\linewidth]{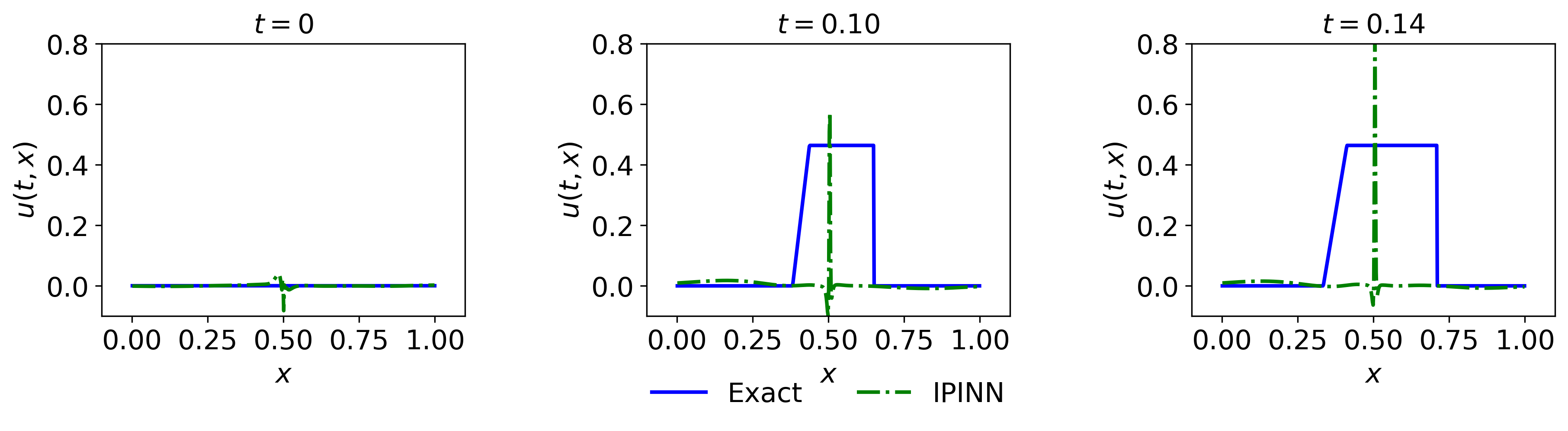} \hfill
    \includegraphics[width=0.48\linewidth]{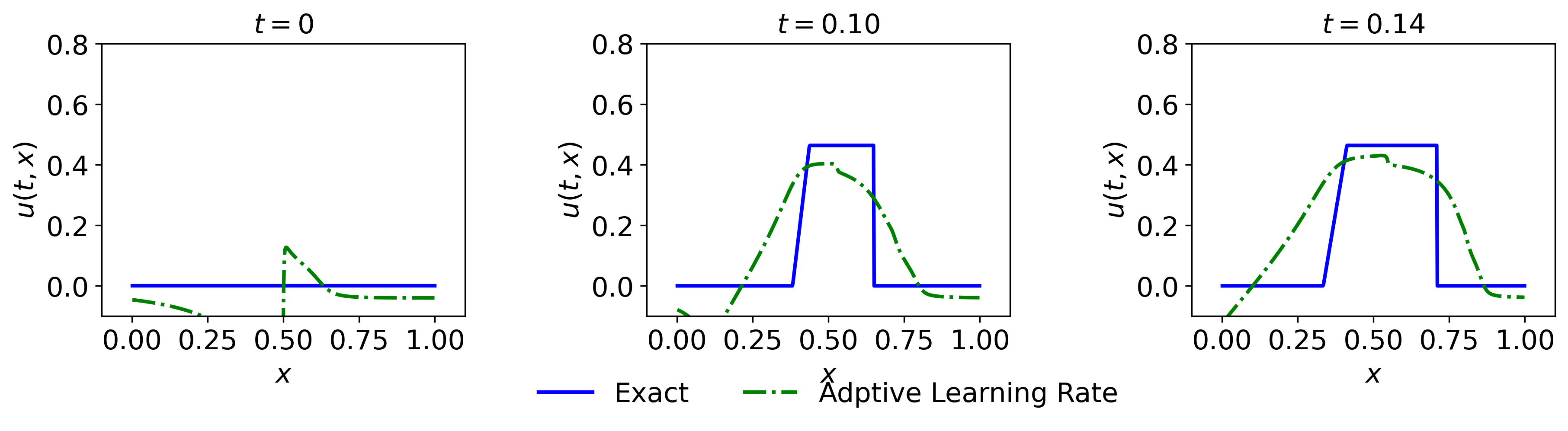}
    \vspace{1mm}
    \includegraphics[width=0.48\linewidth]{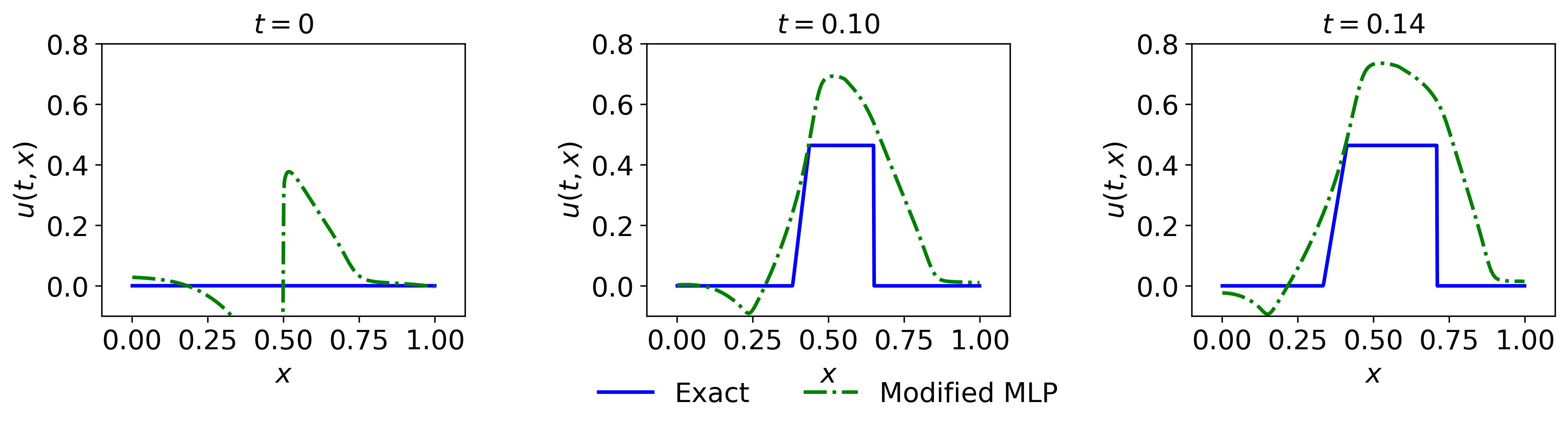} \hfill
    \includegraphics[width=0.48\linewidth]{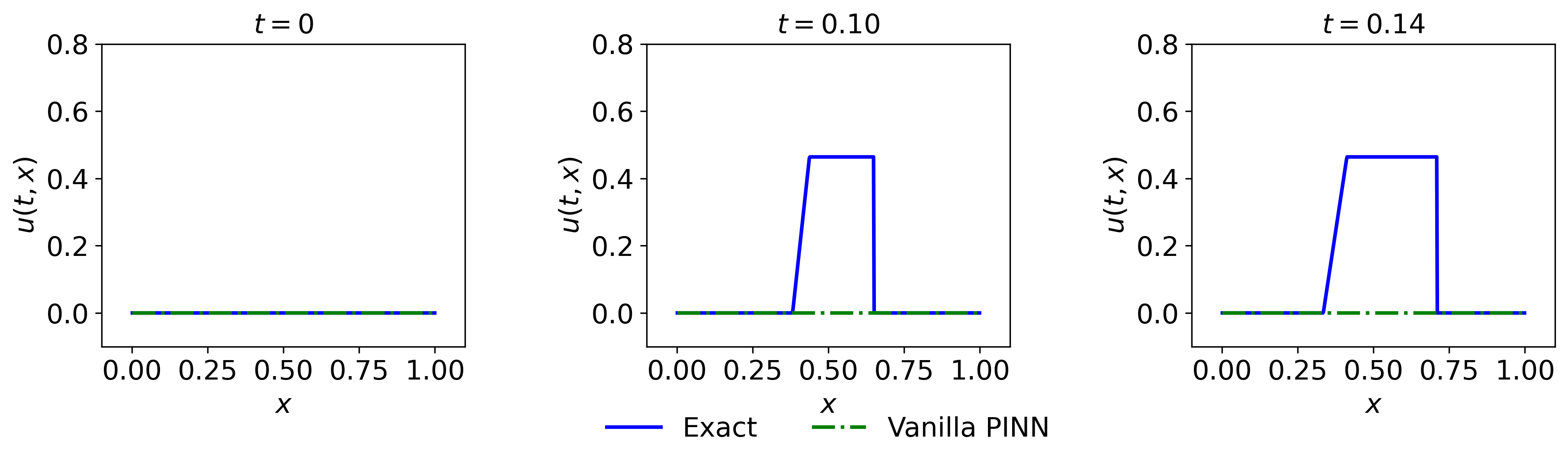}
    \vspace{-2mm}
    \caption{Sod problem: $u$ component.}
    \label{fig:sod_u}
\end{figure}

\clearpage 


\begin{figure}[!h]
    \centering
    \small
    \includegraphics[width=0.48\linewidth]{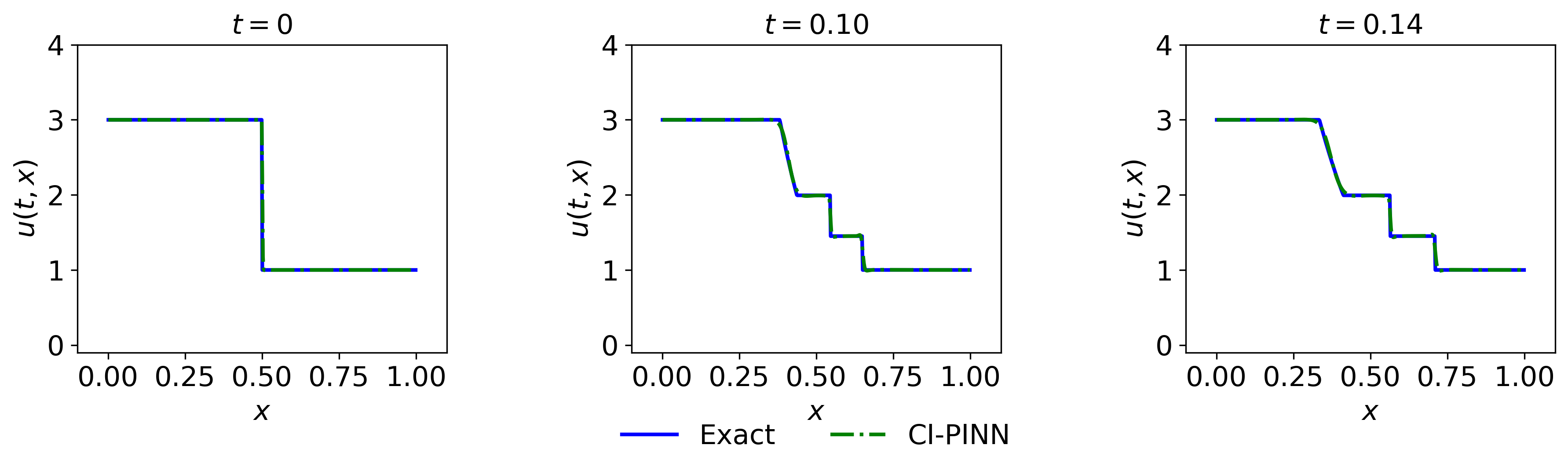} \hfill
    \includegraphics[width=0.48\linewidth]{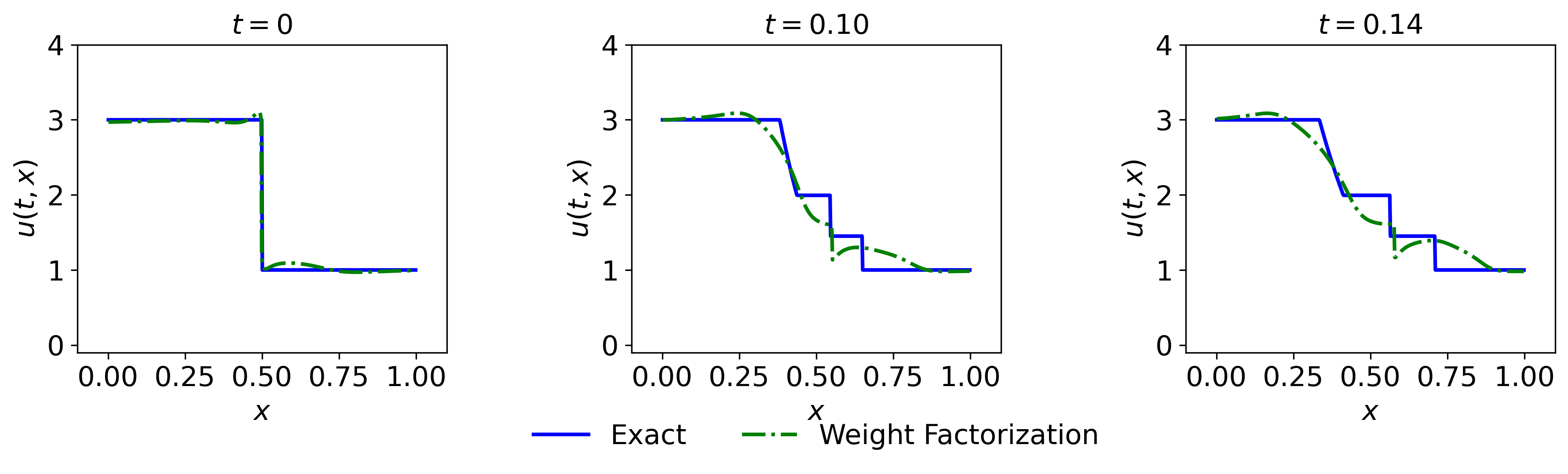}
    \vspace{1mm}
    \includegraphics[width=0.48\linewidth]{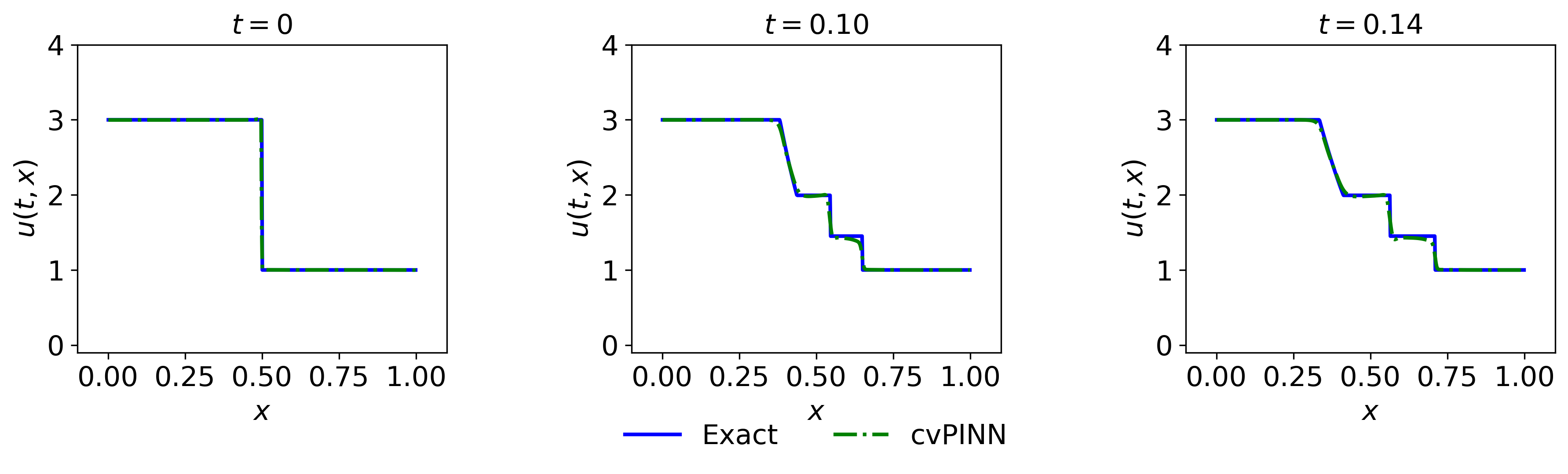} \hfill
    \includegraphics[width=0.48\linewidth]{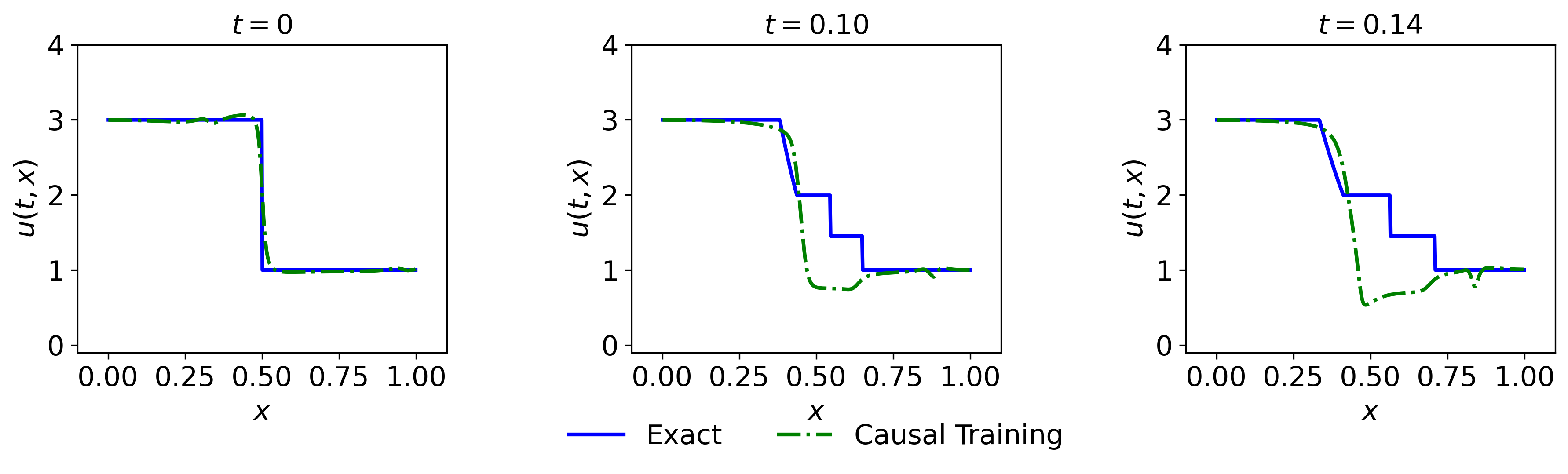}
    \vspace{1mm}
    \includegraphics[width=0.48\linewidth]{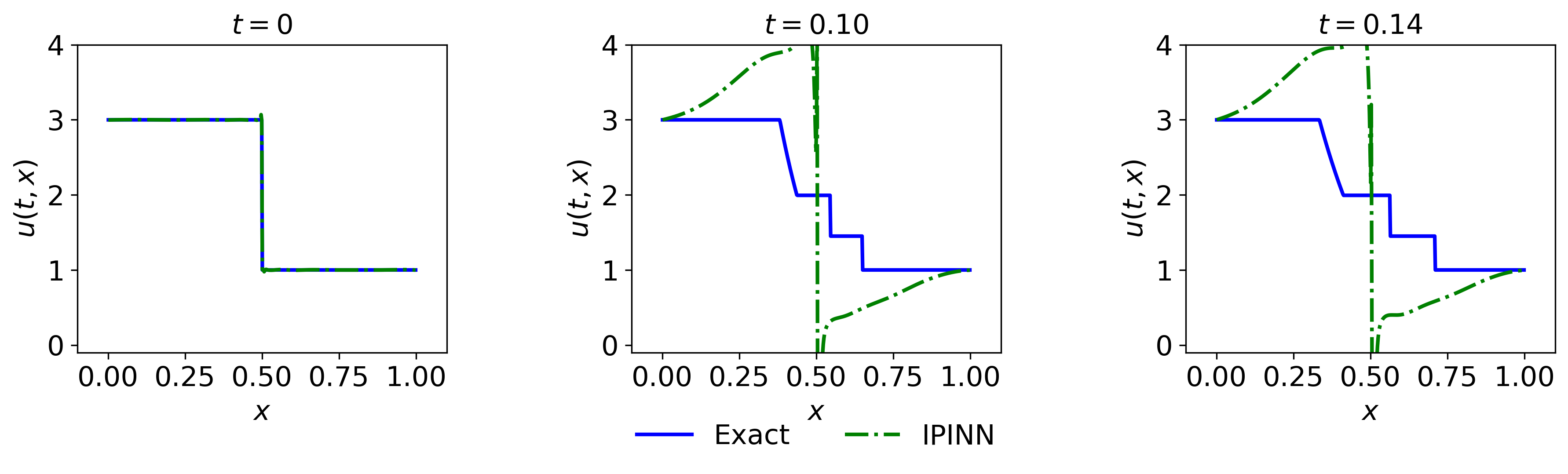} \hfill
    \includegraphics[width=0.48\linewidth]{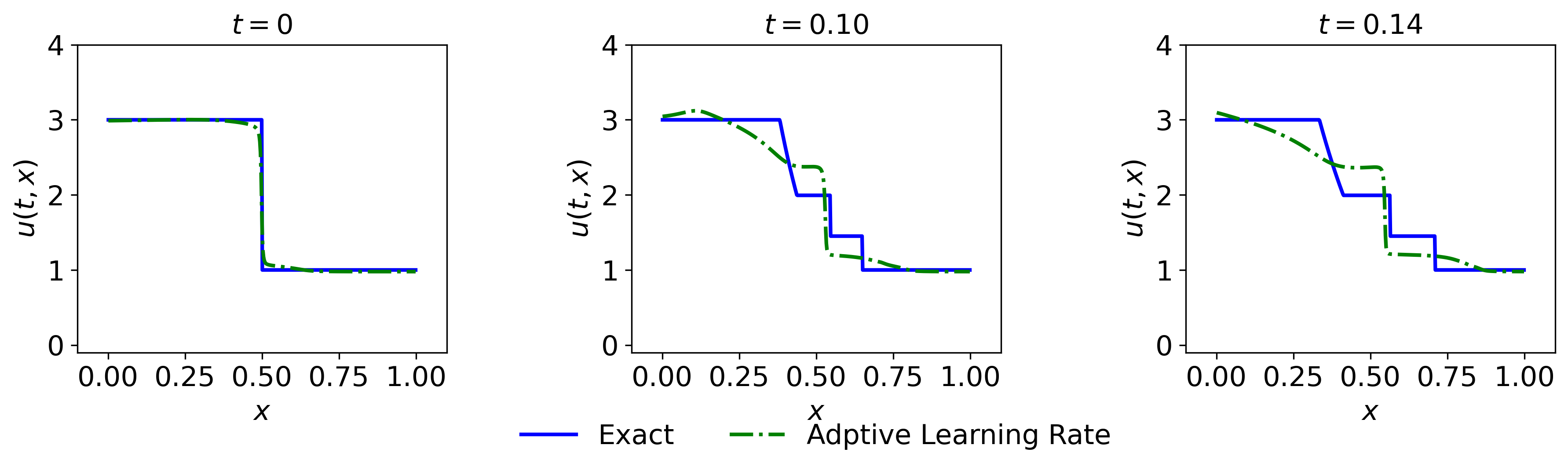}
    \vspace{1mm}
    \includegraphics[width=0.48\linewidth]{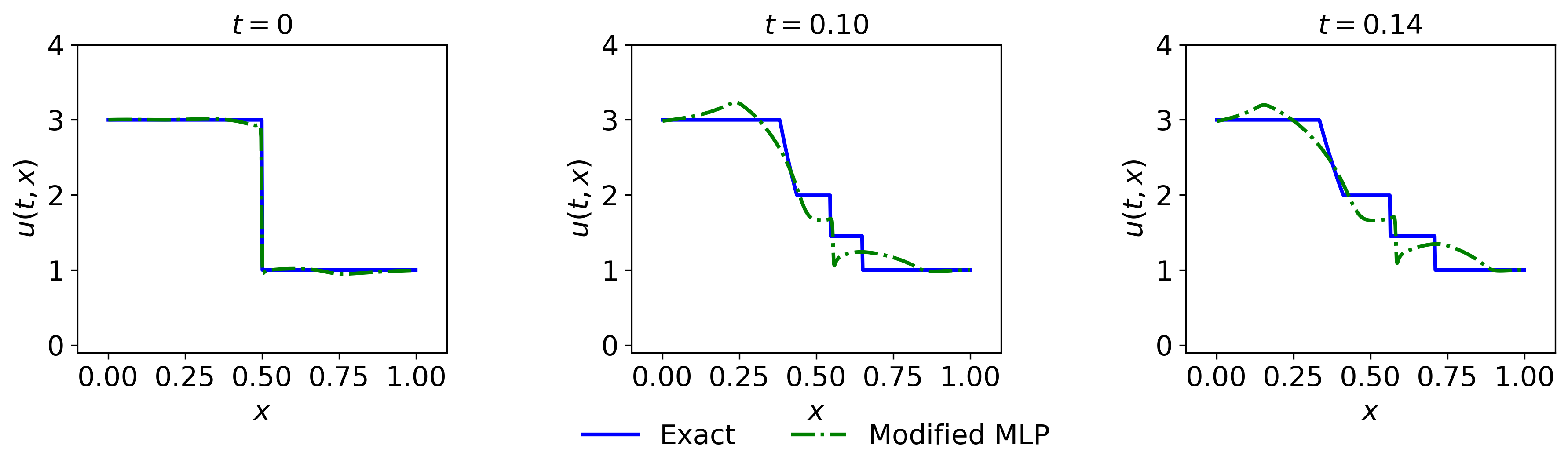} \hfill
    \includegraphics[width=0.48\linewidth]{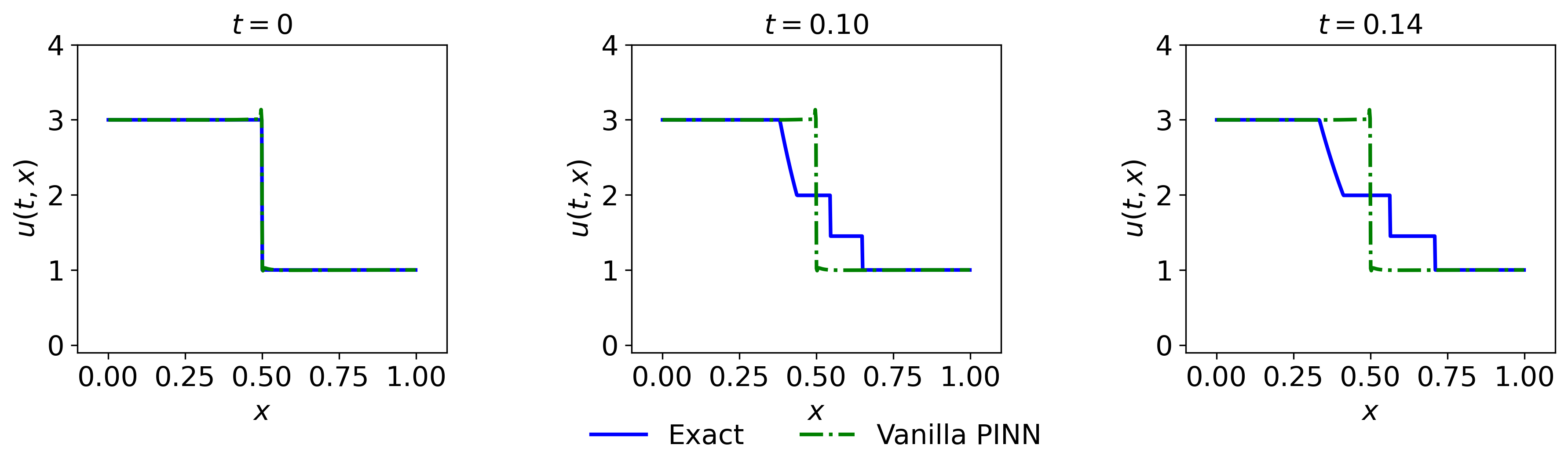}
    \vspace{-2mm}
    \caption{Sod problem: $\rho$ component.}
    \label{fig:sod_rho}
\end{figure}

\vspace{-4mm}

\subsection{Lax Problem}
\label{sec:appendFlax}

\begin{figure}[!ht]
    \centering
    \small
    \includegraphics[width=0.48\linewidth]{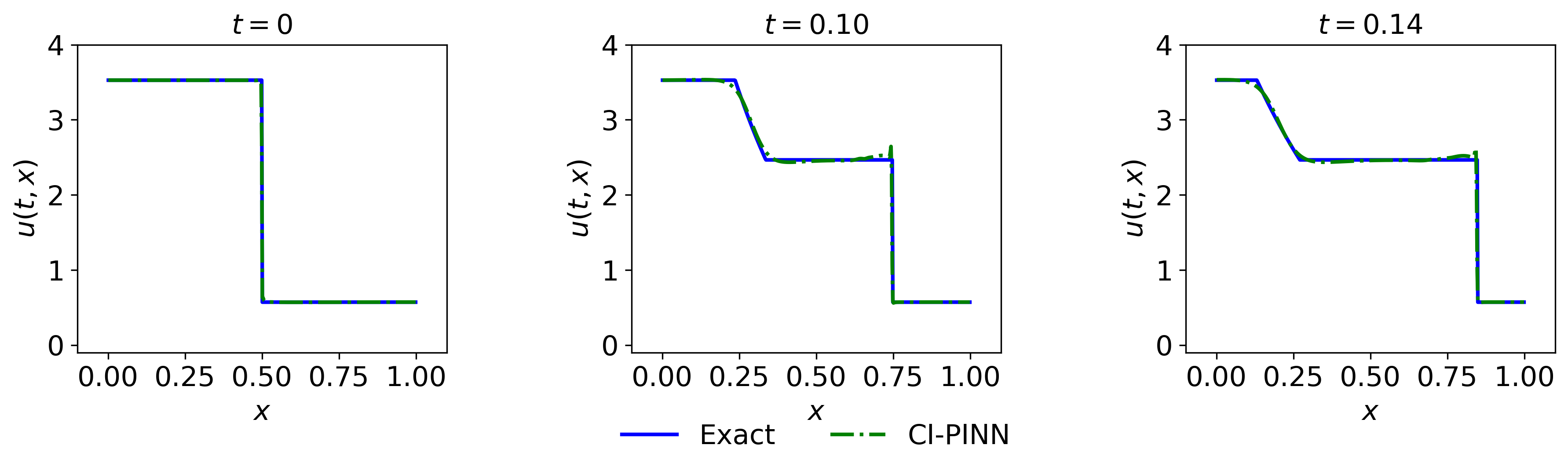} \hfill
    \includegraphics[width=0.48\linewidth]{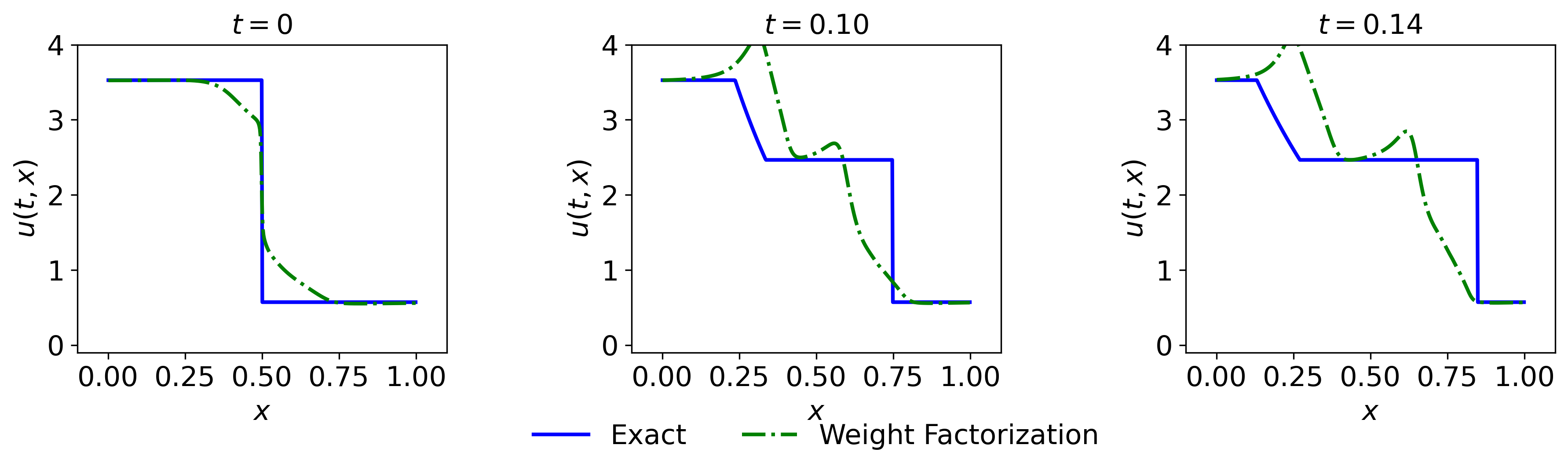}
    \vspace{1mm}

    \includegraphics[width=0.48\linewidth]{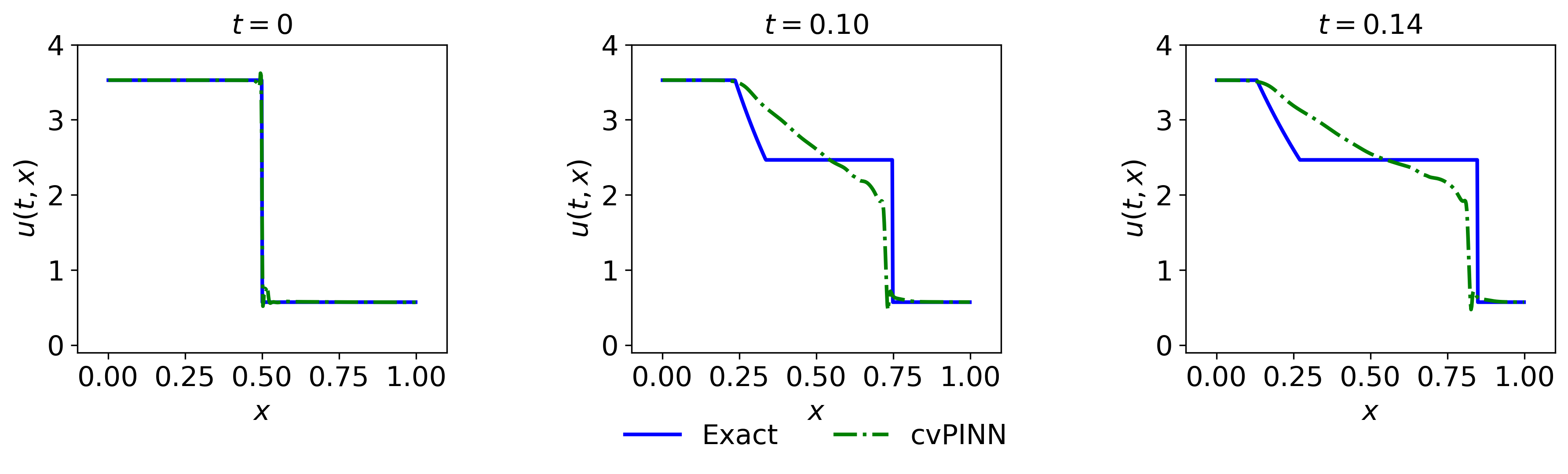} \hfill
    \includegraphics[width=0.48\linewidth]{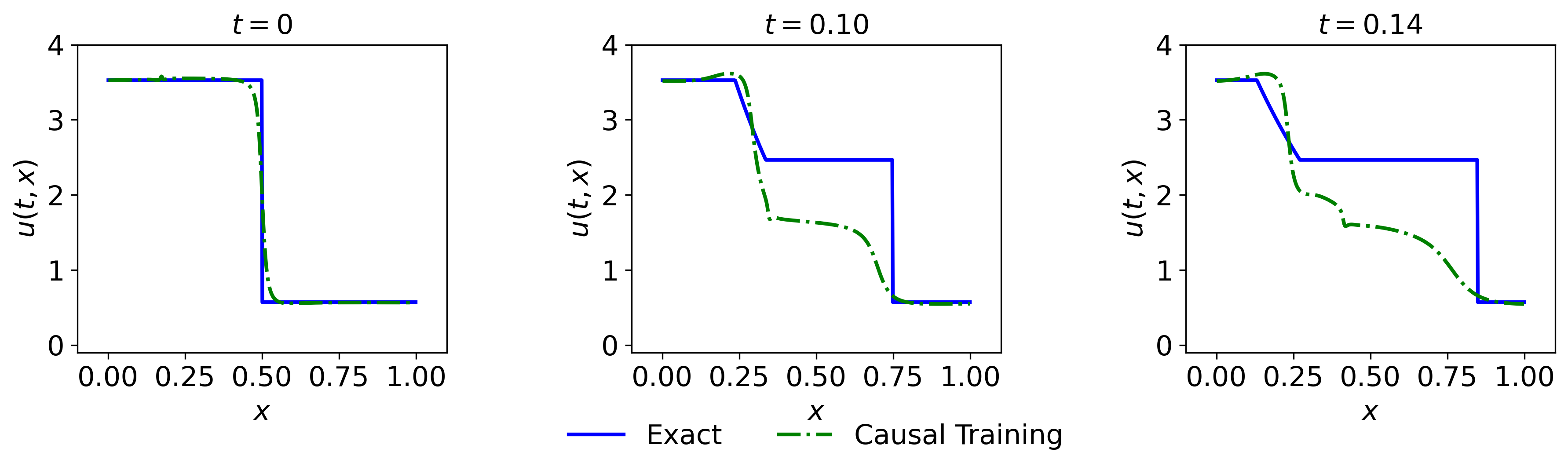}
    \vspace{1mm}

    \includegraphics[width=0.48\linewidth]{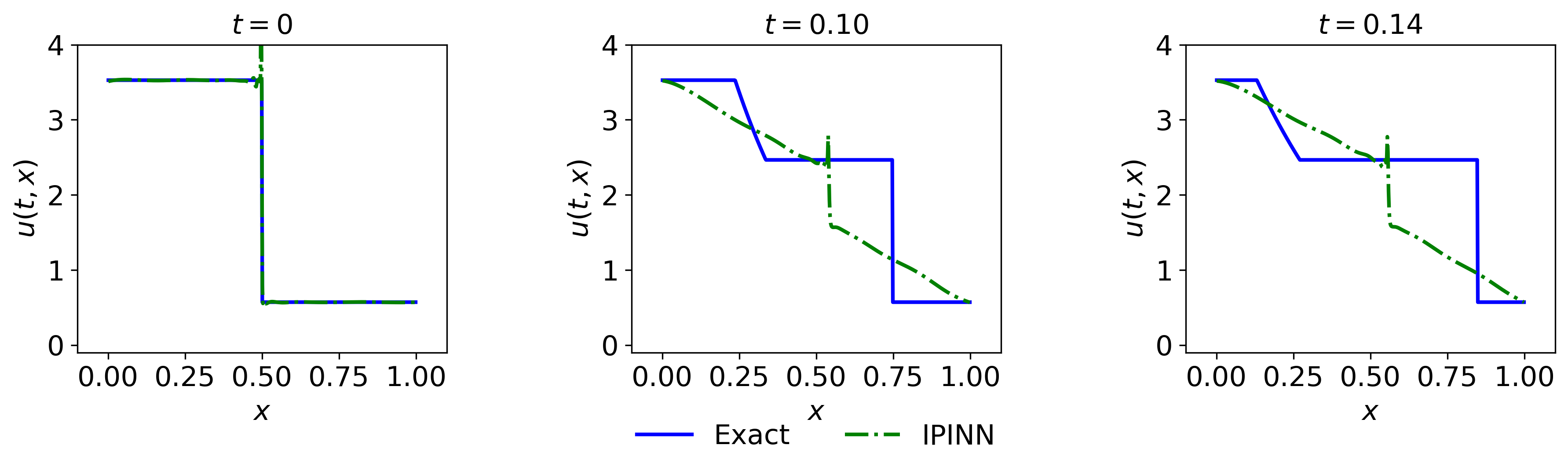} \hfill
    \includegraphics[width=0.48\linewidth]{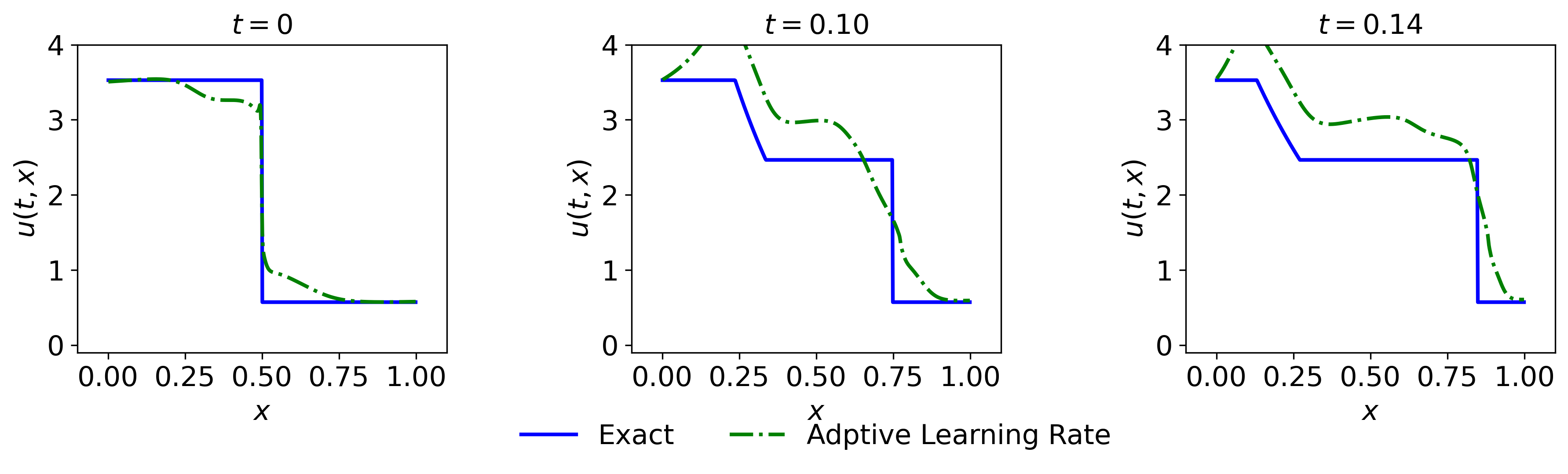}
    \vspace{1mm}

    \includegraphics[width=0.48\linewidth]{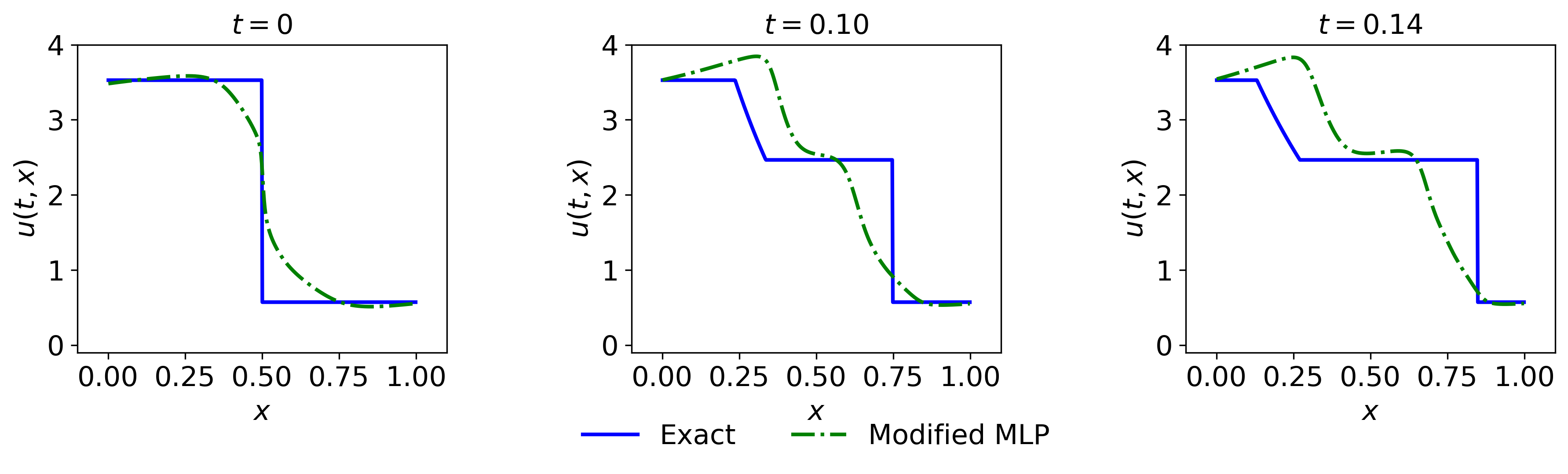} \hfill
    \includegraphics[width=0.48\linewidth]{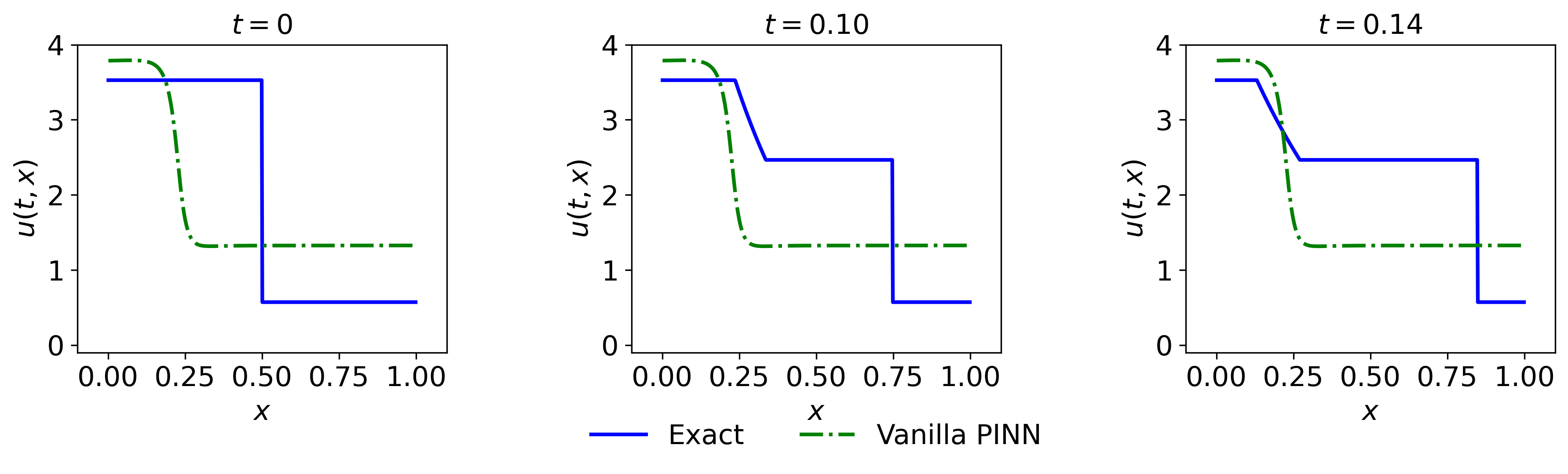}

    \vspace{-2mm}
    \caption{Lax problem: $p$ component.}
    \label{fig:lax_p}
\end{figure}

\clearpage 


\begin{figure}[!ht]
    \centering
    \small
    \includegraphics[width=0.48\linewidth]{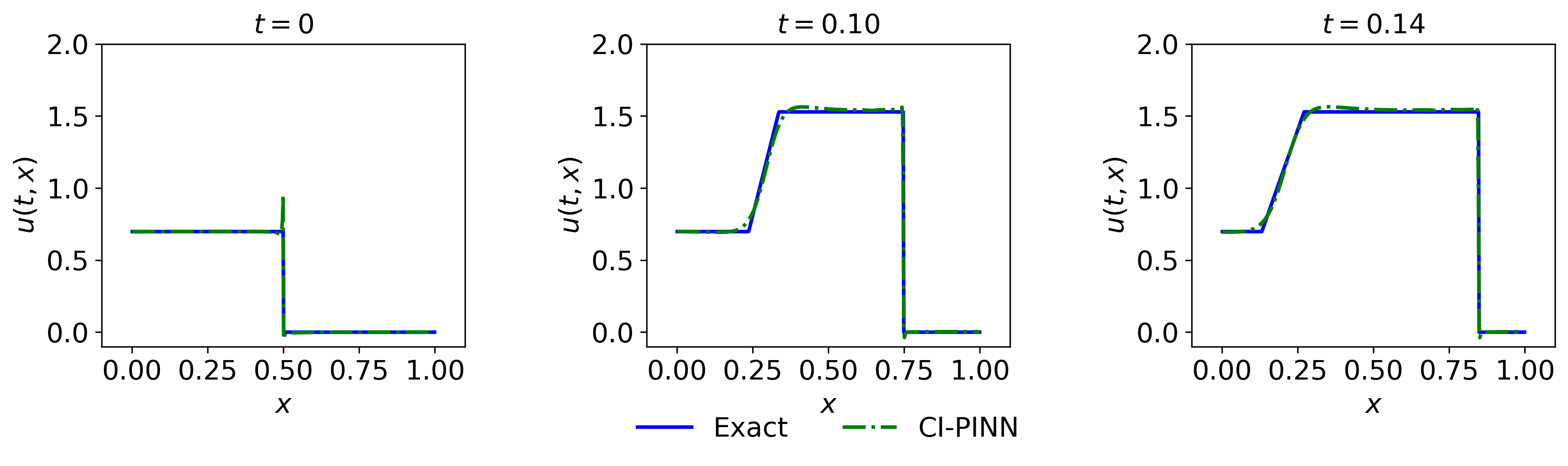} \hfill
    \includegraphics[width=0.48\linewidth]{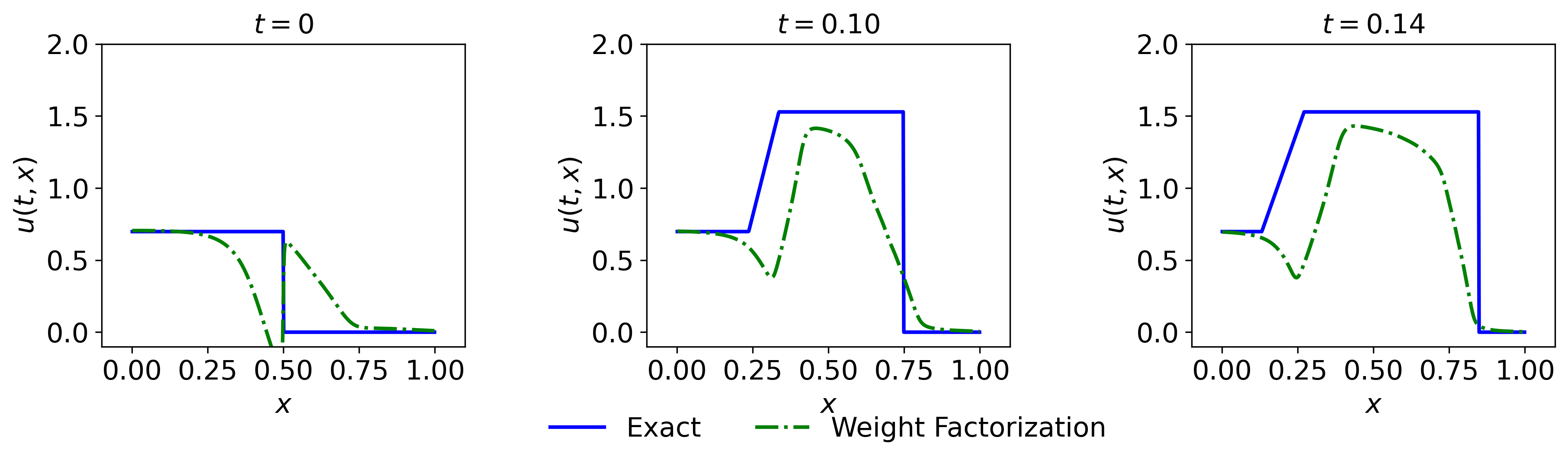}
    \vspace{1mm}

    \includegraphics[width=0.48\linewidth]{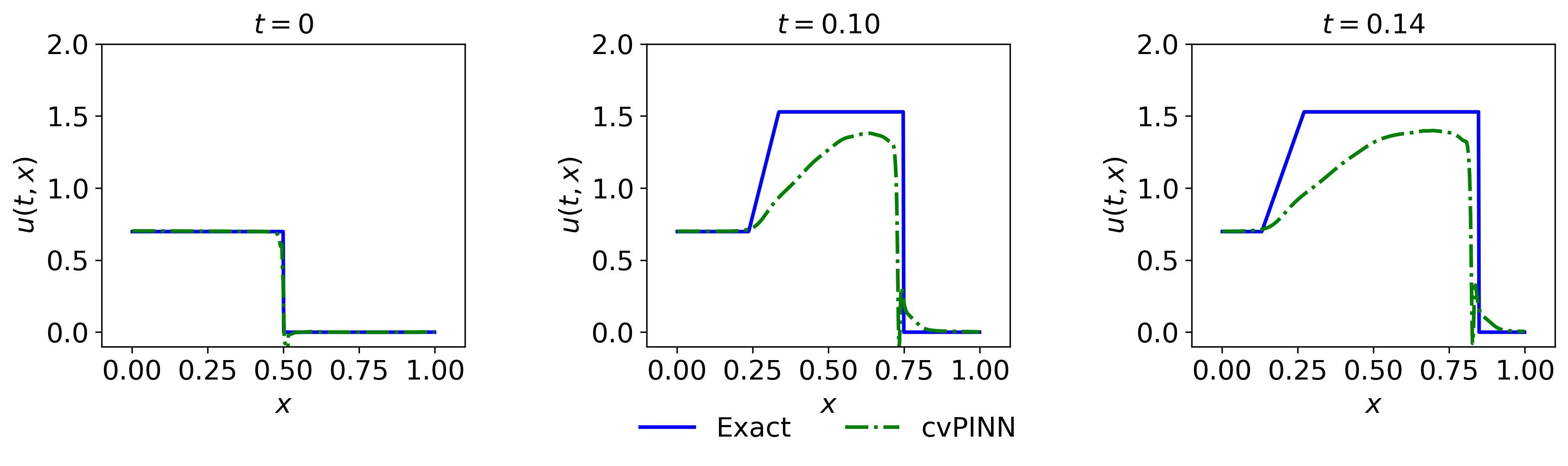} \hfill
    \includegraphics[width=0.48\linewidth]{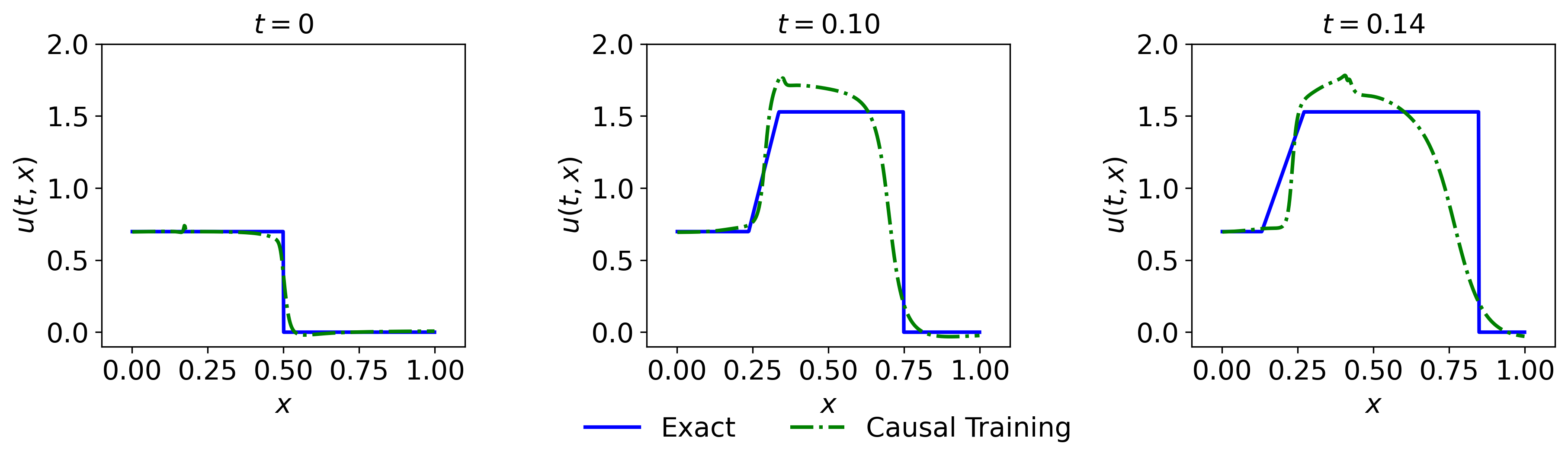}
    \vspace{1mm}

    \includegraphics[width=0.48\linewidth]{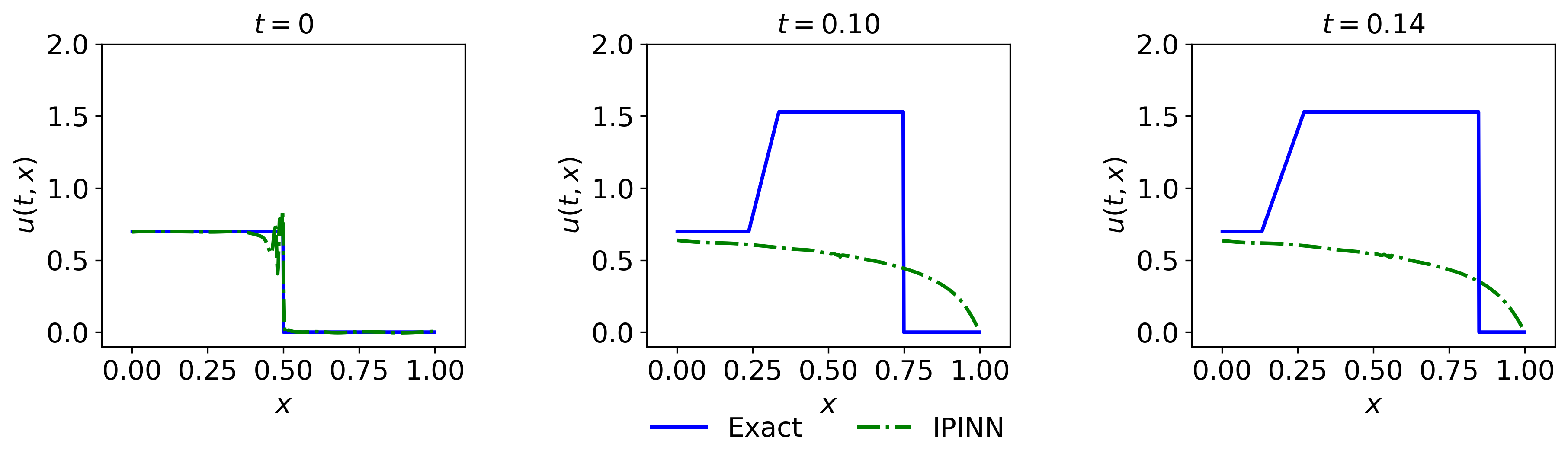} \hfill
    \includegraphics[width=0.48\linewidth]{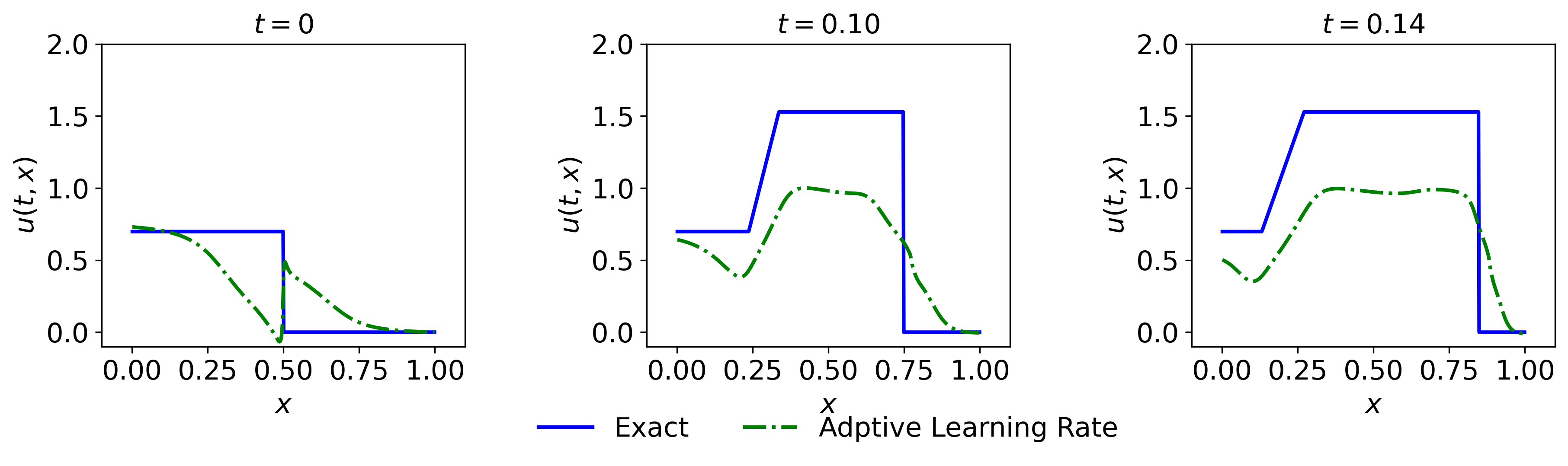}
    \vspace{1mm}

    \includegraphics[width=0.48\linewidth]{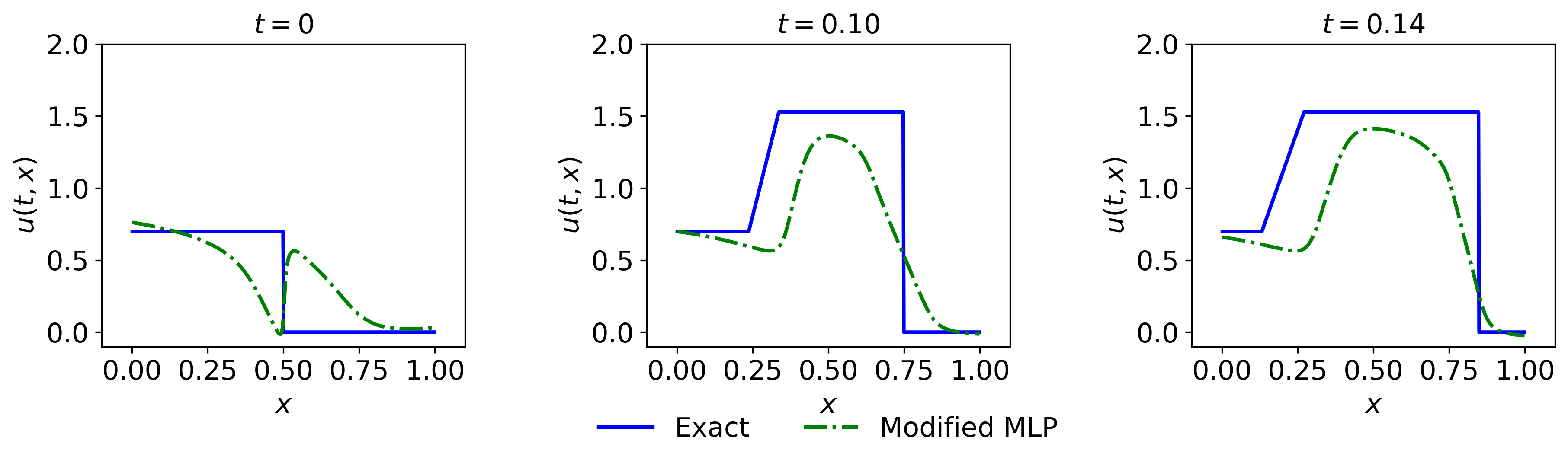} \hfill
    \includegraphics[width=0.48\linewidth]{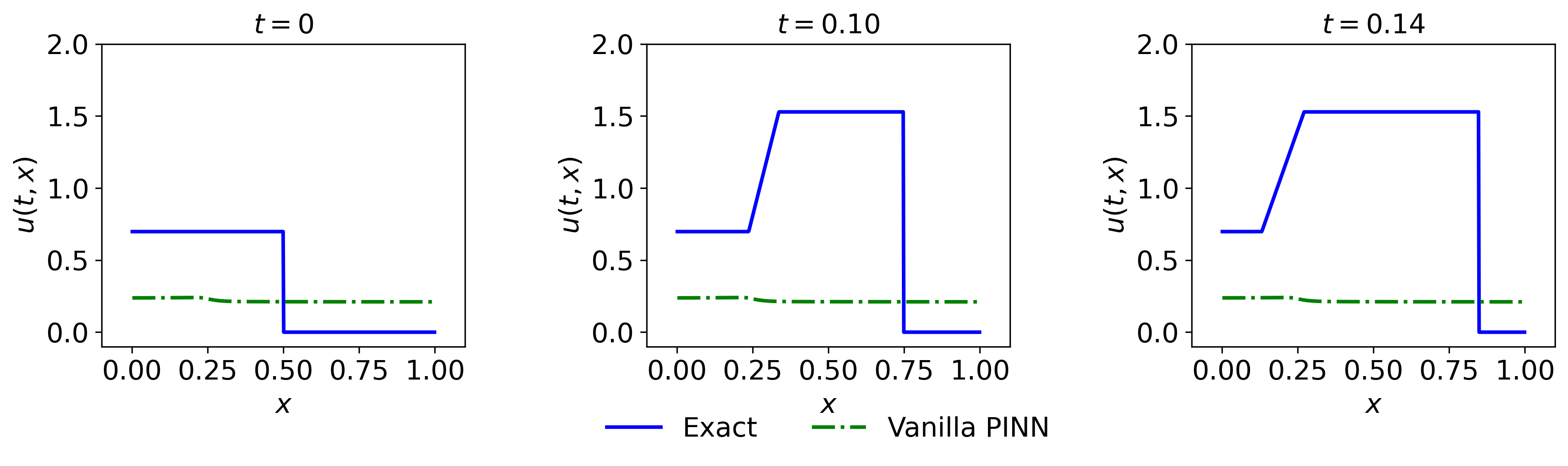}

    \vspace{-2mm}
    \caption{Lax problem: $u$ component.}
    \label{fig:lax_u}
\end{figure}

\vspace{-4mm}

\begin{figure}[!ht]
    \centering
    \small
    \includegraphics[width=0.48\linewidth]{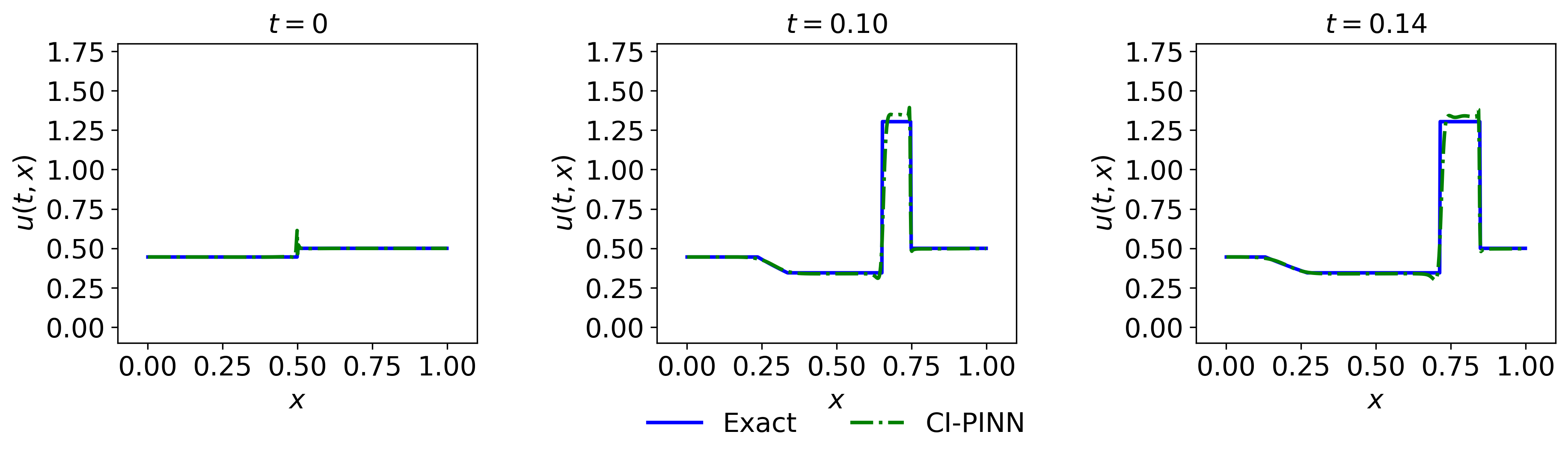} \hfill
    \includegraphics[width=0.48\linewidth]{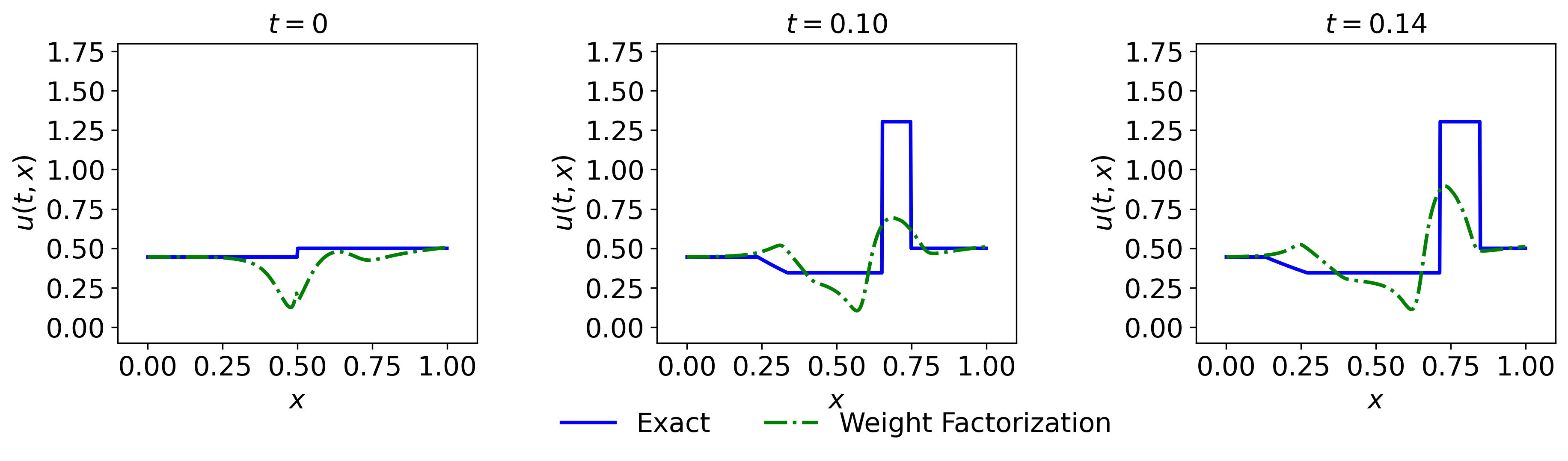}
    \vspace{1mm}

    \includegraphics[width=0.48\linewidth]{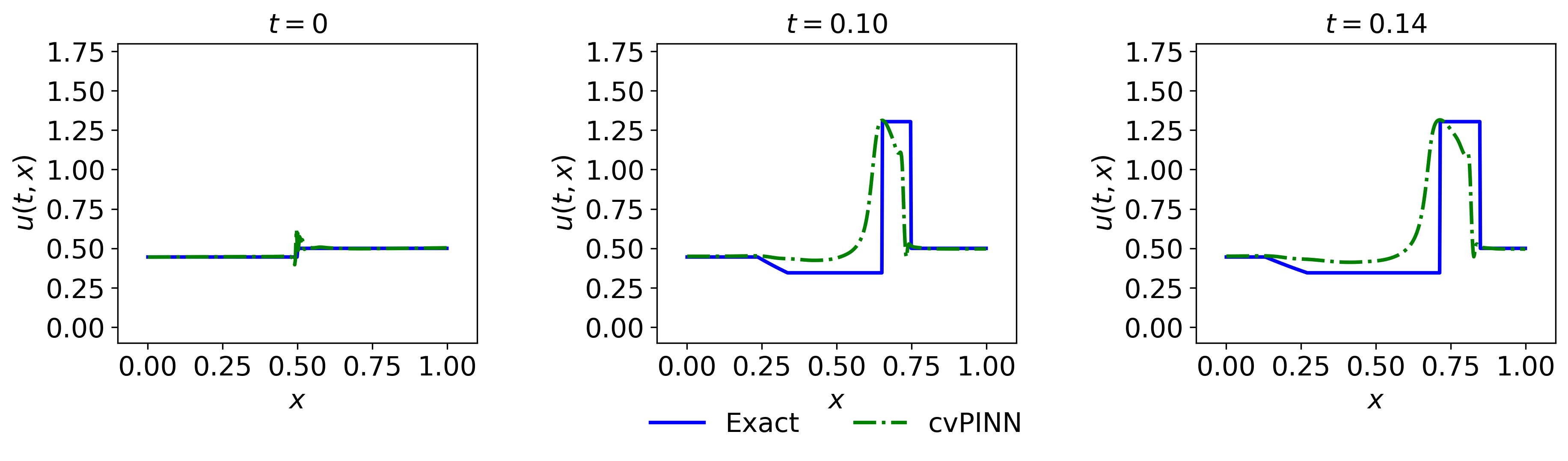} \hfill
    \includegraphics[width=0.48\linewidth]{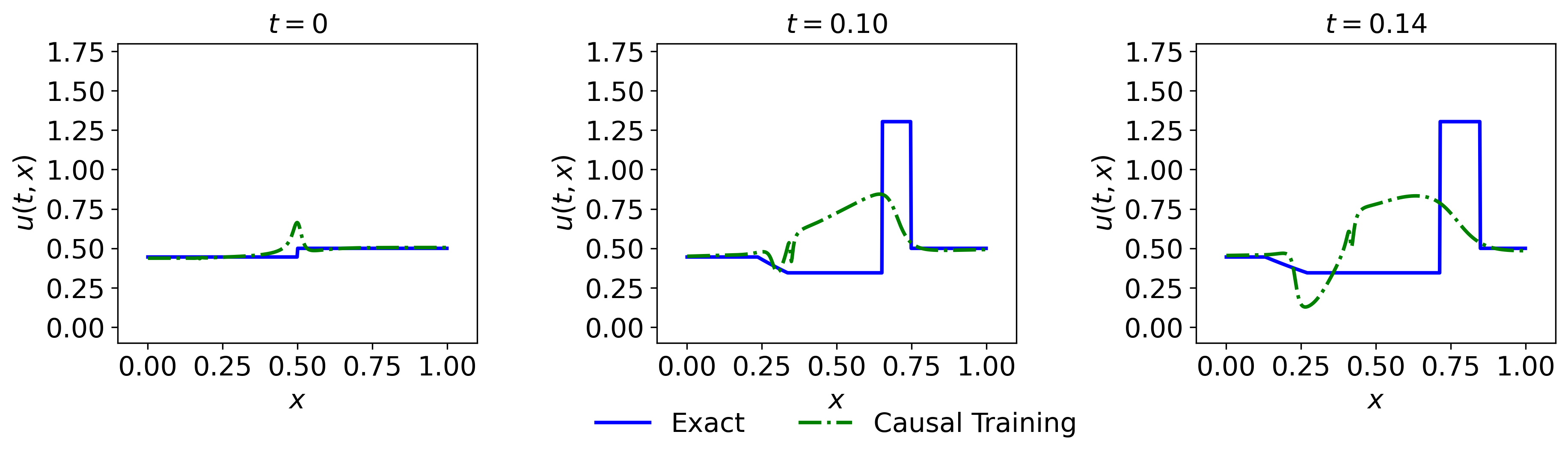}
    \vspace{1mm}

    \includegraphics[width=0.48\linewidth]{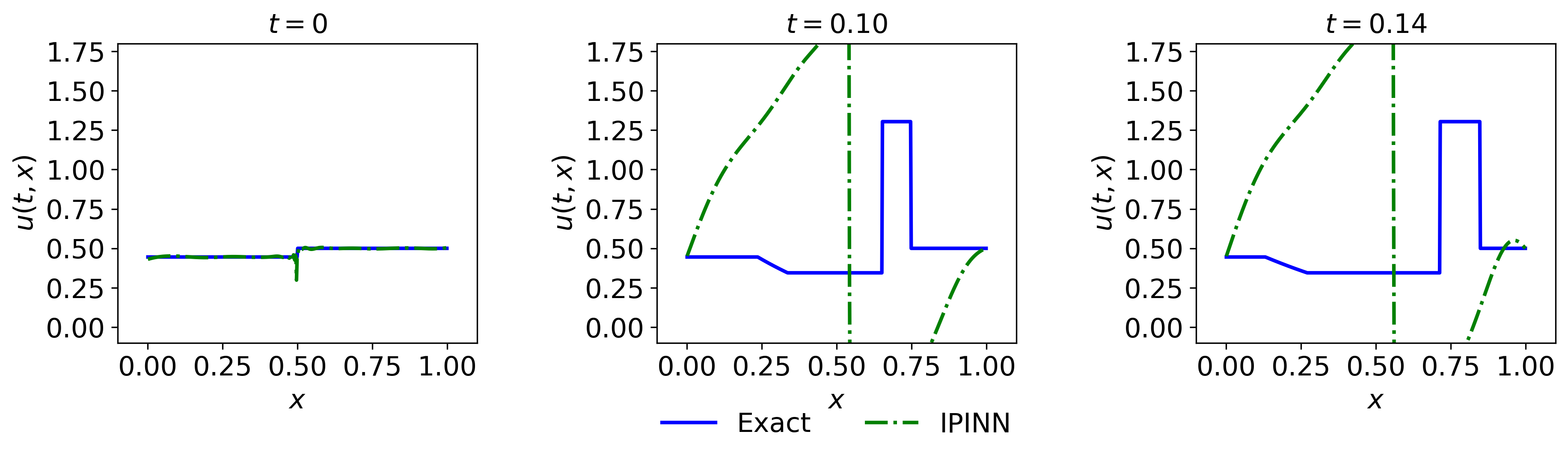} \hfill
    \includegraphics[width=0.48\linewidth]{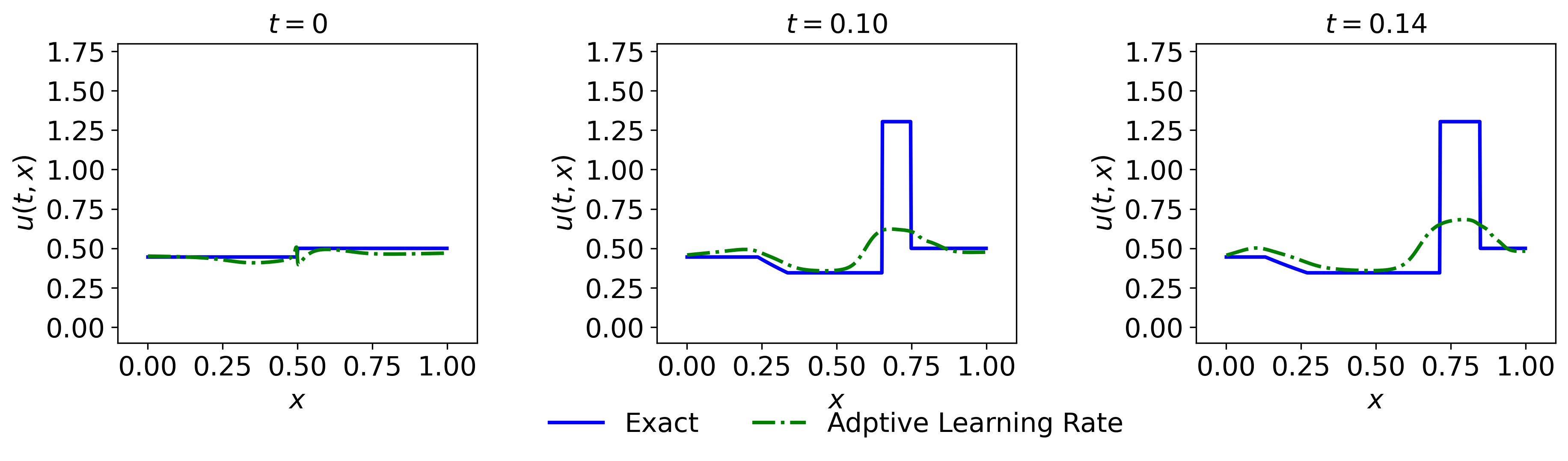}
    \vspace{1mm}

    \includegraphics[width=0.48\linewidth]{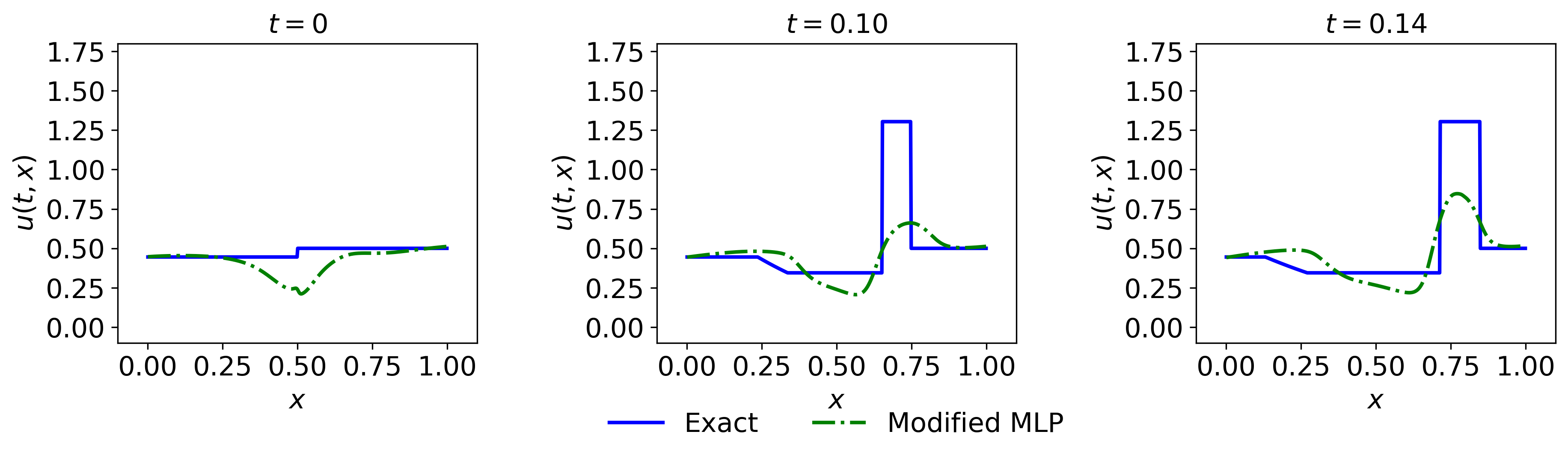} \hfill
    \includegraphics[width=0.48\linewidth]{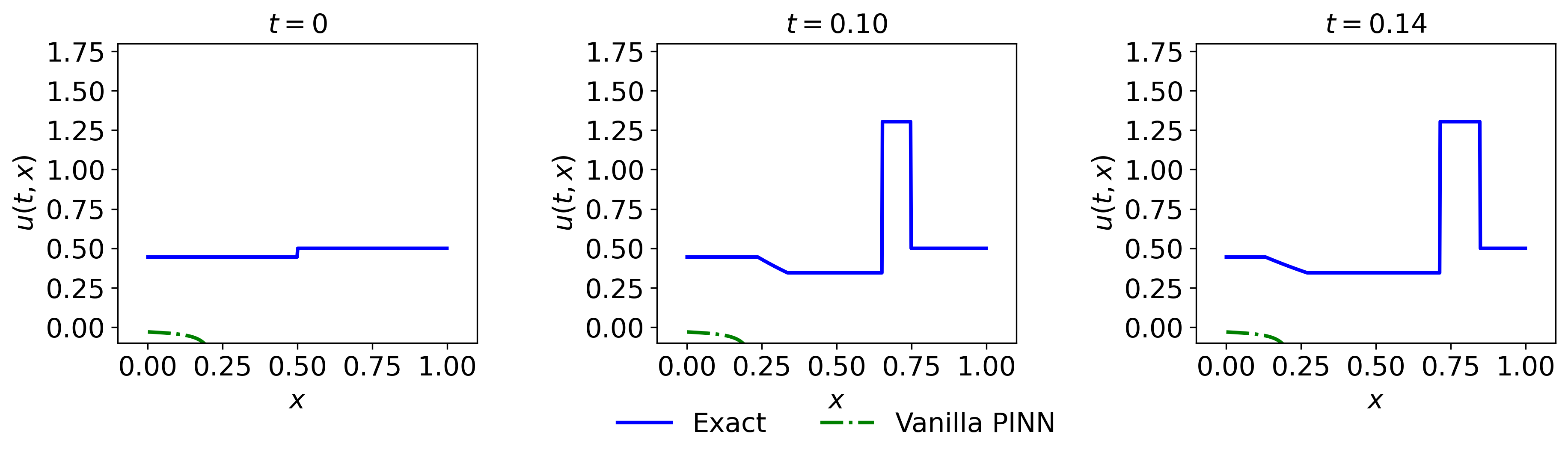}

    \vspace{-2mm}
    \caption{Lax problem: $\rho$ component.}
    \label{fig:lax_rho}
\end{figure}
\section{Domain-Partition}
\label{sec:appendG}
We split the space–time domain 
\(\Omega \times [0,T]\), with \(\Omega\subset\mathbb{R}^d\), into three parts: the shock neighborhood \(\Omega_s\), the rarefaction region \(\Omega_r\), and the full domain \(\Omega\). See appendix for detailed definition.

Let \(\mathbf{x}\in\mathbb{R}^d\) be the spatial coordinate and define the spatial diameter
\[
W \;=\;\mathrm{diam}(\Omega)
=\max_{\mathbf{x},\mathbf{y}\in\Omega}\|\mathbf{x}-\mathbf{y}\|_2.
\]
Denote by \(\mathbf{x}_s(t)\) the shock‐interface location (a curve or surface in \(\Omega\)).  For a fixed radius fraction \(r=0.1\), the shock neighborhood is
\[
\Omega_s 
\;=\;
\bigl\{(\mathbf{x},t)\in\Omega\times[0,T]\;\big|\;
\|\mathbf{x}-\mathbf{x}_s(t)\|_2\le r\,W
\bigr\}.
\]

Let \(\mathcal{R}_t\subset\Omega\) be the rarefaction fan region at time \(t\).  Its space–time closure is
\[
\Omega_r 
\;=\; 
\overline{\bigcup_{t\in[0,T]}\bigl(\mathcal{R}_t\times\{t\}\bigr)}.
\]



\section{Probconserv models}
\label{sec:appendH}
Together with Probconserv models \cite{HANSEN2024133952}, we evaluate our method in the same setting on the 1D inviscid Burgers’ equation where:
\[
\partial_t u + u\,\partial_x u = 0,
\]
defined on the spatial domain $x \in [-1, 1]$ and time interval $t \in [0, 1]$. The initial condition is
\[
u(x, 0) = 
\begin{cases}
-ax, & x < 0, \\
0, & x \geq 0,
\end{cases}
\]
and Dirichlet boundary conditions are enforced: $u(-1, t) = a$ and $u(1, t) = 0$. The parameter $a > 0$ controls the initial shock strength. In our experiments, we test on both $a = 1$ (moderate rarefaction) and $a = 3$ (strong shock).

\begin{figure*}[H]
  \centering

  \begin{subfigure}{\textwidth}
    \includegraphics[width=\linewidth]{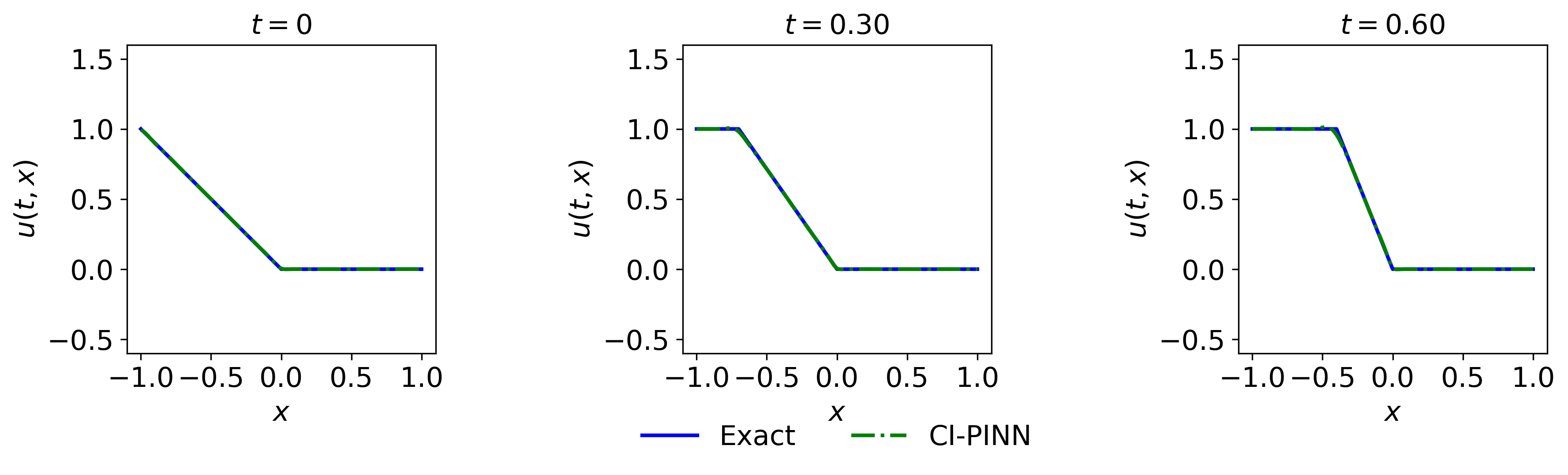}
    \caption{param = 1}
    \label{fig:image1}
  \end{subfigure}
  \hfill
  \begin{subfigure}{\textwidth}
    \includegraphics[width=\linewidth]{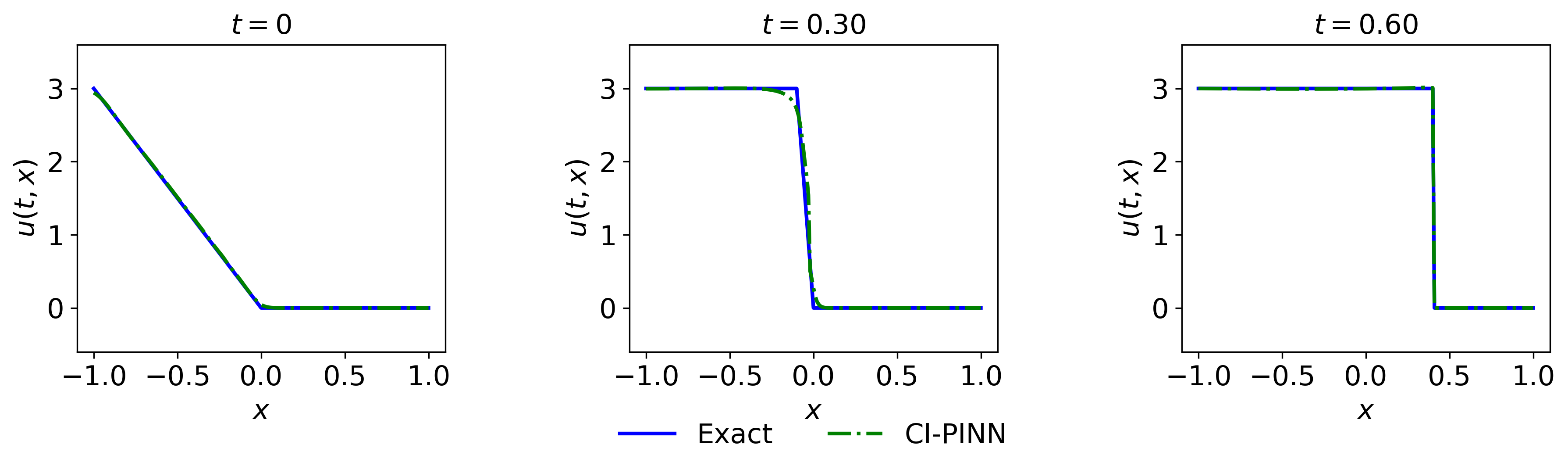}
    \caption{param = 3}
    \label{fig:image2}
  \end{subfigure}

  \caption{Burgers Result}
  \label{fig:burgers_snapshots}
\end{figure*}

\begin{figure}[H]
  \centering

  \begin{subfigure}{0.48\textwidth}
    \includegraphics[width=\linewidth]{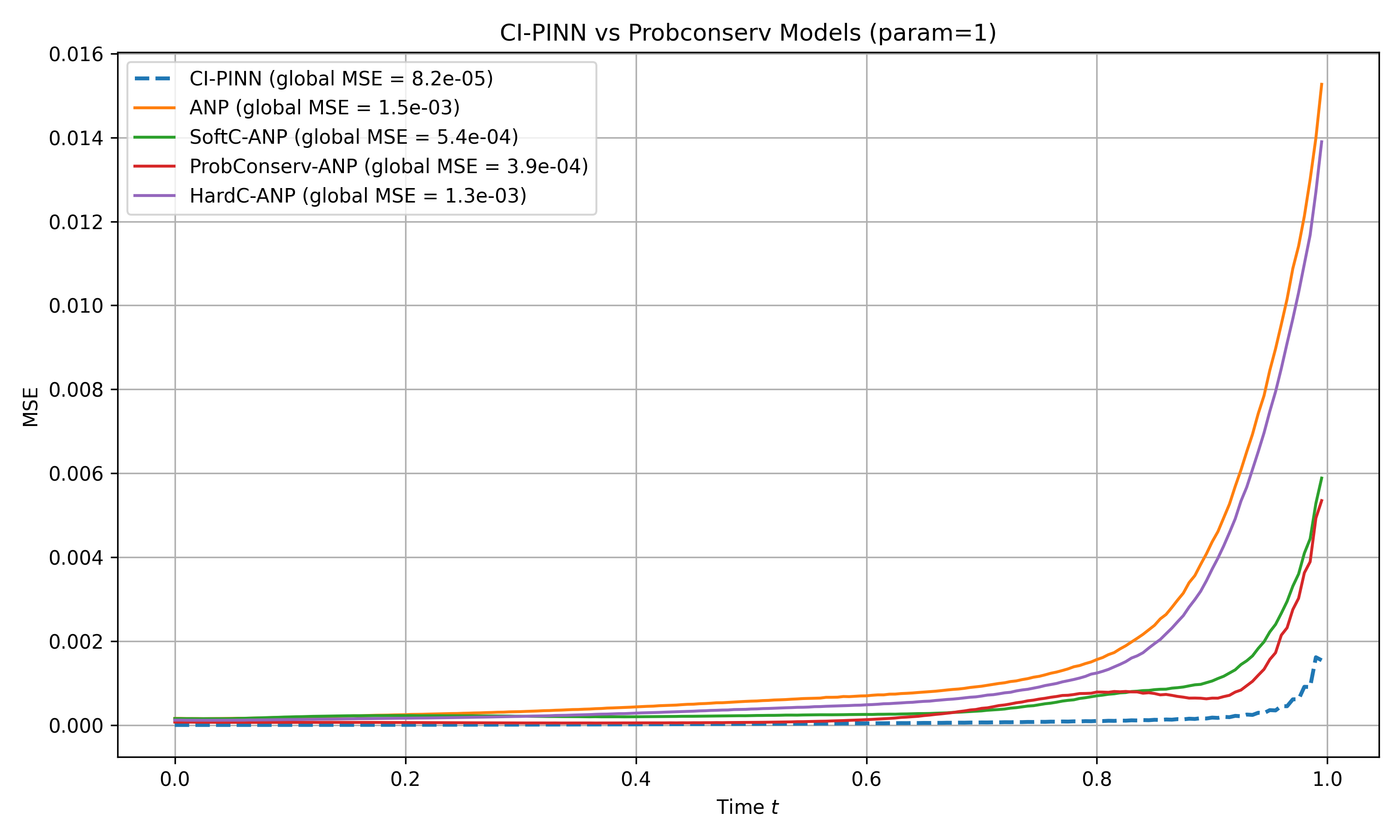}
 
    \label{fig:image1}
  \end{subfigure}
  \hfill
  \begin{subfigure}{0.48\textwidth}
    \includegraphics[width=\linewidth]{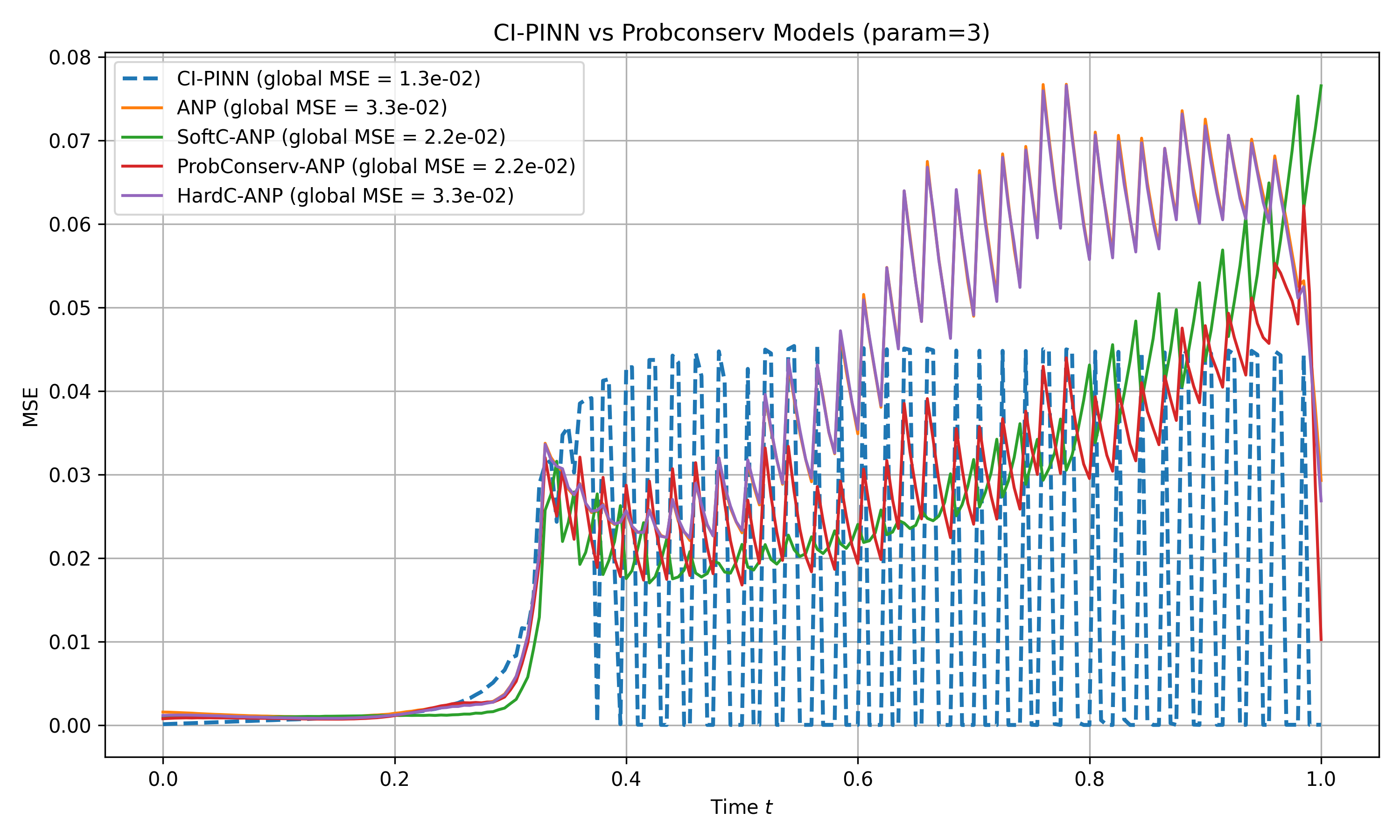}
 
    \label{fig:image2}
  \end{subfigure}

  \caption{MSE by Time}
  \label{fig:msebytime}
\end{figure}

Figure~\ref{fig:msebytime} compares the mean squared error (MSE) over time for CI-PINN and probabilistic baselines on two regimes: param=1 (rarefaction-dominated) and param=3 (shock-dominated). CI-PINN consistently achieves the lowest global MSE in both settings, with approximately half the error of the best-performing competitor. Notably, in the simpler param=1 case, CI-PINN maintains a smooth and low MSE throughout the time horizon, demonstrating its strong generalization in smooth regimes.

In the shock-dominated case (param=3), CI-PINN exhibits small-amplitude, high-frequency oscillations in MSE after the shock forms. This pattern suggests that while the model may slightly misalign with the exact shock location at individual time steps, it quickly corrects itself—indicating stable tracking behavior. In contrast, other models accumulate error over time, especially near discontinuities. Overall, CI-PINN exhibits both low error and resilient behavior in the presence of shocks.




\section{Introduction and formulation of benchmark equations}
\label{sec:formulation}

\subsection{Burgers' Equation}

The Burgers' equation is a fundamental partial differential equation classified as a convection–diffusion equation. It plays a significant role in various fields of applied mathematics, including fluid mechanics, gas dynamics, and traffic flow\cite{doi:10.1080/10618562.2010.523518, https://doi.org/10.1002/cpa.3160030302}. The equation is expressed as:

\begin{equation}
    \frac{\partial u}{\partial t} + u \frac{\partial u}{\partial x} = \nu \frac{\partial^2 u}{\partial x^2},
\end{equation}  

where \( u = u(x,t) \) represents the velocity field, \( t \) denotes time, \( x \) is the spatial coordinate, and \( \nu \) is the viscosity coefficient. When \(\nu = 0\), the equation simplifies to the inviscid Burgers’ equation, which can develop sudden changes (shock) in the solution. These shock arise because the characteristic wave speeds of this equation are given by \(f'(u) = u\). To understand how shocks (sharp jumps) or rarefaction waves (smooth spreading) form, consider a point in the initial condition where \(u\) takes one value on the immediate left (call it \(u_L\)) and another value on the immediate right (call it \(u_R\)). If \(u_L > u_R\), then the wave speed on the left side is higher than on the right side. The faster-moving characteristics from the left eventually overtake the slower ones on the right, causing the solution to “pile up” into a shock. Conversely, if \(u_L < u_R\), then the wave speed on the left is lower than on the right. The characteristics spread out or “fan apart,” resulting in a rarefaction wave \cite{LeVeque1992}. A simple way to illustrate both cases is to use a square-wave initial condition. We initialize the solution with a piecewise-constant profile, setting $u(x,0)=1$ for $x\in(-0.5,0)$ and $u(x,0)=0$ elsewhere.

Here, there are two transition points in the initial data (near \(x = -0.5\) and \(x = 0\)). Depending on which side has the larger or smaller value of \(u\), one observes either a shock or a rarefaction wave in the solution, see Figure \ref{fig:burgers_split}.


It is also important to emphasize that the higher error values observed for methods such as Modified MLP, RWF, Causal, and Adaptive lr do not necessarily indicate poor performance. On the contrary, as illustrated in Appendix~\ref{sec:appendF}, several of these methods exhibit promising behavior by closely approximating the shock structure.





\subsection{Buckley-Leverett Equation}

The displacement of two immiscible fluids is a significant challenge in the study of fluid flow within porous media. This phenomenon is often modeled by a PDE known as the Buckley-Leverett (BL) equation\cite{muskat1938flow}. The BL equation provides a mathematical framework for studying the dynamics and interactions of these fluids under varying conditions in porous substrates. It is particularly relevant in secondary oil recovery, where water is injected into underground rock formations to displace additional unrecovered oil. The Buckley-Leverett equation is given by:
$
u_t + f(u)_x = 0,
$
where
\[
f(u) = \frac{u^2}{u^2 + M \times (1-u)^2}.
\]

Here, \(u(x,t)\) represents the saturation of the injected fluid, and \(M\) is the mobility ratio, defined as the ratio of the mobility of the displacing fluid to the mobility of the displaced fluid. Since \(f(u)\) is a non-convex function in the range \([0,1]\), the solution of the B-L equation exhibits both shocks and rarefaction waves, which are connected.

As an example, we consider \(M = \frac{1}{4}\). The initial condition is prescribed as a piecewise-constant state with $u(x,0)=1$ for $x\in[-1,-0.5]$ and $u(x,0)=0$ for $x\in(-0.5,1]$.
visualized of clips are in the leftmost panel of Appendix~\ref{sec:appendFbl}:








.
\subsection{Euler System}

The Euler equations of gas dynamics form a system of hyperbolic partial differential equations that govern the motion of an inviscid fluid. Derived from the fundamental principles of conservation of mass, momentum, and energy, these equations are extensively used in fluid dynamics and aerodynamics to model gas behavior \cite{toro2013riemann}. The Euler equations also serve as a benchmark for evaluating the accuracy of numerical methods.

The Euler system is expressed as:
$
\frac{\partial \mathbf{U}}{\partial t} + \nabla \cdot \mathbf{F} = 0,
$
where
\[
\mathbf{U} = \begin{pmatrix} \rho \\ \rho u \\ E \end{pmatrix}, \quad 
\mathbf{F} = \begin{pmatrix} \rho u \\ \rho u^2 + p \\ u(E + p) \end{pmatrix},
\]
and
$
E = \frac{1}{2} \rho u^2 + \frac{p}{\gamma - 1}.
$

Here, \(\rho\) denotes the density, \(u\) the velocity, \(p\) the pressure, \(E\) the total energy, and \(\gamma\) the ratio of specific heats.  
We benchmark our method on two canonical Euler shock-tube problems—Sod’s and Lax’s tests.The Sod and Lax shock tube problems are classical Riemann benchmarks that are widely used to assess the accuracy and shock-capturing robustness of numerical methods for the Euler equations~\citep{SOD19781,Lax1954,toro2009}.

The strong-form PINN baseline is omitted in the main text, as previous sections have shown its poor performance on conservation laws; full error tables, including that baseline, are reported in Appendix~\ref{fulltable}.



\subsubsection{Sod Problem}

The Sod shock tube problem considers an ideal gas in a one-dimensional tube initially separated by a diaphragm, with the left state having higher density and pressure and the right state having lower density and pressure, while both sides are initially at rest~\citep{SOD19781}. Upon removal of the diaphragm, the system evolves under the Euler equations and generates a sequence of characteristic waves; we initialize the system with a left state $(\rho,u,p)=(3,\,0,\,3)$ for $x\in[0,0.5]$ and a right state $(\rho,u,p)=(1,\,0,\,1)$ for $x\in(0.5,1]$.


\subsubsection{Lax Problem}

The Lax problem is a classical Riemann benchmark featuring a strong shock and a pronounced contact discontinuity; we initialize the system with a left state
$(\rho,u,p)=(0.445,\,0.698,\,3.528)$ for $x\in[0,0.5]$ and a right state
$(\rho,u,p)=(0.5,\,0,\,0.571)$ for $x\in(0.5,1]$.
Owing to the coexistence of a strong shock and a contact discontinuity, the solution has limited regularity, rendering strong-form residual objectives ill-posed near discontinuities and making this problem a stringent test for PINN-based solvers

\subsection{Two-Dimensional Euler Equations}

The 2D Euler equations describe the dynamics of a compressible, inviscid fluid in two spatial dimensions. This system is governed by the conservation of mass, momentum, and energy (Li, 2002). The conservative form is given by:

\begin{equation}
\frac{\partial \mathbf{U}}{\partial t} + \frac{\partial \mathbf{F}}{\partial x} + \frac{\partial \mathbf{G}}{\partial y} = 0,
\end{equation}

where the state vector $\mathbf{U}$ and the flux vectors $\mathbf{F}, \mathbf{G}$ are:

\begin{equation}
\mathbf{U} = \begin{pmatrix} \rho \\ \rho u \\ \rho v \\ E \end{pmatrix}, \quad 
\mathbf{F} = \begin{pmatrix} \rho u \\ \rho u^2 + p \\ \rho uv \\ u(E + p) \end{pmatrix}, \quad
\mathbf{G} = \begin{pmatrix} \rho v \\ \rho uv \\ \rho v^2 + p \\ v(E + p) \end{pmatrix}.
\end{equation}

Here, $p$ is the pressure, $\rho$ is the density, $(u,v)$ are velocity components, and $E$ is the total energy. The system is closed by the ideal gas law $p = (\gamma - 1)(E - \frac{1}{2}\rho(u^2+v^2))$. 2D Euler problems are highly challenging for neural solvers due to the formation of oblique shocks and Mach stems.

We define initialize the system with:
\begin{equation}
(\rho, u, v, p)(x, y, 0) = 
\begin{cases} 
(2.0, 0.75, 0.5, 1.0) & \text{if } x < 0.5, y > 0.5 \text{ (LT)}, \\
(1.0, 0.75, -0.5, 1.0) & \text{if } x > 0.5, y > 0.5 \text{ (RT)}, \\
(1.0, -0.75, 0.5, 1.0) & \text{if } x < 0.5, y < 0.5 \text{ (LB)}, \\
(3.0, -0.75, -0.5, 1.0) & \text{if } x > 0.5, y < 0.5 \text{ (RB)}.
\end{cases}
\end{equation}

\subsection{Shallow-Water Equations (SWE)}

The Shallow-Water Equations are derived by depth-averaging the Navier-Stokes equations, assuming the horizontal length scale is much larger than the vertical depth (Vreugdenhil, 1994). They are widely used for modeling tsunami propagation and atmospheric flows. The 2D conservative form is:

\begin{equation}
\frac{\partial}{\partial t} \begin{pmatrix} h \\ hu \\ hv \end{pmatrix} + \frac{\partial}{\partial x} \begin{pmatrix} hu \\ hu^2 + \frac{1}{2}gh^2 \\ huv \end{pmatrix} + \frac{\partial}{\partial y} \begin{pmatrix} hv \\ huv \\ hv^2 + \frac{1}{2}gh^2 \end{pmatrix} = 0,
\end{equation}

where $h$ is the fluid depth, $g$ is the acceleration due to gravity, and $(hu, hv)$ are the momentum components in the $x$ and $y$ directions, respectively.

\subsubsection{Problem Setup}
To evaluate shock-capturing performance in the presence of hydraulic jumps, we simulate a radial dam break scenario. The fluid depth is initialized with a discontinuous profile: $h=2$ within a central circle of radius $0.5$, and $h=1$ in the surrounding domain.

\section{2D experiments}
\label{Appendix:2dexp}
\begin{figure}[H]
    \centering
    \begin{subfigure}[b]{0.48\textwidth}
        \centering
        \includegraphics[width=\textwidth]{EULER_rho_NUM_vs_CIPINN_vs_CVPINN_3x4.png}
        \caption{Density ($\rho$)}
    \end{subfigure}
    \hfill
    \begin{subfigure}[b]{0.48\textwidth}
        \centering
        \includegraphics[width=\textwidth]{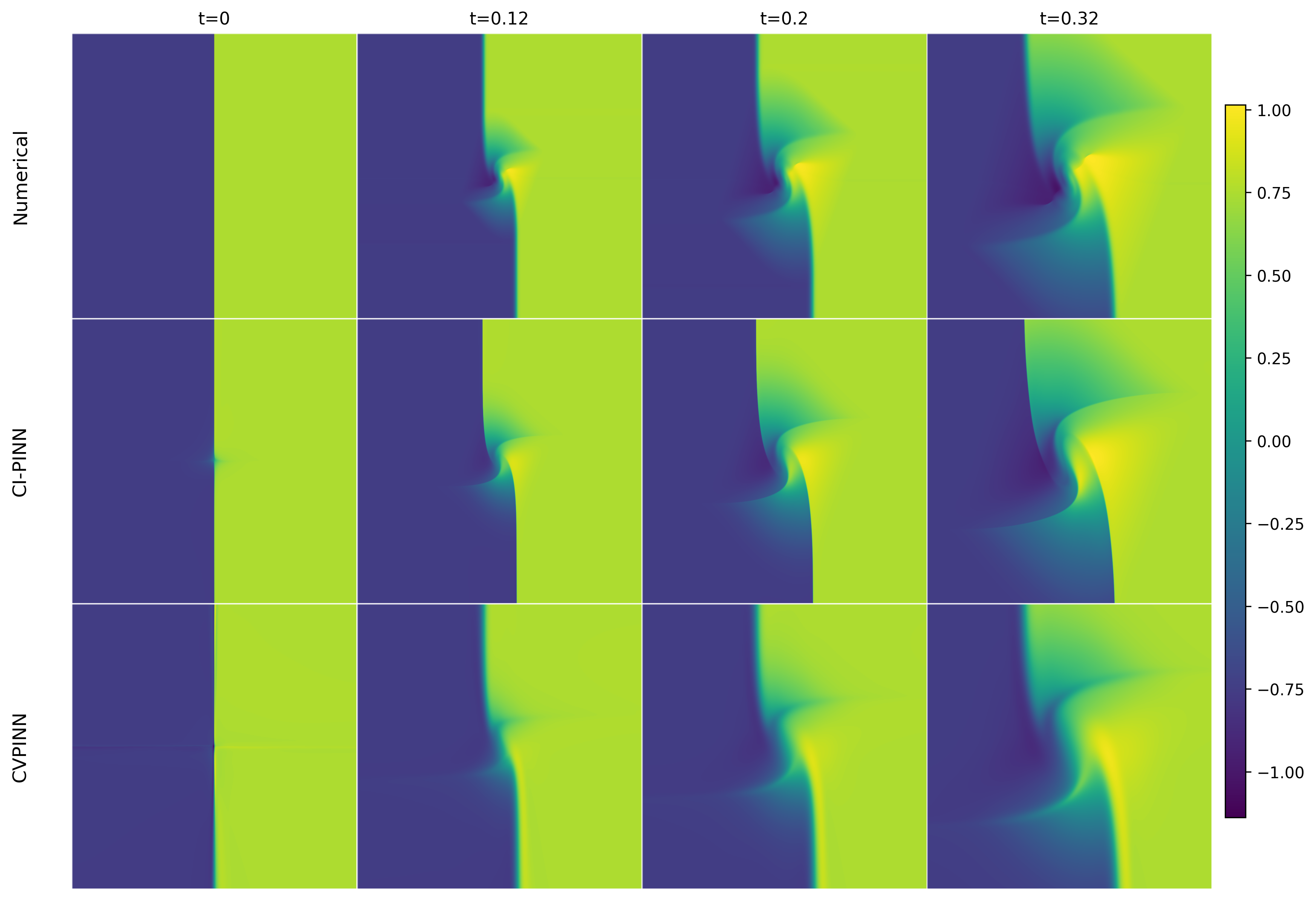}
        \caption{Velocity ($u$)}
    \end{subfigure}
    \vspace{0.5cm} 
    \begin{subfigure}[b]{0.48\textwidth}
        \centering
        \includegraphics[width=\textwidth]{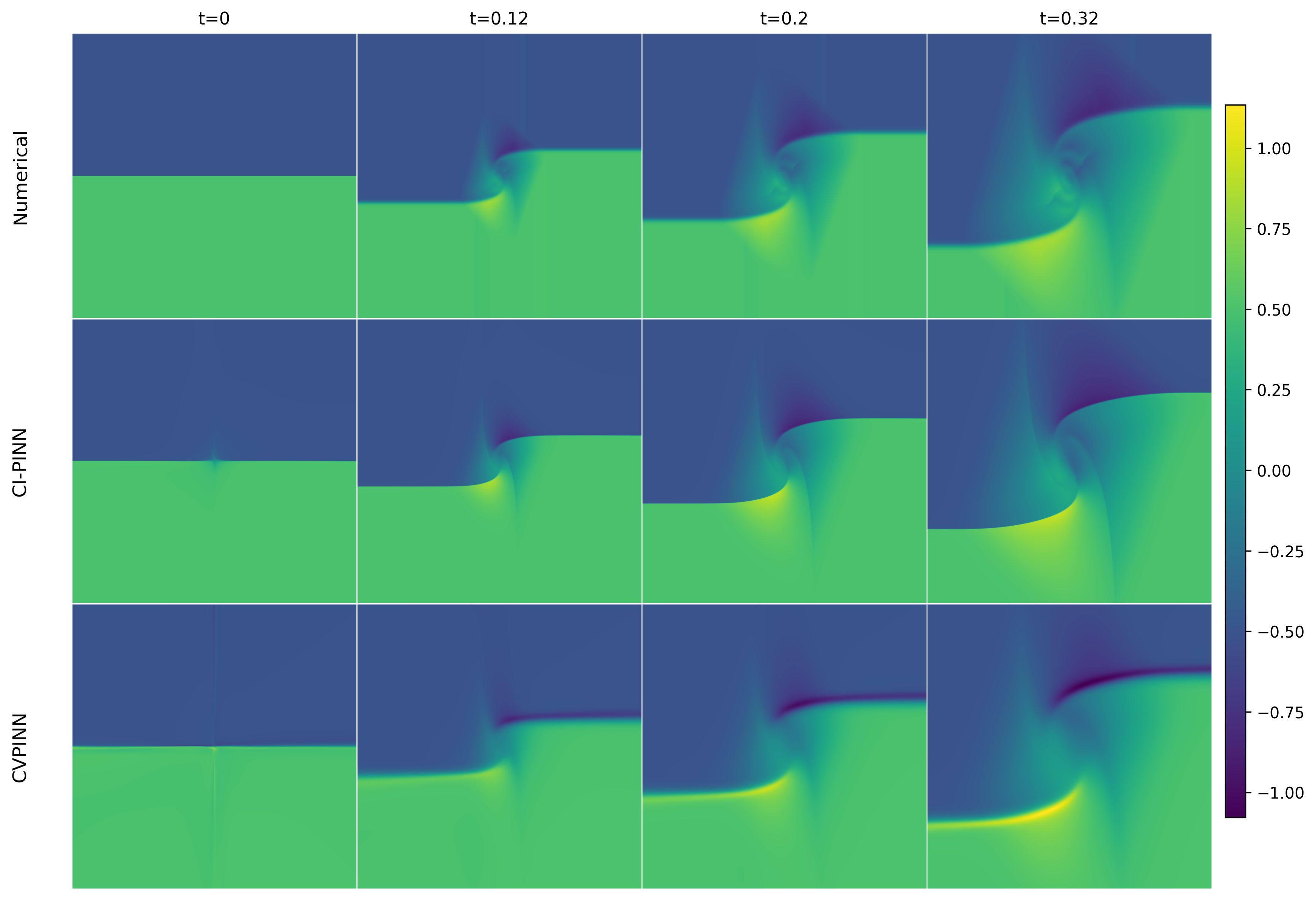}
        \caption{Velocity ($v$)}
    \end{subfigure}
    \hfill
    \begin{subfigure}[b]{0.48\textwidth}
        \centering
        \includegraphics[width=\textwidth]{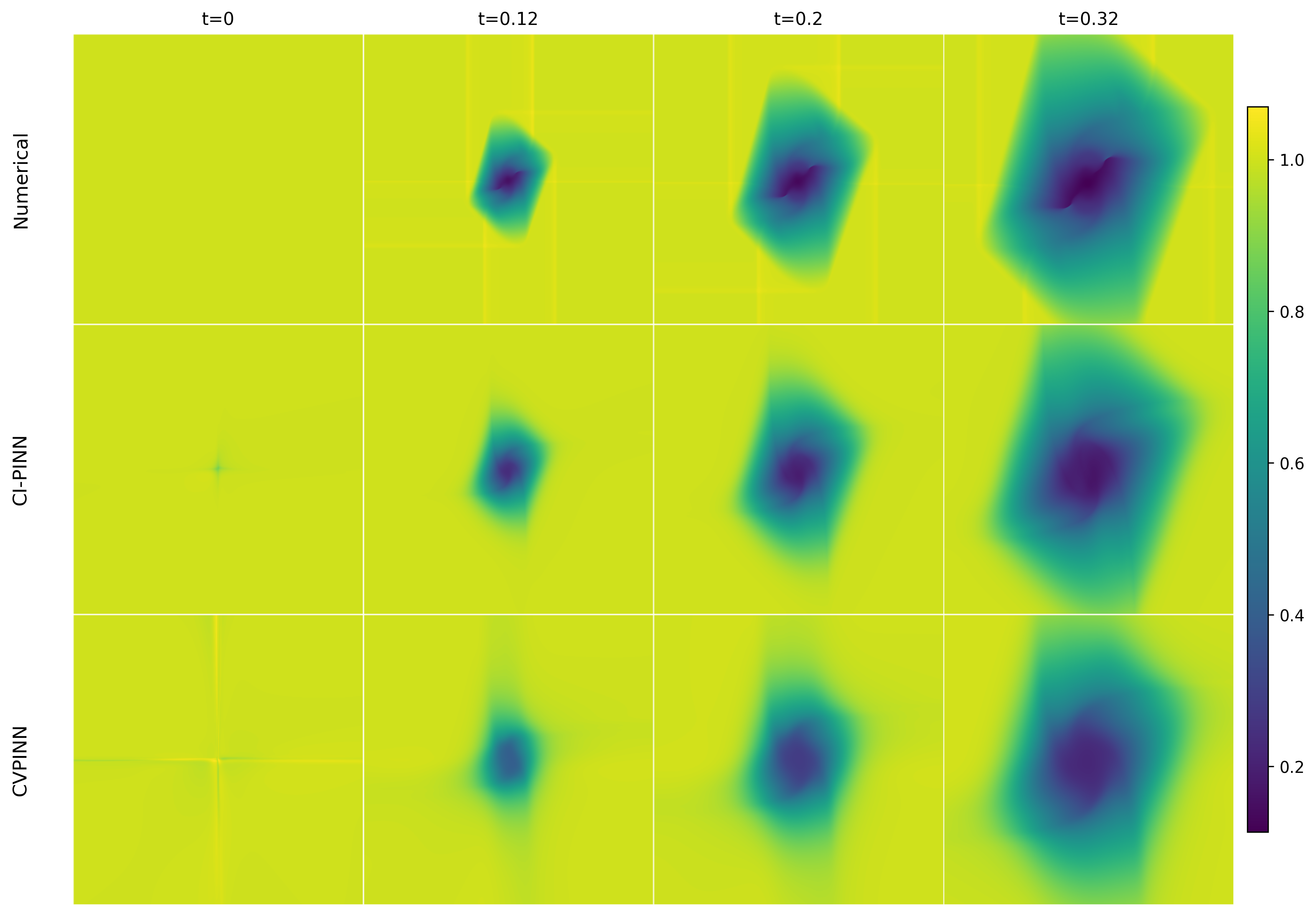}
        \caption{Pressure ($p$)}
    \end{subfigure}

    \caption{2D Euler system}
    \label{fig:ci_pinn_2d_results}
\end{figure}

\begin{figure}[H]
    \centering
    \begin{subfigure}[b]{0.48\textwidth}
        \centering
        \includegraphics[width=\textwidth]{SWE_h_Numerical_vs_CIPINN_vs_CVPINN_3x4.png}
        \caption{Density ($h$)}
    \end{subfigure}
    \hfill
    \begin{subfigure}[b]{0.48\textwidth}
        \centering
        \includegraphics[width=\textwidth]{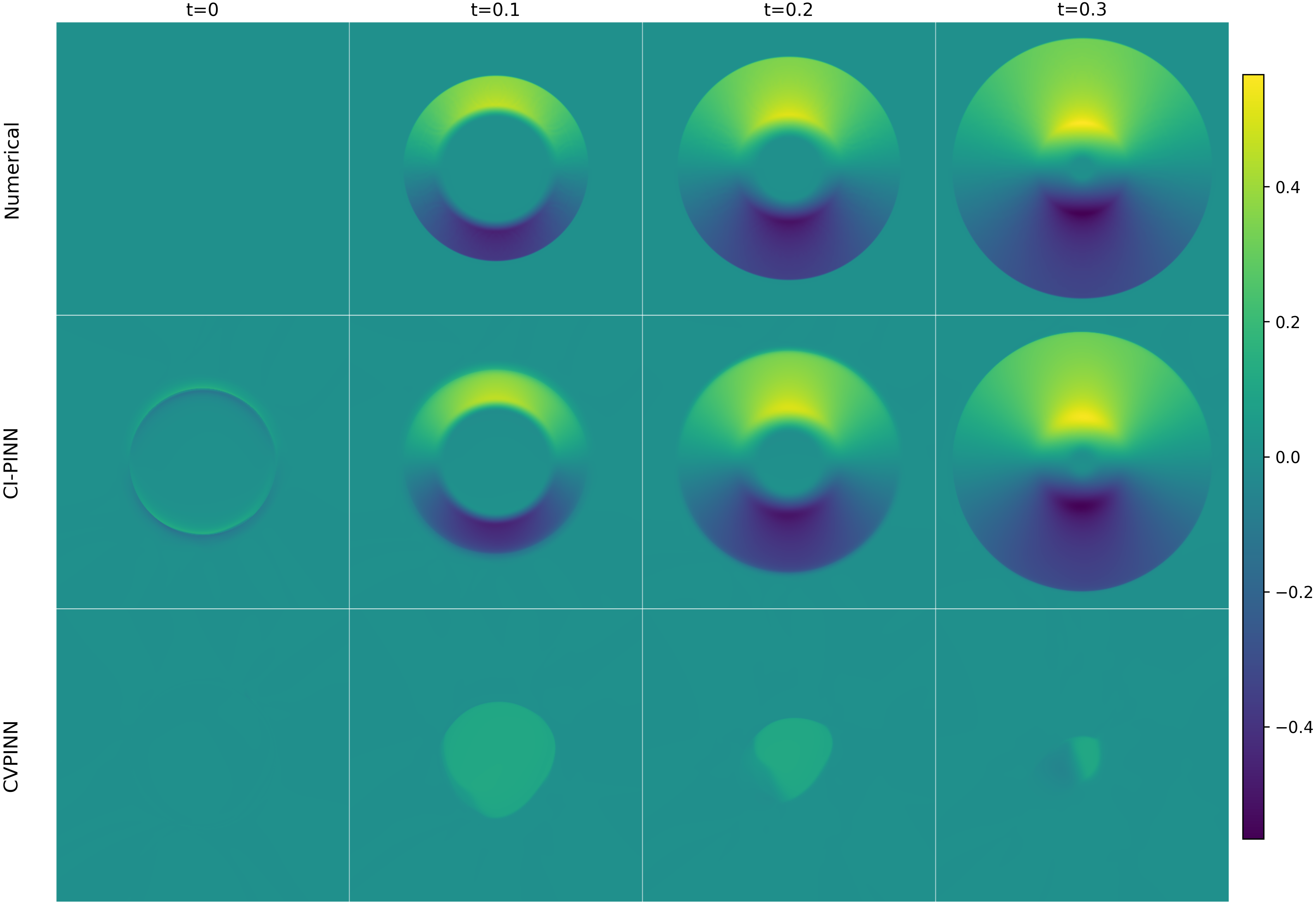}
        \caption{Velocity ($u$)}
    \end{subfigure}  
    \vspace{0.5cm} 
    \begin{subfigure}[b]{0.48\textwidth}
        \centering
        \includegraphics[width=\textwidth]{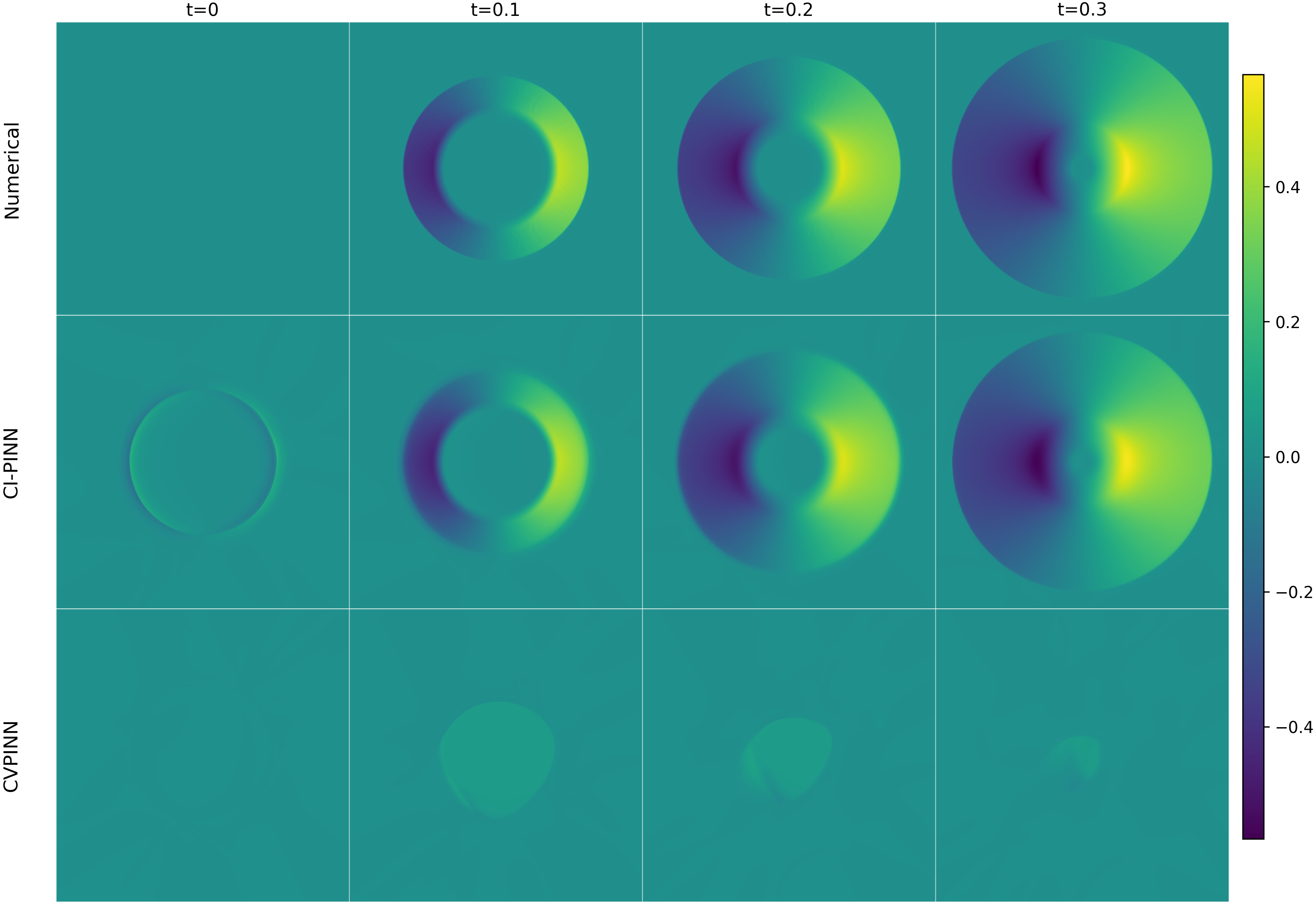}
        \caption{Velocity ($v$)}
    \end{subfigure}
    \hfill
    \caption{Shallow-Water Equations}
    \label{fig:ci_pinn_2d_results}
\end{figure}
\end{document}